\documentclass[12pt]{article}
\pdfoutput=1
\usepackage{amssymb,amsmath,dsfont,mathrsfs}
\usepackage{physics}
\usepackage{slashed}
\usepackage[normalem]{ulem}
\usepackage{epsfig}
\usepackage{epstopdf}
\usepackage{latexsym}
\usepackage{tikz}
\usepackage{tikz-cd, graphicx}
\usetikzlibrary{calc, shapes.geometric, positioning, arrows.meta, decorations.pathreplacing}
\usepackage{graphicx}
\usepackage{booktabs}
\usepackage{bbm}
\usepackage{xspace}
\usepackage{cancel}
\usepackage[utf8]{inputenc}
\usepackage{csquotes}
\usepackage{enumitem}
\usepackage{braket}
\usepackage[numbers,compress]{natbib}
\usepackage[T1]{fontenc}
\usepackage[margin=20pt,small]{caption}
\usepackage[toc]{appendix}
\usepackage{color}
\usepackage{float}
\usepackage{yfonts}
\usepackage{datetime}
\usepackage[normalem]{ulem}
\usepackage[
colorlinks=false,
linkcolor=darkblue,  
urlcolor=blue,    
filecolor=blue,     
citecolor=red,
linktocpage=true,
pdfstartview=FitV,
bookmarksopen=true,
pdfencoding=auto,
pagebackref]{hyperref}
\numberwithin{equation}{section}
\usepackage{anyfontsize}
\usepackage{subcaption}

\usepackage[scr=dutchcal]{mathalpha}

\definecolor{cardinal}{rgb}{0.6,0,0}
\definecolor{darkgreen}{rgb}{0,0.5,0}
\definecolor{golden}{rgb}{0.92, 0.7, 0}
\definecolor{midnight}{rgb}{0, 0, 0.5}
\definecolor{darkblue}{rgb}{0.2, 0, 0.8}

\usepackage[most]{tcolorbox}

\begin{document}  

	\begin{titlepage}
	\begin{flushright}
	SISSA  11/2025/FISI
	\end{flushright}
		\bigskip
		\begin{center} 
			{\fontsize{22pt}{0pt} \bf Symmetry extension by condensation defects \vskip 15pt in five-dimensional gauge theories}

			\vskip30pt

			{\large Matteo Bertolini$^{a,b}$, Lorenzo Di Pietro$^{b,c}$,\\
            \vskip 2mm
            Stefano C. Lanza$^{a,b}$, Pierluigi Niro$^{a,b}$, Antonio Santaniello$^{b,c}$\\}
			\bigskip
                    ${}^{a}$ SISSA, Via Bonomea 265, 34136 Trieste, Italy \\
                    \vskip 2mm
                    ${}^{b}$
            	INFN, Sezione di Trieste, Via Valerio 2, 34127 Trieste, Italy\\
                \vskip 2mm
           		${}^{c}$
            	Dipartimento di Fisica, Universit\`{a} di Trieste, Strada Costiera 11, 34151 Trieste, Italy\\
                    \vskip 5mm

			\texttt{bertmat@sissa.it,~ldipietro@units.it,\\
            slanza@sissa.it,~pniro@sissa.it,~antonio.santaniello@phd.units.it}\\
		\end{center}
		
		\bigskip
		\bigskip

        
		\begin{abstract}
\noindent 
We investigate the symmetry structure of five-dimensional Yang-Mills theories with $\mathfrak{su}(N)$ gauge algebra. These theories feature intertwined 0-, 1-, and 2-form symmetries, depending on the global variant one is considering. In the $SU(N)$ theory, there is a mixed 't~Hooft anomaly between the instantonic 0-form symmetry and the electric 1-form symmetry. We show that in the $PSU(N)$ theory this translates into a $\mathbb{Z}_N$ extension of the instantonic symmetry, generated by an invertible condensation defect of the magnetic 2-form symmetry. We identify the charged configurations as linked 't~Hooft surfaces, while pointlike instanton operators remain insensitive to the extension. We generalize our analysis to the $SU(N)/\mathbb{Z}_k$ global form and show that similar results hold, embedded now in a 3-group structure for generic $k$. We then apply our findings to $SO(3)$ supersymmetric Yang-Mills theory. We determine the global form of the enhanced instantonic symmetry of its superconformal UV completion, showing that it arises through a similar symmetry extension mechanism from the parent $E_1$ theory, which is the UV completion of $SU(2)$ supersymmetric Yang-Mills theory. Finally, we recast our results in the language of the symmetry topological field theory. As a warm-up, we also analyze Maxwell theory, highlighting analogous features involving continuous symmetries and composite currents.

		\end{abstract}

	\end{titlepage}
	
	
	\setcounter{tocdepth}{2}	

\tableofcontents

\section{Introduction and summary of results}
\label{Section:Intro}

Five-dimensional gauge theories provide us with a rich theoretical laboratory to explore several interesting phenomena in quantum field theory (QFT). They are distinguished from the lower-dimensional gauge theories by the property that the gauge coupling $g^2$ has negative mass dimension. As a result, these theories are weakly-coupled at low energies, and they should be interpreted as effective field theories that are only valid up to a UV cutoff $\sim 1/g^2$. To push their definition above these energy scales, some UV completion needs to be specified. In the supersymmetric case, starting from \cite{Seiberg:1996bd}, interacting five-dimensional superconformal field theories (SCFT) that UV complete gauge theories have been found in several cases. While they are well-studied, many aspects of these SCFTs remain unknown. In the non-supersymmetric case, a natural possibility is that a gauge theory could be UV completed by a yet-unknown five-dimensional non-supersymmetric interacting conformal field theory (CFT). For these reasons, five-dimensional gauge theories are natural candidates to look for examples of non-supersymmetric interacting CFTs in dimension greater than four, whose existence remains a basic open question in QFT (see~\cite{
Li:2020bnb,BenettiGenolini:2019zth, Bertolini:2021cew,Bertolini:2022osy,Akhond:2023vlb, DeCesare:2021pfb, DeCesare:2022obt, Florio:2021uoz} for recent progress in this quest, using a variety of approaches).

Symmetries are a basic and powerful non-perturbative tool to constrain the strong-coupling behavior of QFTs. When trying to make progress on the questions mentioned above, it is therefore crucial to understand the symmetries of five-dimensional gauge theories. A universal feature of gauge theories in five dimensions is the presence of a $U(1)^{(0)}_I$ instantonic 0-form symmetry,\footnote{Throughout the paper, we indicate a $p$-form symmetry with group $G$ as $G^{(p)}$.} with current $J_I \sim \star\,\mathrm{Tr}(f\wedge f)$, which is conserved due to the Bianchi identity. The pointlike operators charged under this symmetry are disorder operators, defined by prescribing a non-zero instanton charge on four-spheres linking with the insertion point: these are the so-called instanton operators. In addition, depending on the global form of the gauge group, five-dimensional gauge theories can enjoy electric 1-form symmetries, under which Wilson line operators are charged, or magnetic 2-form symmetries, under which 't~Hooft surface operators are charged, or both. These symmetries can have non-trivial interplays, for instance in the form of mixed ’t Hooft anomalies or higher-group structures. Given the possible existence of 0-,1-, and 2-form symmetries, the symmetry structures are often richer than in the lower-dimensional case, even in the simplest possible gauge theories such as Maxwell or pure Yang-Mills (YM). Exploring these structures, with the aim of discovering new phenomena that can be applied more generally, is another motivation to study these theories.

The symmetries of pure $SU(N)$ YM in five dimensions were studied in \cite{BenettiGenolini:2020doj, Gukov:2020btk, Genolini:2022mpi}. It was found that there is a mixed 't~Hooft anomaly between the $\mathbb{Z}_N^{(1)}$ electric 1-form symmetry and the $U(1)_I^{(0)}$ instantonic 0-form symmetry, with anomaly inflow given by
\begin{equation}\label{eq:BGTanointro} 
    \mathcal{A} =
    \exp \left( \frac{2\pi i}{N} \int 
    \left[\frac{dA_1}{2\pi}\right]_N
    \cup
    \frac{\mathcal{P}(\mathcal{B}_2)}{2} \right) \ ,
\end{equation}
where $\mathcal{B}_2$ and $A_1$ are background gauge fields for the 1- and 0-form symmetries, respectively, $\cup$ is the cup product in cohomology theory, $\mathcal{P}$ denotes a natural ``square'' operation on discrete gauge fields (more details are provided in section \ref{Section:Yang-Mills}), and throughout the paper we use $[\cdot]_p$ to denote reduction mod $p$. This anomaly expresses the fact that the normalization of the instanton charge gets smaller by a factor of $1/N$ in the presence of a background $\mathcal{B}_2$.\footnote{This is the $5d$ analogue of the extension of the periodicity of the theta angle in $4d$ $SU(N)$ YM, in the presence of a background field for the electric symmetry. This phenomenon can also be expressed in terms of an ``anomaly inflow in the space of coupling constants'' \cite{Cordova:2019jnf,Cordova:2019uob} that takes a form similar to \eqref{eq:BGTanointro}.} A notable application of this result is to the case of $SU(2)$ supersymmetric Yang-Mills (SYM). This theory was discovered in \cite{Seiberg:1996bd} to admit a UV completion by an SCFT, the so-called $E_1$ theory, in which $U(1)_I^{(0)}$ enhances to a non-Abelian symmetry with $\mathfrak{su}(2)_I$ algebra. Under the assumption that the $\mathbb{Z}_2^{(1)}$ is also a symmetry of the UV SCFT, \cite{BenettiGenolini:2020doj,Genolini:2022mpi} argued that the global form of the 0-form symmetry of the $E_1$ SCFT is $SO(3)_I^{(0)}$, because only with this global form one can write a UV anomaly inflow,
\begin{equation}\label{eq:UVanoint}
   \mathcal{A}^{\text{UV}} = \exp \left( \frac{2\pi i}{2} \int 
    w_2(SO(3)_I)
    \cup
    \frac{\mathcal{P}(\mathcal{B}_2)}{2} \right) 
    \ ,
\end{equation}
that matches \eqref{eq:BGTanointro} in the IR. Here $w_2$ is the second Stiefel-Whitney class of the $SO(3)$ bundle.\footnote{The symmetries as well as other aspects of the $E_1$ SCFT have been studied also in, e.g., \cite{Aharony:1997ju,Aharony:1997bh,DeWolfe:1999hj,Kim:2012gu, Bashkirov:2012re, Rodriguez-Gomez:2013dpa,Tachikawa:2015mha,Cremonesi:2015lsa,Closset:2018bjz,Apruzzi:2019vpe,Apruzzi:2021vcu,DelZotto:2022joo,Cvetic:2022imb,Closset:2023pmc,Acharya:2024bnt,Akhond:2024nyr}.}

In this paper we study other global forms of five-dimensional YM theory with $\mathfrak{su}(N)$ gauge algebra, with a particular focus on the case of $PSU(N) = SU(N)/\mathbb{Z}_N$. The latter can be obtained from $SU(N)$ YM by gauging the $\mathbb{Z}_N^{(1)}$ symmetry, i.e.~promoting $\mathcal{B}_2$ to a dynamical gauge field $\mathscr{b}_2$. This operation gives rise to a dual magnetic 2-form symmetry $\mathbb{Z}_N^{(2)}$, whose topological charge defects are generated by the surface operators
\begin{equation}\label{eq:Zn2fint}
U[\Sigma_2] = \exp \left( \frac{2\pi i}{N} \oint_{\Sigma_2} \mathscr{b}_2 \right) \ .
\end{equation}
The operators charged under this symmetry are 't~Hooft surfaces, labeled by $SU(N)$ representations, whose $N$-ality gives the $\mathbb{Z}_N^{(2)}$ charge. 

A natural question in $PSU(N)$ YM is: what is the consequence of the anomaly \eqref{eq:BGTanointro} after the gauging of $\mathbb{Z}_N^{(1)}$? One finds the short exact sequence
\begin{equation}
1 \to \mathbb{Z}_N \to \widetilde{U(1)}_I \to U(1)_I \to 1 \ ,
\end{equation}
expressing that the instantonic symmetry $\widetilde{U(1)}_I^{(0)}$ of $PSU(N)$ YM is a $\mathbb{Z}_N$ extension of the $U(1)_I^{(0)}$ symmetry of $SU(N)$ YM. This means that, due to the gauging procedure, the periodicity of the parameter $\alpha$ appearing in the symmetry generator $U_I^\alpha = \exp \left(i \alpha \oint \star J_I\right)$ is extended from $\alpha \sim \alpha + 2\pi$ to $\alpha \sim \alpha + 2\pi N$ in $PSU(N)$. While this is a clear consequence of the change of normalization in the instanton charge in the presence of $\mathscr{b}_2$, it cannot be the whole story. There must exist some structure relating the 0-form and the 2-form symmetry in $PSU(N)$, reflecting the existence of the 't~Hooft anomaly in the parent $SU(N)$ theory. The key to uncovering this structure is the expression of the generator of the $\mathbb{Z}_N^{(0)}$ subgroup that extends the instantonic symmetry, namely
\begin{equation}\label{eq:ZNextint}
\mathcal{C}[\Sigma_4] = \exp\left(-\frac{2\pi i}{N}
    \oint_{\Sigma_4} \frac{\mathcal{P}(\mathscr{b}_2)}{2}\right)~.
\end{equation}
Intuitively, a relation between this $\mathbb{Z}_N^{(0)}$ and $\mathbb{Z}_N^{(2)}$ is clear from the fact that the operator in the exponent of \eqref{eq:ZNextint} is the ``square'' of the operator in the exponent of \eqref{eq:Zn2fint}. We make this intuition precise by showing that $\mathcal{C}[\Sigma_4]$ is actually a {\it condensation defect} \cite{Roumpedakis:2022aik} for the magnetic 2-form symmetry, with the inclusion of a discrete torsion term that is crucial to obtain an invertible defect. The realization that $\mathcal{C}[\Sigma_4]$ is a condensation defect has important consequences for the type of objects charged under the instantonic symmetry. It implies that pointlike instanton operators are not charged under the symmetry extension. As a result, their instanton charges are quantized in the same units in $SU(N)$ and $PSU(N)$ YM. We show that the configurations charged under $\mathcal{C}[\Sigma_4]$ are links of 't~Hooft surface operators. These are the configurations that are sensitive to the extension, i.e.~they carry fractional instanton charge, in a sense that we specify.\footnote{The fact that magnetically charged objects can source an instanton charge was also discussed recently in \cite{Garcia-Valdecasas:2024cqn,GarciaGarcia:2025uub}.}

We claim that this symmetry extension, realized via a condensation defect of the dual symmetry, is precisely the structure that arises from the 't~Hooft anomaly \eqref{eq:BGTanointro} upon gauging. To our knowledge, this symmetry structure is distinct from other well-understood possibilities for duals of mixed 't~Hooft anomalies, such as standard extensions, higher-groups, or non-invertible symmetries \cite{Cordova:2019jnf, Cordova:2019uob,  Gaiotto:2017yup, Bhardwaj:2017xup,  Tachikawa:2017gyf, Bergman:2020ifi, Kaidi:2021xfk}.\footnote{In the context of $4d$ gauge theories, an anomaly inflow very similar to \eqref{eq:BGTanointro} is well-known to give rise to a non-invertible symmetry upon gauging the 1-form symmetry \cite{Kaidi:2021xfk}. The crucial difference is the presence in \eqref{eq:BGTanointro} of an additional derivative on the 0-form background gauge field.} This particular kind of extension can be seen as a generalization to higher-form symmetries of the duals of nonstandard anomalies analyzed in section 4 of \cite{Tachikawa:2017gyf} for 0-form symmetries.
A full characterization of this phenomenon would require an embedding in a specific higher-categorical symmetry structure, which however we do not explore in this paper.

We also consider the more general case of the global form $SU(N)/\mathbb{Z}_k$, with $k$ a divisor of $N$, obtained from $SU(N)$ by gauging a $\mathbb{Z}^{(1)}_k$ subgroup of $\mathbb{Z}^{(1)}_N$. This theory has both an electric $\mathbb{Z}^{(1)}_{N/k}$ symmetry and a magnetic $\mathbb{Z}^{(2)}_k$ one. In this case we find an extension of the 0-form symmetry when $N$ is not a multiple of $k^2$, and an anomalous 3-group, involving also the 1- and 2-form symmetries, when $k$ and $N/k$ are not coprime. The presence of an anomalous 3-group was noted previously in \cite{Gukov:2020btk}.

As a concrete example of the use of symmetries to understand UV behaviors, we apply our results to $PSU(2)=SO(3)$ SYM theory. Under the assumption that the $E_1$ SCFT that UV completes $SU(2)$ SYM retains the $\mathbb{Z}^{(1)}_2$ 1-form symmetry, with anomaly \eqref{eq:UVanoint}, the UV completion of $SO(3)$ SYM can be obtained from $E_1$ by gauging $\mathbb{Z}^{(1)}_2$. The resulting SCFT, that we denote with $\widehat{E}_1$, has the dual 2-form symmetry $\mathbb{Z}^{(2)}_2$, that in the IR becomes the magnetic 2-form symmetry of $SO(3)$ SYM. One can use the same logic used for the gauge theory to derive that the instantonic symmetries of $E_1$ and $\widehat{E}_1$ are in the short exact sequence
\begin{equation}
1 \to \mathbb{Z}_2 \to SU(2)_I \to SO(3)_I \to 1~.
\end{equation}
This is also required by consistency with the gauge theory result upon relevant deformation. The global form of the instantonic symmetry in $\widehat{E}_1$ is thus $SU(2)_I^{(0)}$. Also in this case the $\mathbb{Z}_2^{(0)}$ symmetry that extends the instantonic symmetry is a condensation defect of the 2-form symmetry, implying that only configurations of linking surface operators are charged (in an appropriate sense that we explain) under the symmetry extension.

The different global forms of $\mathfrak{su}(N)$ YM can be neatly organized in a unified framework using the language of the symmetry topological field theory (SymTFT) \cite{Freed:2012bs, Gaiotto:2020iye, Apruzzi:2021nmk, Freed:2022qnc}. We discuss in details the form of the resulting 6d TFT, and use it to provide an alternative derivation of the symmetry structure with gauge group $PSU(N)$, and more generally $SU(N)/\mathbb{Z}_k$. The same is done for the UV-completion of the $\mathfrak{su}(2)$ SYM theory discussed above, providing the SymTFT describing the $E_1$ and $\widehat{E}_1$ theories.

As a warm-up for the analysis of non-Abelian gauge theories, we start by studying the symmetry structure of five-dimensional Maxwell theory, which is rather rich and interesting on its own (see e.g.~\cite{Brauner:2020rtz,Grassi:2025tfp} for related work). Some of the salient features that we mentioned above, such as an instantonic symmetry under which the only ``charged'' operators are links of 't~Hooft surfaces, and an anomalous 3-group structure, are realized in this example, only with continuous rather than discrete symmetries. The instantonic symmetry in this case takes the form of a so-called {\it composite symmetry} \cite{Brauner:2020rtz,Heidenreich:2020pkc}, i.e.~its current is obtained from the square of the 2-form symmetry current. We show that this is tightly related to the fact that the instantonic charge defect can also be viewed as a condensation defect of the 2-form symmetry, like in the non-Abelian case.

The remainder of the paper is organized as follows. In section \ref{Section:Maxwell} we discuss the symmetry structure of five-dimensional Maxwell theory. We then move to $\mathfrak{su}(N)$ YM in section \ref{Section:Yang-Mills}, discuss its different global forms, and apply our findings to the UV completion of $SO(3)$ SYM. Finally, in section \ref{Section:SymTFT}, we rederive our main results in the language of the SymTFT. We have included some technical details in an appendix.

\section{Five-dimensional Maxwell Theory}
\label{Section:Maxwell}

We consider five-dimensional Maxwell theory with action
\begin{equation}
S = \int_{\mathcal{M}_5} \frac{1}{2e^2} f \wedge \star f \ ,     
\end{equation}
where $f=da$ is the field strength of a $U(1)$ connection $a$ subject to gauge transformations $a \rightarrow a + d\chi$, with $\oint d\chi \in 2\pi\mathbb{Z}$,  $\mathcal{M}_5$ is a five-dimensional manifold, and $e^2$ is the gauge coupling, which has dimension of a length.  Note that we are not including any Chern-Simons term.\footnote{For a discussion on the effects of the Chern-Simons term in 5$d$ Maxwell theory see e.g.~\cite{Damia:2022bcd,Arbalestrier:2025poq}.} 

The theory has three continuous global symmetries: a $U(1)^{(2)}_M$ magnetic 2-form symmetry, a $U(1)^{(1)}_E$ electric 1-form symmetry, and a $U(1)^{(0)}_I$ instantonic 0-form symmetry.
Their currents $J^{(p+1)}$, which we normalize so that the corresponding charges $\oint_{\Sigma_{d-p-1}}\star J^{(p+1)}$ are integers, are given by\footnote{Here and elsewhere, we adopt Euclidean signature and we assume for simplicity that the spacetime manifold $\mathcal{M}_5$ is orientable and spin. Charges associated to 0-form symmetries are defined on orientable four-cycles $\Sigma_4\subset\mathcal{M}_5$, which -- under these assumptions -- are necessarily closed spin (sub)manifolds \cite{MathStack}. This fixes the normalization of the $J_I^{(1)}$ instantonic current.\label{footnotespin}}
\begin{equation}\label{Maxwell_currents}
\star J^{(3)}_M = \frac{1}{2\pi} f \ , \qquad
\star J^{(2)}_E = \frac{i}{e^2} \star f \ , \qquad
\star J^{(1)}_I = \frac{1}{8\pi^2}  f \wedge f  \ .
\end{equation}
The $U(1)$ electric and magnetic symmetry defects are topological operators defined on orientable closed cycles $\Sigma_i$, 
\begin{equation}
U_E^{\alpha_E} [\Sigma_3] = \exp\left(i\alpha_E \oint_{\Sigma_3} \star J^{(2)}_E\right) \ , \  U_M^{\alpha_M} [\Sigma_2] = \exp\left(i\alpha_M \oint_{\Sigma_2} \star J^{(3)}_M\right)  \ , \hspace{3mm} \alpha_{E,M} \sim \alpha_{E,M}+ 2\pi \ .
\end{equation}
They measure the integer charges of (non-topological) Wilson lines and 't~Hooft surfaces, respectively,
\begin{equation}
\label{W&tH}
W^n[\gamma_1] = \exp\left(i n \oint_{\gamma_1} a \right)  \ , \qquad  H^m[\gamma_2] = \exp\left(i m \oint_{\gamma_2} \tilde{a} \right) \ , \qquad m,n \in \mathbb{Z} \ .
\end{equation}
Here, $\gamma_k$ are orientable closed $k$-cycles, and $\tilde{a}$ is the dual 2-form gauge field, $d\tilde{a}= \frac{2\pi i}{e^2} \star da$.\footnote{Equivalently, we can define 't~Hooft surfaces on $\gamma_2$ as disorder operators. Insertions of one such operator in any correlation function is equivalent to imposing in the path integral the boundary condition $\oint da=2\pi m$, on a two-cycle $\Sigma_2$ with unit linking with the surface $\gamma_2$.} In particular, the charges of $W^n$ and $H^m$ are detected thanks to the relations\footnote{These relations hold in any correlation function as long as the other operator insertions do not link or intersect $\Sigma_i$. This also applies below, when discussing similar relations involving symmetry defects and charged operators.}
\begin{align}
\begin{split}
U_E^{\alpha_E} [\Sigma_3] W^n[\gamma_1]  &= \exp\left(i\alpha_E \,n L_2(\Sigma_3,\gamma_1)\right)  W^n[\gamma_1] \ , \\
U_M^{\alpha_M} [\Sigma_2] H^m[\gamma_2]  &= \exp\left(i\alpha_M m L_2(\Sigma_2,\gamma_2)\right) H^m[\gamma_2] \ .
\label{Maxwell_chargemeasuring}
\end{split}
\end{align}
The integer $L_2$ is the two-component linking (double-linking for short), defined, on a generic spacetime $\mathcal{M}_d$, as 
\begin{equation}
L_2 (\Sigma_{q},\gamma_{r}) \equiv \int_{\mathcal{M}_d} \delta^{(d-q-1)}(D_{q+1}) \wedge d\delta^{(d-r-1)}(M_{r+1}) \ , \qquad q+r=d-1 \ .
\label{doublelink}
\end{equation}
Here, $D_{q+1}$ and $M_{r+1}$ are the Seifert surfaces of the closed manifolds $\Sigma_{q}$ and $\gamma_{r}$, respectively, namely compact, connected, and oriented surfaces such that $\partial D_{q+1} = \Sigma_{q}$ and $\partial M_{r+1} = \gamma_{r}$. 
The delta form $\delta(\cdot)$ is the Poincaré dual of the corresponding argument.\footnote{We define $\int_{X_p}\omega_p=\int_{\mathcal{M}_d}\omega_p \wedge \delta^{(d-p)}(X_p)$, such that $d\delta^{(d-p)}(X_p)=(-1)^p\delta^{(d-p+1)}(\partial X_p)$, by Stokes theorem.}  $L_2$ can be also interpreted as the intersection number between $D_{q+1}$ and $\gamma_{r}$, that counts their (oriented) intersection points.
Integrating by parts, we have the following symmetry property of the double-linking,
\begin{equation}
L_2 (\Sigma_{q},\gamma_{r}) = (-1)^{d(q+1)}L_2 (\gamma_{r},\Sigma_{q}) \ .
\end{equation}
Notice that this implies that the double-linking between two-surfaces in 5$d$ is antisymmetric, and thus self-statistics is trivial in 5$d$ \cite{Chen:2021xuc}.

The instantonic symmetry, whose symmetry defects are 
\begin{equation}
\label{symdef_U}
U^{\alpha}_I[\Sigma_4] = \exp \left( i\alpha\oint_{\Sigma_4} \star J^{(1)}_I \right)
= \exp \left( \frac{i\alpha}{8\pi^2} \oint_{\Sigma_4} f \wedge f \right) \ , \qquad \alpha \sim \alpha + 2\pi \ ,
\end{equation}
does not act on pointlike operators, simply because there are no $U(1)$ instantons on $S^4$, namely $U_I^\alpha[S^4]=\mathds{1}$.
Despite the absence of charged pointlike operators, the $U(1)_I^{(0)}$ symmetry acts on extended configurations composed of 't~Hooft surfaces (charged under the magnetic symmetry) that can be wrapped by a four-cycle $\Sigma_4$ where
\begin{equation}
\frac{1}{8\pi^2} \oint_{\Sigma_4} f \wedge f \neq 0 \ .
\label{u1instnumb}
\end{equation}
This peculiarity is due to the fact that the instantonic symmetry can be interpreted as a special case of a \textit{composite symmetry}\footnote{Given a $d$-dimensional free theory with continuous $p_1$- and $p_2$-form symmetries, with currents $J^{(p_1+1)}$ and $J^{(p_2+1)}$, one can construct a $(p_1+p_2+1-d)$-form symmetry whose current is proportional to $\star(\star J^{(p_1+1)} \wedge \star J^{(p_2+1)})$. The composite symmetry arises when $p_1+p_2 \geq d-1$. In $d$-dimensional Maxwell theory, with $d \geq 3$, there are an electric 1-form and a magnetic $(d-3)$-form $U(1)$ symmetries. The composite of the magnetic symmetry with itself appears for $d\geq 5$, and it is a $(d-5)$-form symmetry with current $\star J^{(d-4)} \sim f \wedge f$, which is precisely $U(1)_I^{(0)}$ in 5$d$. Note that, while in free theories the operation of multiplying two composite operators at the same point is always well-defined, because one can use normal ordering, this operation is not defined in general in interacting theories. In the latter case, the closest operation is an operator product expansion (OPE). It would be interesting to explore whether one can always find a non-singular term in the OPE of conserved currents that allows to define composite symmetries even in interacting theories. For instance, one can try to use the OPE in interacting CFTs. In this case, however, one would have to relax unitarity, because conformal invariance together with unitarity only allows $p$-form continuous symmetries if $p\leq \frac{d}{2}-1$, forbidding composite symmetries.}  of the magnetic symmetry with itself \cite{Brauner:2020rtz,Heidenreich:2020pkc}, since $\star J^{(1)}_I = \frac{1}{2} \star J^{(3)}_M \wedge \star J^{(3)}_M$. This implies that certain configurations of the operators that are charged under $U(1)^{(2)}_M$ will carry ``charge'' (in a specific sense that we will explain below) under $U(1)^{(0)}_I$ (see also \cite{Gukov:2020btk,Nakajima:2022feg,Grassi:2025tfp}). Another possibility to activate the defect of the $U(1)^{(0)}_I$ symmetry is to consider a space with non-trivial topology, such that $U(1)$ instantons can be supported on some codimension-one submanifolds, e.g.~$S^1 \times S^2 \times S^2$. In the following, we concentrate on configurations of extended operators in backgrounds with trivial topology.

\subsection{Action of the instantonic symmetry and triple-linking} 
\label{subsec:Max-Action-Cond}

To understand the action of the instantonic symmetry on 't~Hooft surfaces, let us consider two 't~Hooft surfaces $H^m[\gamma_2]$ and $H^{m'}[\gamma_2']$. To express these operators in terms of electric variables, it is convenient to extend them on three-surfaces $M_3$ and $M_3'$ such that $\partial M_3=\gamma_2$ and $\partial M_3'=\gamma_2'$, so that one can rewrite \eqref{W&tH} for the 't~Hooft surface as
\begin{equation}
H^m[\gamma_2] = \exp\left(-\frac{2\pi m}{e^2} \int_{M_3} \star f \right) \ , \quad 
H^{m'}[\gamma_2'] = \exp\left(-\frac{2\pi m'}{e^2} \int_{M_3'} \star f \right) \ .
\end{equation}
Notice that these operators are genuine because $m,m'\in\mathbb{Z}$ and the quantization of the electric charge imply that they depend only on the boundaries of $M_3$ and $M_3'$. Inserting these operators in the action, we get that the modified equations of motion are solved by
\begin{equation}\label{eq:tripMax}
f = 2\pi \left( m \, \delta^{(2)}(M_3) + m' \, \delta^{(2)}(M_3') \right) + \text{ (smooth) } \ ,
\end{equation}
corresponding to the fact that inserting 't~Hooft surface operators activates a magnetic flux. We then get
\begin{equation}
U^\alpha_I[\Sigma_4] H^m[\gamma_2] H^{m'}[\gamma_2'] = 
\exp\left(i\alpha m m' \, L_3(\Sigma_4,\gamma_2,\gamma_2') \right)
H^m[\gamma_2] H^{m'}[\gamma_2'] \ .
\label{Maxwell_triplelinking}
\end{equation}
The integer $L_3$ is a type-1 three-component linking (triple-linking for short),\footnote{Generically, it is possible to define three different three-component linking numbers, referred to as type-$k$, with $k=0,1,2$, between three closed manifolds with dimensions $p$, $q$, and $r$. They satisfy $p+q+r=2d-3+k$, see e.g.~appendix A of \cite{Kaidi:2023maf}.} defined as
\begin{align}
\begin{split}
L_3&\left(\Sigma_p,\gamma_q,\gamma_r'\right) = \int_{\mathcal{M}_d} \delta^{(d-q-1)}(M_{q+1}) \wedge \delta^{(d-r-1)}(M_{r+1}') \wedge d \delta^{(d-p-1)} (D_{p+1}) \\
&=(-1)^{p+1} \oint_{\Sigma_p} \delta^{(d-q-1)}(M_{q+1}) \wedge \delta^{(d-r-1)}(M_{r+1}') \ ,  \quad \text{with }\,p+q+r=2d-2 \ ,  
\end{split}
\label{triplelink}
\end{align}
where $\partial D_{p+1} = \Sigma_p$, $\partial M_{q+1}=\gamma_q$, and $\partial M'_{r+1}=\gamma'_r$. The triple-linking formula can be expressed in terms of the double-linking formula for instance as
\begin{equation}\label{eq:tl=dl}
L_3(\Sigma_p,\gamma_q,\gamma_r')=(-1)^{d-r-1} L_2(\gamma_q \cap M'_{r+1},\Sigma_p)+L_2(M_{q+1}\cap \gamma'_r, \Sigma_p) \ .
\end{equation}
The relation \eqref{Maxwell_triplelinking} implies that $U(1)_I^{(0)}$ acts on a generic configuration of 't~Hooft surfaces that activates $L_3(\Sigma_4,\gamma_2,\gamma_2')$. The integer instanton charge $mm'$ of these configurations is measured by a $U(1)_I^{(0)}$ symmetry defect supported on a suitable $\Sigma_4$ -- triple-linking with $\gamma_2$ and $\gamma_2'$ -- where \eqref{u1instnumb} is indeed non-trivial.\footnote{For $N$ 't~Hooft surfaces of magnetic charges $m_i$ supported on $\gamma_2^i$, the factor in the parenthesis of \eqref{Maxwell_triplelinking} generalizes to $\sum_{i\neq j}^N m_i m_j L_3(\Sigma_4,\gamma_2^i,\gamma_2^j)$.} Integrating by parts, we can write the triple-linking in \eqref{Maxwell_triplelinking} as
\begin{align}
\begin{split}
L_3(\Sigma_4,\gamma_2,\gamma_2') &=  - \oint_{\Sigma_4} \delta^{(2)}(M_3) \wedge \delta^{(2)}(M_3') \\
&= - \int_{D_5} \delta^{(2)}(M_3) \wedge d\delta^{(2)}(M_3')
- \int_{D_5}\delta^{(2)}(M_3') \wedge d\delta^{(2)}(M_3) \ .
\label{Maxwell_alternativetlk}
\end{split}
\end{align}
It follows from this formula that a necessary (but not sufficient) condition to activate the triple-linking is that the two 't~Hooft surfaces are in a geometric configuration such that they are double-linking in the sense of \eqref{doublelink}. (Notice that this is possible for two 2-surfaces in 5$d$, as they satisfy $q+r=d-1=4$.)
The surface $\Sigma_4$ is codimension-one and separates spacetime into two connected components, which we refer to as the interior and the exterior, where the former corresponds to $D_5$. To gain a geometric intuition of the triple-linking in \eqref{Maxwell_alternativetlk}, we now discuss some typical configurations of the instantonic symmetry defect with support $\Sigma_4$ and the two double-linking 't~Hooft surfaces with support $\gamma_2$ and $\gamma_2'$. 

\begin{figure}[t]
	\centering
	\includegraphics[width=0.5\textwidth]{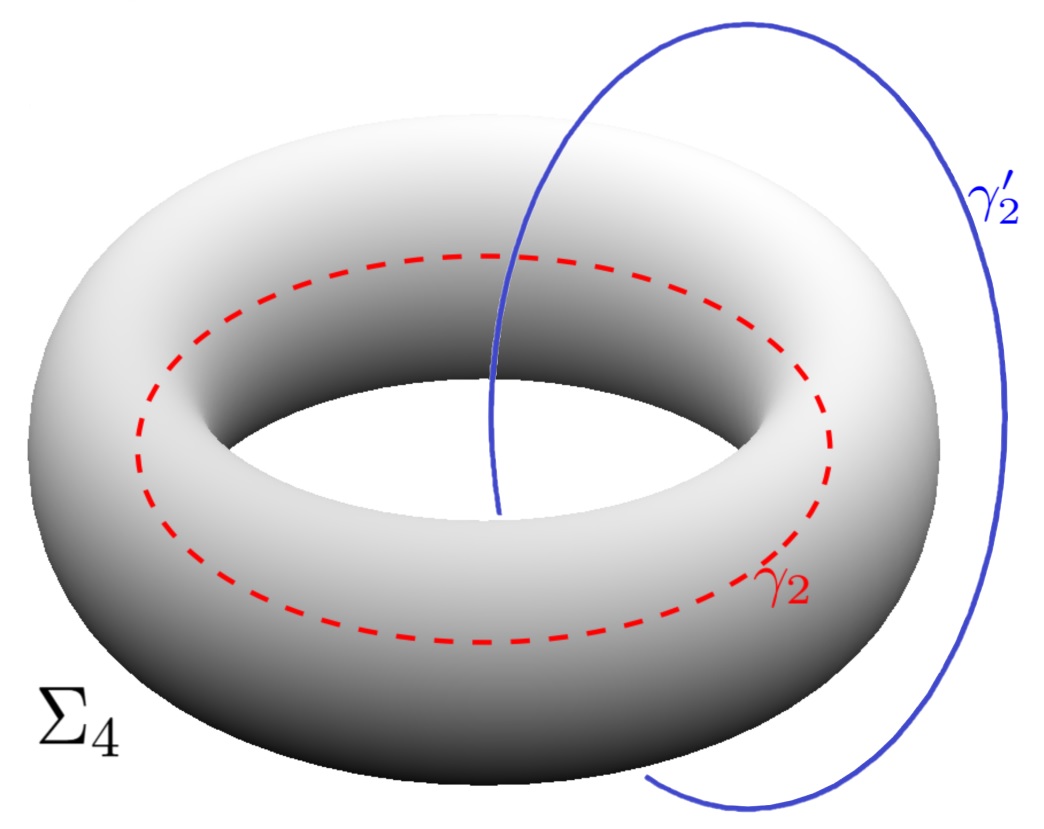}
	\caption{Three-dimensional depiction of a configuration activating the triple-linking in \eqref{Maxwell_alternativetlk}, between the instantonic operator defined on $\Sigma_4$ (2-torus in the figure) and two double-linking 't~Hooft surfaces $\gamma_2$ and $\gamma_2'$ (circles in the figure). The surface $\gamma_2$ (dashed red in the figure) is inside $\Sigma_4$, while the surface $\gamma_2'$ (thick blue in the figure) is outside $\Sigma_4$. The simplest five-dimensional configuration with triple-linking is the one in which the circles in the figure are interpreted as two-spheres, i.e.~$\gamma_2$ and $\gamma_2'$ are linked two-spheres, and $\Sigma_4=S^2\times S^2$.}
	\label{fig:triplelinking}
\end{figure}

\begin{enumerate}
    \item[1)] Consider the cases where both $\gamma_2$ and $\gamma_2'$ are either outside or inside $\Sigma_4$, so that $M_3$ and $M_3'$ can be chosen to avoid intersecting $\Sigma_4$. Then, this geometric configuration does not activate the linking in \eqref{Maxwell_alternativetlk}, as the integral over $\Sigma_4$ identically vanishes. In the outside case, this is obvious as the topological surface $\Sigma_4$ can be shrunk to zero. In the inside case, this follows from the fact that the topology of $\Sigma_4$ is irrelevant and can hence be chosen to be the sphere $S^4$. As the 't~Hooft surfaces are inside $S^4$, the latter can be taken to be at infinity in spacetime without crossing any operator, so that $D_5$ is actually the whole spacetime manifold $\mathcal{M}_5$. In this case, the formula \eqref{Maxwell_alternativetlk} reads
    \begin{equation}
    L_3(\Sigma_4,\gamma_2,\gamma_2') = - L_2(\gamma_2,\gamma_2') - L_2(\gamma_2',\gamma_2) = 0 \ ,  
    \end{equation}
    as $L_2(\gamma_2,\gamma_2')$ defined in \eqref{doublelink} is antisymmetric in its arguments. This result has a clear interpretation in terms of absence of pointlike operators charged under $U(1)_I$: since there are no $U(1)$ instantons on $S^4$, one cannot measure a non-trivial instanton charge in any configuration where $\Sigma_4$ can be taken to be $S^4$.

    Notice that these scenarios can be reformulated in terms of the two intersection points $P$ (between $\gamma_2$ and $M_3'$) and $P'$ (between $\gamma_2'$ and $M_3$), defined by the two double-linking 't~Hooft surfaces. In the first case, $P$ and $P'$ are both outside $\Sigma_4$ and each integral on the right-hand side of \eqref{Maxwell_alternativetlk} separately vanishes. In the second case, $P$ and $P'$ are both inside $\Sigma_4$ and the two integrals, despite being non-vanishing, sum up to zero.

    \item[2)] Consider now the case where $\gamma_2$ lies inside $\Sigma_4$ and $\gamma_2'$ lies outside $\Sigma_4$, such that both $M_3$ and $M_3'$ must intersect $\Sigma_4$. In this case, the integral \eqref{Maxwell_alternativetlk} is non-vanishing if $\gamma_2$ and $\gamma_2'$ have a non-trivial double-linking. In terms of the two intersection points $P$ and $P'$ defined above, we have that $P$ is inside and $P'$ is outside $\Sigma_4$, so that the first integral on the right-hand side of \eqref{Maxwell_alternativetlk} vanishes while the second is 
    non-zero. This is a typical configuration where the triple-linking is activated. Notice that, as expected, this configuration cannot be realized if $\Sigma_4$ has a trivial topology (namely, we need $\Sigma_4 \neq S^4$). In figure \ref{fig:triplelinking} we depict a 3$d$ version of this triple-linking configuration. The codimension-one $\Sigma_4$ is depicted as a 2$d$ torus, while the supports of the 't~Hooft surfaces $\gamma_2$ and $\gamma_2'$ are depicted as lines with a non-trivial double-linking, see \eqref{doublelink}. These three objects have the correct dimensionality such that they can triple-link in 3$d$, as in \eqref{triplelink}.
    
\end{enumerate}

In the previous discussion, we considered the relation \eqref{Maxwell_triplelinking} where we discarded potentially divergent terms of the form $m^2 L_3(\Sigma_4,\gamma_2,\gamma_2)$ and $m'^2 L_3(\Sigma_4,\gamma'_2,\gamma'_2)$, that arise when computing the action of the instantonic symmetry defect on 't~Hooft surfaces. The reason is that such terms vanish, once they are regularized.\footnote{In the more familiar case of 3$d$ TQFTs, as Chern-Simons theories, the regularized self-linking does not vanish and it is related to the choice of framing for the line operators, that leads to the definition of their spin.} A simple argument for this is that the regularization of $L_3(\Sigma_4,\gamma_2,\gamma_2)$ can be heuristically thought of as $L_3(\Sigma_4,\gamma_2,\gamma^\varepsilon_2)$, where $\gamma^\varepsilon_2$ is a small deformation of $\gamma_2$. Then, the relation \eqref{Maxwell_alternativetlk} instructs us to consider two intersections points, the one between $\gamma_2$ and $M_3^\varepsilon$, and the one between $\gamma_2^\varepsilon$ and $M_3$. These two points lie either both inside or both outside $\Sigma_4$ and, using the same arguments as in point 1) above, give a total vanishing result for $L_3(\Sigma_4,\gamma_2,\gamma^\varepsilon_2)$. Equivalently, the relation \eqref{Maxwell_alternativetlk} instructs us to count (with orientation) the intersection points between $\Sigma_4$ and the line obtained by intersecting $M_3$ and $M_3^\varepsilon$. For each such intersection point (when it exists), there is always another one that is oppositely oriented. The net result is again zero.

\subsubsection{Instantonic symmetry defects by topological manipulations}
\label{cond_defect}

We now show that the instantonic symmetry defect $U_I^\alpha[\Sigma_4]$ can be obtained by topological manipulations.\footnote{Defects defined with similar manipulations of continuous symmetries have been studied in \cite{Arbalestrier:2024oqg} in the context of the $U(1)^{(0)}$ axial symmetry of 4$d$ QED.}
These topological manipulations correspond to higher $p$-gauging, namely to the gauging of a $q$-form symmetry on a codimension-$p$ manifold of spacetime, first introduced in \cite{Roumpedakis:2022aik}. In contrast to 0-gauging (i.e.~ordinary gauging) that changes the theory, this operation does not affect the bulk dynamics; instead, it defines codimension-$p$ topological defects.

In the case of Maxwell theory, we consider the codimension-one defect defined by
\begin{equation}
\mathcal{D}[\alpha;\Sigma_4] \equiv \int \mathcal{D}\Phi_2 \mathcal{D}\varphi_2 \mathcal{D}\Phi_1 \mathcal{D}\varphi_1 \exp\left(S_4[\alpha;\Sigma_4] \right) \ ,
\label{doublegaugingdefect}
\end{equation}
where
\begin{equation}
S_4[\alpha;\Sigma_4] = -\frac{i}{2\pi} \oint_{\Sigma_4}  \left[ f \wedge \Phi_2 + \Phi_2 \wedge d\varphi_1 + \Phi_2 \wedge \varphi_2 + \varphi_2 \wedge  d\Phi_1 - \frac{\alpha}{4\pi} \varphi_2 \wedge \varphi_2 \right] \ . 
\label{S4action}
\end{equation}
Here, $\alpha \in \mathbb{R}$ parametrizes a torsion term, $f=da$ is the Maxwell field strength (which is path-integrated over in the 5$d$ bulk, but not in \eqref{doublegaugingdefect}), $\Phi_2$ and $\Phi_1$ are $U(1)$ gauge fields, $\varphi_2$ and $\varphi_1$ are $\mathbb{R}$ gauge fields, all living on $\Sigma_4$ and subject to the gauge transformations
\begin{equation}
\Phi_2 \rightarrow \Phi_2 + d\Theta_1 \ , \ \varphi_2 \rightarrow \varphi_2 + d\theta_1 \ ,  \ \varphi_1 \rightarrow \varphi_1 + d\theta_0 - \theta_1 \ , \ \Phi_1 \rightarrow \Phi_1 + d\Theta_0 - \Theta_1 + \frac{\alpha}{2\pi} \theta_1 \ ,  
\end{equation}
where $\oint d\Theta_i \in 2\pi\mathbb{Z}$ and $\oint d\theta_i = 0$.

This action corresponds to performing two subsequent gaugings on $\Sigma_4$, in presence of a torsion term. First, we perform a flat 1-gauging\footnote{Flat gauging a $U(1)$ symmetry with current $J$ means that the dynamical $U(1)$ gauge field $A$ coupling to $J$ is flat, $dA=0$, and has no kinetic term. This corresponds to adding to the action $i \star J \wedge A + i \varphi \wedge dA$, where $\varphi$ is an $\mathbb{R}$ gauge field that acts as a Lagrange multiplier enforcing $dA=0$, and path integrating over $A$ and $\varphi$. Flat gauging is a topological operation \cite{Brennan:2024fgj,Antinucci:2024zjp}.} of the $U(1)^{(2)}_M$ magnetic symmetry, by coupling its current $\frac{f}{2\pi}$ to a $U(1)$ 2-form gauge field $\Phi_2$ living on $\Sigma_4$, via the first term. The flatness of $\Phi_2$ is enforced by the second term, which couples it to the $\mathbb{R}$ Lagrange multiplier $\varphi_1$.
This operation produces on $\Sigma_4$ a 1-form symmetry $\mathbb{Z}^{(1)}$, the Pontryagin dual of $U(1)^{(2)}_M$, whose symmetry defects are $\exp\left(iq\oint \Phi_2\right)$ with $q\in\mathbb{Z}$. Then, with the third term, we perform a gauging of this dual symmetry on $\Sigma_4$ by coupling it to a $\mathbb{Z}$ 2-form gauge field. The latter is obtained as an $\mathbb{R}$ gauge field $\varphi_2$, which is flat and whose holonomies are quantized. These conditions are enforced by the fourth term, which couples $\varphi_2$ to the $U(1)$ Lagrange multiplier $\Phi_1$, and makes it effectively a $\mathbb{Z}$ gauge field, namely $d\varphi_2=0$ and $\oint \varphi_2 \in 2\pi\mathbb{Z}$. Finally, the last term is a torsion term parametrized by $\alpha \in \mathbb{R}$. This is crucial, otherwise one would get a trivial result, as the second gauging is just undoing the first one.

The path integral over $\varphi_2$ and $\Phi_2$ yields (with a slight abuse of notation, we write only the exponential of the integrand in \eqref{doublegaugingdefect})
\begin{equation}
S_4[\alpha;\Sigma_4] = \frac{i}{2\pi} \oint_{\Sigma_4}  \left[ d\widetilde\Phi_1 \wedge \left( f + d\varphi_1 \right) - \frac{\alpha}{4\pi} d\varphi_1 \wedge d\varphi_1 \right] + \frac{i\alpha}{8\pi^2} \oint_{\Sigma_4} f \wedge f \ ,    
\end{equation}
where $\widetilde{\Phi}_1 = \Phi_1 + \frac{\alpha}{2\pi} \varphi_1$ is still a $U(1)$ gauge field. Notice that there are two trivial terms: the first simply enforces on $\Sigma_4$ the Bianchi identity $df=0$ and the quantization condition $\oint f \in 2\pi\mathbb{Z}$, which are both consistent with the 5$d$ bulk Maxwell theory; the second is a theta-term for an $\mathbb{R}$ gauge field, which is trivial. This makes \eqref{doublegaugingdefect} coincide with \eqref{symdef_U}, that is 
\begin{equation}
\mathcal{D}[\alpha;\Sigma_4] = U_I^\alpha[\Sigma_4] \ .    
\end{equation}
Therefore, the defect defined by the topological manipulations in \eqref{doublegaugingdefect} and \eqref{S4action} is precisely the symmetry defect operator for the $U(1)_I^{(0)}$ instantonic symmetry. 
As pointed out above, in absence of the torsion term (namely if $\alpha=0$) we get the trivial operator.

The result discussed in this section is tied to the fact that the instantonic symmetry is a composite symmetry of the magnetic symmetry with itself, and that it does not act on pointlike operators. Consistently, its topological operators are (invertible) condensation defects that can be obtained by a topological gauging of the magnetic symmetry (and its Pontryagin dual) on a codimension-one surface.

\subsection{Background fields, anomalies, and 3-group}
In this section, we discuss the structure of the (mixed) anomalies in 5$d$ Maxwell theory and show that, as a consequence of the composite nature of the instantonic symmetry, a 3-group structure emerges. This is an instance of the general analysis carried out in \cite{Brauner:2020rtz}. To this end, we couple the 5$d$ Maxwell theory to $U(1)$ background gauge fields $C_3$, $B_2$, and $A_1$ for the $U(1)_M^{(2)}$, $U(1)_E^{(1)}$, and $U(1)_I^{(0)}$ symmetries, respectively. The action reads
\begin{align}\label{Maxwell_5d_action}
\begin{split}
S  = \int_{{\cal M}_5} &\left[\frac{1}{2e^2} (f-B_2)\wedge \star (f- B_2) + \frac{i}{2\pi} C_3 \wedge (f-\ell B_2) \right. \\
 & \left.+ \frac{i}{8\pi^2} A_1 \wedge (f- \kappa B_2)\wedge (f- \kappa B_2)\right]~,   
\end{split}
\end{align}
where $\ell$ and $\kappa$ are integer coefficients. Notice that they introduce three additional terms in the action. Two of them are counterterms that do not depend on the dynamical field $f$. Instead, the term
\begin{equation}
-\frac{i\kappa}{4\pi^2} A_1 \wedge B_2 \wedge f \ ,
\label{Maxwell_dynamicalcounterterm}
\end{equation}
can be interpreted, in the theory \eqref{Maxwell_5d_action} with $\ell=0$, as a shift of the background field for the magnetic symmetry as
\begin{equation}\label{B3_redef}
C_3 \qquad \longrightarrow \qquad \widetilde{C}_3 \equiv C_3 - \frac{\kappa}{2\pi} A_1 \wedge B_2 \ ,    
\end{equation}
which has consistent gauge transformations since $\kappa \in \mathbb{Z}$, as we will see below.

In presence of background fields, the Bianchi identity and the equations of motion are
\begin{equation}\label{Maxwell_eoms}
df=0 \ , \qquad \frac{i}{e^2}d\star(f-B_2) = \frac{1}{2\pi} dC_3 + \frac{1}{4\pi^2}dA_1 \wedge (f-\kappa B_2) + \frac{\kappa}{4\pi^2} A_1 \wedge dB_2 \ .
\end{equation}
The second equation can also be rewritten as the same equation with $\kappa=0$ and $C_3 \rightarrow \widetilde{C}_3$. In presence of background gauge fields, the currents -- that we define such that $i\star J$ is the functional derivative of the action with respect to the corresponding gauge field -- read\footnote{Notice that if we want to maintain the `composite symmetry' nature of the instantonic current, $\star J_I^{(1)} = \frac{1}{2} \star J_M^{(3)} \wedge \star J_M^{(3)}$, even in presence of background fields, we must choose $\kappa=\ell$.}
\begin{equation}
\begin{split}
\star J_I^{(1)} &= \frac{1}{8\pi^2} (f - \kappa B_2) \wedge (f - \kappa B_2) \ , \\
\star J_E^{(2)} &= \frac{i}{e^2}\star(f-B_2) - \frac{\ell}{2\pi} C_3 - \frac{\kappa}{4\pi^2} A_1 \wedge (f-\kappa B_2) \ , \\
\star J_M^{(3)} &= \frac{1}{2\pi}(f-\ell B_2) \ ,
\end{split}
\end{equation}
so that using \eqref{Maxwell_eoms} we get
\begin{equation}\label{Maxw:cons-eq}
\begin{split}
d \star J^{(1)}_I &= - \frac{\kappa}{4\pi^2} (f - \kappa B_2) \wedge dB_2 \ , \\
d \star J^{(2)}_E &= \frac{1-\kappa}{4\pi^2} \left( dA_1 \wedge (f - \kappa B_2)  + \kappa A_1 \wedge dB_2 \right) + \frac{1-\ell}{2\pi} dC_3 \ , \\
d \star J^{(3)}_M &= - \frac{\ell}{2\pi} dB_2 \ .
\end{split}
\end{equation}
The counterterm proportional to $\ell$ is the one associated to the usual mixed 't~Hooft anomaly between the electric and magnetic symmetries in Maxwell theory (in any dimensions). There is no choice of $\ell$ such that we can preserve both the electric and the magnetic symmetries. Instead, $\kappa$ is associated with a 3-group symmetry, which is a direct manifestation of the composite nature of $U(1)_I^{(0)}$, as we now show. 

Let us distinguish the two relevant cases where either the magnetic and the instantonic, or the electric currents are preserved even in presence of background fields, which correspond to $\kappa=\ell=0$ and $\kappa=\ell=1$, respectively.

\begin{itemize}
    \item If $\kappa=\ell=0$, both the magnetic and the instantonic symmetries are preserved, whereas the conservation law of the electric symmetry reads 
    \begin{equation}\label{non-cons-JE}
     d \star J^{(2)}_E = \frac{1}{4\pi^2} dA_1 \wedge f + \frac{1}{2\pi} dC_3 = \frac{1}{2\pi} \star J_M^{(3)} \wedge dA_1 + \frac{1}{2\pi} dC_3 \ .
    \end{equation}
    The second term on the right-hand side signals the usual mixed electric-magnetic 't~Hooft anomaly, whereas the first term depends on the dynamical current operator of the magnetic symmetry. This signals a 3-group structure of the theory \cite{Cordova:2018cvg}. Indeed, in this case the background gauge transformations are
    \begin{equation}
    a \rightarrow a + \Lambda_1 \,, \quad B_2 \rightarrow B_2 + d\Lambda_1 \,, \quad
    A_1 \rightarrow A_1 + d\Lambda_0 \,, \quad C_3 \rightarrow C_3 + d\Lambda_2 - \frac{1}{2\pi} d\Lambda_1 \wedge A_1 \ ,
    \label{Maxwell_gauge1}
    \end{equation}
    with $\oint d\Lambda_i\in 2\pi\mathbb{Z}$, which is a 3-group transformation law.\footnote{Notice that these gauge transformations are well-defined. Indeed, the ambiguity of $A_1$ is a gauge transformation $d\Lambda_0$ with periods that are integer multiples of $2\pi$. This produces a shift $C_3 \rightarrow C_3 - \frac{1}{2\pi} d\Lambda_1 \wedge d\Lambda_0$, which is correctly quantized to be reabsorbed in a $d\Lambda_2$ gauge transformation.} These gauge transformations lead to a variation of the action that depends on the background fields only, and it signals the presence of mixed 't~Hooft anomalies. Indeed, such a variation can be canceled by the 6$d$ anomaly-inflow action
    \begin{equation}
    \int_{\mathcal{Y}_6} \left[-\frac{i}{2\pi} dC_3 \wedge B_2 - \frac{i}{8\pi^2} dA_1 \wedge B_2 \wedge B_2 \right] \ ,
    \label{Maxwell_anomalyinflow1}
    \end{equation}
    with $\mathcal{Y}_6$ defined such that $\partial \mathcal{Y}_6 = {\cal M}_5$.
    The first term captures the mixed anomaly between the electric and the magnetic symmetries, whereas the second term captures the one between the electric and the instantonic symmetries.
    
    \item If $\kappa=\ell=1$, the electric symmetry is preserved, the magnetic symmetry is violated by the usual mixed electric-magnetic 't~Hooft anomaly, whereas the conservation law of the instantonic symmetry reads 
    \begin{equation}
     d \star J^{(1)}_I = - \frac{1}{4\pi^2} (f - B_2) \wedge dB_2 = - \frac{1}{2\pi} \star J^{(3)}_M \wedge dB_2 \ .
    \end{equation}
    Again, this depends on the dynamical current operator of the magnetic symmetry, and it is a different but equivalent presentation of the 3-group structure of the theory. Indeed, in this case the modified background gauge transformations are
    \begin{equation}
    a \rightarrow a + \Lambda_1 \ , \quad B_2 \rightarrow B_2 + d\Lambda_1 \,, \quad
    A_1 \rightarrow A_1 + d\Lambda_0 \ , \quad C_3 \rightarrow C_3 + d\Lambda_2 + \frac{1}{2\pi} d\Lambda_0 \wedge B_2  \ ,    
    \label{Maxwell_gauge2}
    \end{equation}
    with $\oint d\Lambda_i\in 2\pi\mathbb{Z}$, which is again a 3-group transformation law. These gauge transformations lead to a variation of the action that is now canceled by the 6$d$ anomaly-inflow action
    \begin{equation}
    \int_{\mathcal{Y}_6} \left[\frac{i}{2\pi} dB_2 \wedge C_3 \right] \ . 
    \label{Maxwell_anomalyinflow2}
    \end{equation}
    Notice that in this description there is no mixed term involving $dA_1$, as instead in \eqref{Maxwell_anomalyinflow1}.
    \end{itemize}
Finally, we observe that the anomaly inflow in \eqref{Maxwell_anomalyinflow1} is reproduced by plugging the shift \eqref{B3_redef} of $C_3$, with $\kappa=1$, in \eqref{Maxwell_anomalyinflow2}. This redefinition also maps the gauge transformations in \eqref{Maxwell_gauge1} to the ones in \eqref{Maxwell_gauge2}. Indeed, as observed in \cite{Brauner:2020rtz}, the two forms of the anomalies are actually equivalent: the two 6$d$ anomaly-inflow actions arise from the same 7$d$ anomaly polynomial,
    \begin{equation}
    \mathcal{P}_7 = - \frac{i}{2\pi} H_4 \wedge dB_2 \ ,    
    \end{equation}
    where $H_4$ is the gauge invariant field strength of $C_3$, satisfying $dH_4=\frac{1}{2\pi}dB_2 \wedge dA_1$ so that $d\mathcal{P}_7=0$. For instance, in the cases where $\kappa=\ell=0$ and $\kappa=\ell=1$, it is given by $H_4 = dC_3 + \frac{1}{2\pi} B_2 \wedge dA_1$ and $H_4 = dC_3 + \frac{1}{2\pi} A_1 \wedge dB_2$, respectively.

\section{Global forms of five-dimensional \texorpdfstring{$\mathfrak{su}(N)$}{su(N)} Yang-Mills}
\label{Section:Yang-Mills}

We now turn to Yang–Mills theories, focusing on the $\mathfrak{su}(N)$ gauge algebra and its various global forms.

\subsection{Review of \texorpdfstring{$SU(N)$}{SU(N)} Yang-Mills and its mixed 't~Hooft anomaly}

Let us consider five-dimensional Yang-Mills with gauge group $SU(N)$ and action
\begin{equation}\label{eq:actionYM}
    S = \int_{\mathcal{M}_5} \frac{1}{2g^2} \Tr \left(f\wedge \star f\right) \ ,
\end{equation}
where $f$ is the non-Abelian field strength and $g^2$ is the gauge coupling.\footnote{If $N=2$, one can consider a discrete theta-term with $\theta=\pi$, which is possible because $\pi_4(SU(2))=\mathbb{Z}_2$. For $N > 2$, one can introduce a Chern-Simons term at level $k$ for the $SU(N)$ theory. We do not consider these possibilities here.} The theory has a $\mathbb{Z}_N^{(1)}$ electric 1-form symmetry, generated by codimension-two topological Gukov-Witten operators, whose charged operators are Wilson lines in a representation $\rho$ of $SU(N)$, with $\mathbb{Z}_N^{(1)}$ charge given by the $N$-ality of $\rho$. Moreover, the theory has a 0-form instantonic symmetry $U(1)_I^{(0)}$, with conserved current 
\begin{equation}\label{instJ}
J^{(1)}_I = \frac{1}{8\pi^2} \star \Tr\left(f\wedge f\right) \ ,
\end{equation}
that satisfies (under the same assumptions discussed in footnote \ref{footnotespin})
\begin{equation}
Q_I[\Sigma_4] = \oint_{\Sigma_4}\,\star J^{(1)}_I  \in \mathbb{Z} \ .
\end{equation}
An important difference from the Abelian case is that the $U(1)^{(0)}_I$ instantonic symmetry now acts on pointlike operators, the instanton operators.

Let us now review the results of \cite{BenettiGenolini:2020doj,Genolini:2022mpi}. If one turns on a background gauge field $\mathcal{B}_2 \in H^2(\mathcal{M}_5,\mathbb{Z}_N)$ for the 1-form symmetry $\mathbb{Z}_N^{(1)}$, the quantization condition for the instantonic charge gets modified to
\begin{align} \label{ch-fract}
\begin{split}
   &  \oint_{\Sigma_4}  \star J^{(1)}_I = \frac{N-1}{2N}\oint_{\Sigma_4} \mathcal{P}(\mathcal{B}_2) \quad\text{mod }1~,\\
  \mathcal{P}(\mathcal{B}_2) &=    \begin{cases}  \mathcal{B}_2\cup \mathcal{B}_2 \quad \text{for odd $N$}~,   \vspace{2mm} \\
     \mathfrak{P}(\mathcal{B}_2)\quad \,\,\text{ for even $N$}~,
    \end{cases}
\end{split}    
\end{align}
where $\cup$ is the product between cochains valued in the same Abelian ring (in this work we only consider $\mathbb{Z}$ and $\mathbb{Z}_r$ for some $r\in\mathbb{N}$), and $\mathfrak{P}$ denotes the Pontryagin square, that is a map\footnote{See e.g.~the appendix of \cite{Kapustin:2013qsa} for a review for physicists.} 
\begin{equation}
\mathfrak{P}: H^2(\mathcal{M}_5,\mathbb{Z}_N)\to 
        H^4(\mathcal{M}_5,\mathbb{Z}_{2N})~.
\end{equation}
With our assumptions on $\mathcal{M}_5$ (see again footnote \ref{footnotespin}), we have 
\begin{align}
\begin{split}
\frac 12 \oint_{\Sigma_4} \mathcal{P}(\mathcal{B}_2) & \in \mathbb{Z}_{N} \ .
\end{split}
\end{align}
Therefore, in the presence of a background $\mathcal{B}_2$ one gets that
\begin{equation}\label{chargefractionalization}
Q_I[\Sigma_4,\mathcal{B}_2] \in \frac{1}{N}\mathbb{Z} \ .
\end{equation}
As a result, if we also turn on a $U(1)_I^{(0)}$ background gauge field $A_1$ for the instantonic symmetry, the partition function $Z[A_1,\mathcal{B}_2]$ is multiplied by a phase under large gauge transformations $A_1 \to A_1 + d\Lambda_0$, with $\oint d\Lambda_0 \in 2\pi\mathbb{Z}$. This means that the two symmetries $\mathbb{Z}_N^{(1)}$ and $U(1)_I^{(0)}$ have a mixed 't~Hooft anomaly. The anomaly is captured by the following anomaly inflow  \cite{BenettiGenolini:2020doj} 
\begin{equation} \label{BGT-anomaly}
    \mathcal{A}_6[A_1,
    \mathcal{B}_2] =
    \exp \left( \frac{2\pi i}{N} \int_{\mathcal{Y}_6} 
    \left[\frac{dA_1}{2\pi}\right]_N
    \cup
    \frac{\mathcal{P}(\mathcal{B}_2)}{2} \right) \ ,
\end{equation}
where $\mathcal{Y}_6$ is such that $\partial \mathcal{Y}_6=\mathcal{M}_5$, the background fields have been extended to $\mathcal{Y}_6$, and $[\cdot]_N$ is the reduction modulo $N$.

Defining the topological defect operators
\begin{equation}
U^{\alpha}_I[\Sigma_4,\mathcal{B}_2] = \exp(i\, \alpha \,Q_I[\Sigma_4, \mathcal{B}_2]) \ ,
\end{equation}
the quantization condition \eqref{ch-fract} leads to
\begin{equation}\label{D-ext}
    U^{\alpha+ 2\pi}_I[ \Sigma_4,\mathcal{B}_2] = 
    U^{\alpha}_I[\Sigma_4,\mathcal{B}_2]
    \exp\left(-\frac{2\pi i}{N}
    \oint_{\Sigma_4} \frac{\mathcal{P}(\mathcal{B}_2) }{2}\right)  \ .
\end{equation}
The periodicity of the parameter $\alpha$ is then extended to $\alpha\sim \alpha + 2\pi N$. Clearly, in the $SU(N)$ theory this extension is only visible when $\mathcal{B}_2$ is turned on, while the correct identification in any correlation function without background sources  is still $\alpha\sim\alpha+2\pi$. 

\begin{figure}[t]
    \centering
    \includegraphics[width=0.5\linewidth]{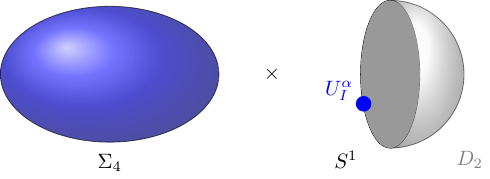}
    \caption{The evaluation of the anomaly-inflow functional on $\mathcal{Y}_6$ with $\partial \mathcal{Y}_6 = \mathcal{M}_5$, where $\mathcal{M}_5 = S^1\times \Sigma_4$ and $\mathcal{Y}_6 = D_2\times \Sigma_4$. A symmetry defect of the instantonic symmetry $U^\alpha_I$ wraps $\Sigma_4$. Shifting $\alpha\to\alpha+2\pi$ corresponds to a large gauge transformation. This activates the anomaly functional evaluated on $\mathcal{Y}_6$. As a result, $U^\alpha_I$ and $U^{\alpha+2\pi}_I$ are not equal, but rather related by a phase factor determined by the anomaly. \label{fig:S1M5}}
\end{figure}

For future reference, it is useful to derive eq.~\eqref{D-ext} directly from the anomaly inflow \eqref{BGT-anomaly}, by noting that the combination $Z \mathcal{A}_6$ is gauge invariant. The insertion of the symmetry defect corresponds to turning on a flat background gauge field for the instantonic symmetry $A_1^{\alpha}= \alpha\, \delta^{(1)}(\Sigma_4)$, where $\Sigma_4$ is a closed four-cycle. The defect with parameter $\alpha + 2\pi$ differs from the one with parameter $\alpha$ by
\begin{equation}
A_1^{\alpha + 2\pi} = A_1^{\alpha} + d\Lambda_0 \ , \qquad
d\Lambda_0 = 2\pi \delta^{(1)}(\Sigma_4) \ ,
\end{equation}
which is a large gauge transformation when $\Sigma_4$ is homologically non-trivial. As a prototypical example, we can take $\mathcal{M}_5 = S^1 \times \Sigma_4$, in which case the symmetry defect is localized at one point on $S^1$, it wraps $\Sigma_4$, and the large gauge transformation is given by a function  $e^{i\Lambda_0}:S^1 \times \Sigma_4 \to U(1)$ with winding along $S^1$, see figure \ref{fig:S1M5}.

To compute the phase picked by $Z$ we evaluate $\mathcal{A}_6$ before and after the gauge transformation. To this end, we need to pick a manifold $\mathcal{Y}_6$ with $\partial\mathcal{Y}_6 =\mathcal{M}_5$ and lift the background fields to $\mathcal{Y}_6$. We take $\Sigma_4$ homologically non-trivial in $\mathcal{Y}_6$ and 
\begin{equation}
d \bar{A}_1^\alpha = \alpha \, \delta^{(2)}(\Sigma_4)~,
\end{equation}
where the bar denotes the lift to $\mathcal{Y}_6$. In the example above, we can take $\mathcal{Y}_6 = D_2 \times \Sigma_4$ where $D_2$ is a two-dimensional disk, see figure \ref{fig:S1M5}. We then have\footnote{Here the partition function can be taken to depend, in addition of $A_1$ and $\mathcal{B}_2$, on any source for  operators of the theory. As a result, the equations we derive capture any correlation function of the symmetry defect operator and are operator equations.}
\begin{equation}
Z[A_1^\alpha, \mathcal{B}_2] \mathcal{A}_6[\bar{A}_1^\alpha,\bar{\mathcal{B}_2}] = Z[A_1^{\alpha+2\pi}, \mathcal{B}_2] \mathcal{A}_6[\bar{A}_1^{\alpha+2\pi},\bar{\mathcal{B}_2}]~. 
\end{equation}
Therefore
\begin{align}\label{extension_IR}
\begin{split}
U^{\alpha+ 2\pi}_I[ \Sigma_4,\mathcal{B}_2] & = 
    U^{\alpha}_I[\Sigma_4,\mathcal{B}_2] \frac{\mathcal{A}_6[\bar{A}_1^\alpha,\bar{\mathcal{B}_2}]}{\mathcal{A}_6[\bar{A}_1^{\alpha+2\pi},\bar{\mathcal{B}_2}]} \\
& = U^{\alpha}_I[\Sigma_4,\mathcal{B}_2] \exp \left( \frac{2\pi i}{N} \int_{\mathcal{Y}_6} 
    \left[\frac{d\bar{A}_1^\alpha - d\bar{A}_1^{\alpha+2\pi}}{2\pi}\right]_N
    \cup
    \frac{\mathcal{P}(\bar{\mathcal{B}_2})}{2} \right) \\
&  = U^{\alpha}_I[\Sigma_4,\mathcal{B}_2]  \exp\left(-\frac{2\pi i}{N}
    \oint_{\Sigma_4} \frac{\mathcal{P}(\mathcal{B}_2) }{2}\right)
~,
\end{split}    
\end{align}
which reproduces eq.~\eqref{D-ext}. Equivalently, the ratio of $\mathcal{A}_6$ functionals can be evaluated by viewing it as the exponential of a single anomaly action integrated on a closed manifold $\mathcal{M}_6$, obtained by gluing $\mathcal{Y}_6$ and its orientation reversal along the common boundary $\mathcal{M}_5$. The gauge field $\bar{A}_1$ on $\mathcal{M}_6$ satisfies
\begin{equation}
\oint_{\mathcal{M}_6} \left[\frac{d\bar{A}_1}{2\pi}\right]_N \cup \frac{\mathcal{P}(\bar{\mathcal{B}_2})}{2} = -\oint_{\Sigma_4} \frac{\mathcal{P}(\mathcal{B}_2) }{2}~,
\end{equation}
giving again the expected result.

\subsection{Gauging the 1-form symmetry: \texorpdfstring{$PSU(N)$}{PSU(N)} Yang-Mills}
\label{gaug-PSU}

In $SU(N)$ Yang-Mills, the coupling to a background field $\mathcal{B}_2$ for the $\mathbb{Z}_N^{(1)}$ 1-form symmetry requires summing over $PSU(N)$ bundles for which the characteristic class $w_2 \in H^2(\mathcal{M}_5,\mathbb{Z}_N)$, i.e.~the obstruction for a $PSU(N)$ bundle to be lifted to an $SU(N)$ one, is $w_2=\mathcal{B}_2$. Gauging the $\mathbb{Z}_N^{(1)}$ symmetry requires a further sum over all possible values of $w_2$, i.e.~all $PSU(N)$ bundles.\footnote{Recall that here we are discussing the theory \eqref{eq:actionYM} without a Chern-Simons term. In the $SU(N)$ theory with a Chern-Simons term the 1-form symmetry would suffer from self-anomalies and generically it could not be gauged. This same phenomenon occurs in $3d$ Yang-Mills-Chern-Simons theories (see e.g.~\cite{Hsin:2018vcg,Argurio:2024oym}).} We can therefore present $PSU(N)$ YM as coupling $SU(N)$ YM to a dynamical $\mathbb{Z}_N^{(1)}$ gauge field, denoted by  $\mathscr{b}_2\in H^2(\mathcal{M}_5,\mathbb{Z}_N)$. The partition function is
\begin{equation}
    Z_{PSU(N)}[A_1,\mathcal{C}_3] = \frac{|H^0(\mathcal{M}_5,\mathbb{Z}_N)|}{|H^1(\mathcal{M}_5,\mathbb{Z}_N)|}{}\sum_{\mathscr{b}_2\in H^2(\mathcal{M}_5,\mathbb{Z}_N)} Z_{SU(N)}[A_1,\mathscr{b}_2 ]\exp\left(\frac{2\pi i}{N}\int_{\mathcal{M}_5} \mathscr{b}_2\cup\mathcal{C}_3\right)\ .
\end{equation}
Here $Z_{SU(N)}[A_1,\mathcal{B}_2 ]$ denotes the partition function of the $SU(N)$ theory coupled to the background gauge fields $A_1$ for the instantonic 0-form symmetry and  $\mathcal{B}_2$ for the electric 1-form symmetry, and $\mathcal{C}_3\in H^3(\mathcal{M}_5,\mathbb{Z}_N)$ is a background gauge field for the magnetic 2-form symmetry. The prefactor is a standard normalization factor that arises in gauging of higher-form discrete symmetries \cite{Gaiotto:2014kfa}. 

The $PSU(N)$ theory has a $\mathbb{Z}_N^{(2)}$ magnetic 2-form symmetry, whose symmetry defects are the topological operators
\begin{equation}\label{SDO-M}
    U_M^n[\Sigma_2] = \exp \left( \frac{2\pi i n}{N} \oint_{\Sigma_2} \mathscr{b}_2 \right) \quad , \qquad n\in\mathbb{Z}_N \ ,
\end{equation}
and whose charged objects are 't~Hooft surfaces, labeled by a magnetic weight up to the action of the Weyl group, or equivalently by an irreducible representation $\rho$ of $SU(N)$ \cite{Kapustin:2005py}. The insertion of the 't~Hooft surface $H^{\rho}[\gamma_2]$ fixes the value of the holonomy of $\mathscr{b}_2$ on two-cycles that link with $\gamma_2$ to coincide with the $N$-ality $m_\rho \in \mathbb{Z}_N$ of the representation, so that
\begin{equation}\label{action-magnetic}
U_M^n[\Sigma_2]H^{\rho}[\gamma_2]  = \exp\left(\frac{2\pi i n m_\rho}{N} L_2(\Sigma_2,\gamma_2)\right) H^{\rho}[\gamma_2]  \ ,
\end{equation}
where $L_2$ is the double-linking defined in \eqref{doublelink}. We can formalize this statement as follows: removing $\gamma_2$ gives rise to an additional 3-cycle in the relative homology $H_3(\mathcal{M}_5,\gamma_2,\mathbb{Z}_N)$, represented by $M_3$ with $\partial M_3 = \gamma_2$. Its dual in cohomology is an additional cocycle in $H^2(\mathcal{M}_5 \backslash \gamma_2,\mathbb{Z}_N)$ represented by $\delta^{(2)}(M_3)$. Inserting a 't~Hooft surface in the representation $\rho$ of $SU(N)$ fixes the component of $\mathscr{b}_2$ along this additional cocycle, which we can represent by the shift  
\begin{equation}
\mathscr{b}_2 \, \longrightarrow  \, \mathscr{b}_2+ \mathscr{b}^H_2 ~, \qquad \mathscr{b}^H_2 = m_\rho \, \delta^{(2)}(M_3) \ .
\end{equation}
The previous relation does not uniquely characterize the 't~Hooft surface. Indeed, we stress that the 't~Hooft surface operator is labeled by $\rho$ and not just by the $N$-ality $m_\rho$. The latter merely determines the $\mathbb{Z}_N^{(2)}$ charge. As a result, 't~Hooft surfaces in different representations with the same $N$-ality are not distinguished at the level of the action of the 2-form symmetry, as shown by \eqref{action-magnetic}. We will show below that this is not the case for the 0-form symmetry.

In addition to the 2-form symmetry, the $PSU(N)$ theory still has an instantonic $U(1)^{(0)}_I$ symmetry generated by the same conserved current \eqref{instJ} as in the $SU(N)$ theory. It is clear that there must be a relation between the $\mathbb{Z}^{(2)}_N$ magnetic symmetry and the instantonic symmetry, which reflects the existence of the 't~Hooft anomaly \eqref{BGT-anomaly} in the parent $SU(N)$ theory. In the following we will explain in detail what this relation is.

In $PSU(N)$ eq.~\eqref{D-ext} reads
\begin{align}\label{D-ext-2}
    U^{\alpha+ 2\pi}_I[ \Sigma_4] = 
    U^{\alpha}_I[\Sigma_4]
    \mathcal{C}[\Sigma_4] \quad \text{where}\quad
    \mathcal{C}[\Sigma_4]  \equiv \exp\left(-\frac{2\pi i}{N}
    \oint_{\Sigma_4} \frac{\mathcal{P}(\mathscr{b}_2)}{2}\right) \ .
\end{align}
$\mathcal{C}[\Sigma_4]$ is a non-trivial topological operator in the $PSU(N)$ theory, generating a $\mathbb{Z}_N^{(0)}$ symmetry that through eq.~\eqref{D-ext-2} is seen to extend the instantonic symmetry. Equivalently, the periodicity of parameter $\alpha$ is modified in the $PSU(N)$ theory to $\alpha \sim \alpha +2\pi N$, and $\mathcal{C} = U_I^{2\pi}$. This additional $\mathbb{Z}_N^{(0)}$ symmetry was discussed previously in \cite{Gukov:2020btk}, where it was claimed to form a direct product structure with $U(1)_I^{(0)}$ in the $PSU(N)$ theory. In fact, because of the mixed 't~Hooft anomaly \eqref{BGT-anomaly}, we have shown here that the instantonic symmetry $\widetilde{U(1)}^{(0)}_I$ in the $PSU(N)$ theory is not a product of $U(1)_I^{(0)}$ -- the instantonic symmetry in the $SU(N)$ theory -- and $\mathbb{Z}_N^{(0)}$, but rather an extension of $U(1)_I^{(0)}$ by $\mathbb{Z}_N^{(0)}$. This happens because the exact sequence
\begin{equation}\label{exactseq_PSU(N)}
1\to \mathbb{Z}_N \to \widetilde{U(1)}_I \to U(1)_I \to 1    
\end{equation}
does not split.\footnote{In the following we denote both symmetries with $U(1)_I^{(0)}$ since ambiguities should not arise.} 

Since there are no fractional $PSU(N)$ instantons on $S^4$ -- because any $PSU(N)$ bundle on $S^4$ is also an $SU(N)$ bundle, as there are no non-trivial 2-cycles on $S^4$ -- there are no pointlike operators in the theory charged under $\mathcal{C}$, namely $\mathcal{C}[S^4]=\mathbbm{1}$. This is similar to what we observed for the instantonic symmetry of Maxwell theory in section \ref{subsec:Max-Action-Cond}. As we saw in that context, also here we can trace the absence of charged pointlike operators back to the fact that $\mathcal{C}$ is a condensation defect, as we show explicitly in section \ref{subsection:Znascondensation} below. Instead, $\mathcal{C}$ acts on a configuration of 't~Hooft surfaces that link with each other. (Indeed, $\mathcal{C}$ can be activated both by a non-trivial topology and by suitable operator insertions.) To show this, let us consider two 't~Hooft surfaces $H^\rho[\gamma_2]$ and $H^{\rho'}[\gamma_2']$. Following the previous discussion for a single surface, in this case we fix the component of $\mathscr{b}_2$ along the two additional cocycles in $H^2(\mathcal{M}_5 \backslash (\gamma_2 \sqcup \gamma_2'),\mathbb{Z}_N)$, namely
\begin{equation}\label{shifted_b}
     \mathscr{b}^H_2 = m_\rho \, \delta^{(2)}(M_3) 
     + m_{\rho'} \, \delta^{(2)}(M_3')  
     \ ,
 \end{equation}
where $\partial M_3 = \gamma_2$ and $\partial M'_3 = \gamma'_2$. The restriction of $\mathscr{b}^H_2$ to $\Sigma_4$ gives
\begin{equation}
\mathscr{b}^H_2\vert_{\Sigma_4} = m_\rho \, \delta^{(2)}_{\Sigma_4}(M_2) 
     + m_{\rho'} \, \delta^{(2)}_{\Sigma_4}(M_2') \ ,
\end{equation}
where $M_2 = M_3 \cap \Sigma_4$, $M_2' = M_3' \cap \Sigma_4$, and the subscript in $\delta^{(2)}_{\Sigma_4}$ denotes that we are considering the cocycle in $H^2(\Sigma_4,\mathbb{Z}_N)$. We can then obtain $\mathcal{C}[\Sigma_4]$ by evaluating $\frac12\mathcal{P}(\mathscr{b}^H_2\vert_{\Sigma_4})$. Assuming that there are no torsion cycles, this evaluation can be done by lifting $\mathscr{b}^H_2\vert_{\Sigma_4}$ to integer cohomology $H^2(\Sigma_4,\mathbb{Z})$, and using the pairing given by the intersection form of $\Sigma_4$. Let us specify this evaluation to the same configurations considered for Maxwell theory in section \ref{subsec:Max-Action-Cond}: (1) for $\Sigma_4 = S^4$ such that both surfaces lie inside or outside, $H^2(\Sigma_4,\mathbb{Z})$ is trivial and $\frac12\mathcal{P}(\mathscr{b}^H_2\vert_{\Sigma_4})$ evaluates to $0$; (2) for $\Sigma_4 = S^2\times S^2$, where each $S^2$ factor links with either one of the 't~Hooft surfaces (see figure \ref{fig:triplelinking}), we take the basis of $H^2(\Sigma_4,\mathbb{Z})$ given by the two $S^2$ factors so that the intersection form is {\tiny{$\begin{pmatrix}0&1\\1&0\end{pmatrix}$}}, and $\frac12\mathcal{P}(\mathscr{b}^H_2\vert_{\Sigma_4})$ evaluates to $m_\rho m_{\rho'}$. In both cases the result can be expressed in terms of the triple-linking in \eqref{triplelink} as
\begin{align}
\begin{split}
\mathcal{C}[\Sigma_4] H^\rho[\gamma_2] H^{\rho'}[\gamma_2']  =
\exp\left(-\frac{2\pi i m_\rho m_{\rho'}}{N}   \, L_3(\Sigma_4,\gamma_2,\gamma_2')\right)
H^\rho[\gamma_2] H^{\rho'}[\gamma_2'] \ ,
\label{YM_triplelinking_1}
\end{split}
\end{align} 
in close analogy with the result \eqref{Maxwell_triplelinking} for Maxwell theory. The evaluation presented here gives a different perspective on the contribution of the self-linking of the surfaces: here we do not get a contribution proportional to $m_\rho^2$ and $m_{\rho'}^2$ because the intersection form on $S^2\times S^2$ does not have diagonal
entries.

Let us now consider a configuration involving a defect $U^{\alpha}_I$ for a generic $U(1)_I^{(0)}$ transformation and two 't~Hooft surfaces. Using equations \eqref{D-ext-2} and \eqref{YM_triplelinking_1} we get 
\begin{equation}
U^{\alpha + 2\pi}_I[\Sigma_4] H^\rho[\gamma_2] H^{\rho'}[\gamma_2'] = \exp\left(-\frac{2\pi i m_\rho m_{\rho'}}{N}   \, L_3(\Sigma_4,\gamma_2,\gamma_2')\right) U^{\alpha}_I[\Sigma_4]H^\rho[\gamma_2] H^{\rho'}[\gamma_2'] \ .
\end{equation}
We can solve this equation to find the action of the instantonic symmetry defect on the link of 't~Hooft surfaces for any  $\alpha \in [0,2\pi N)$. We obtain
\begin{align}
\begin{split}
U^{\alpha}_I[\Sigma_4] H^\rho[\gamma_2] H^{\rho'}[\gamma_2']  = \exp\left(-i \alpha \frac{\widetilde{m}_\rho \widetilde{m}_{\rho'}}{N} \, L_3(\Sigma_4,\gamma_2,\gamma_2')\right)
 H^\rho[\gamma_2] H^{\rho'}[\gamma_2'] \ ,
\label{YM_triplelinking_2}
\end{split}
\end{align}
where $\widetilde{m}_{\rho,\rho'}$ denotes an integer lift of $m_{\rho,\rho'}\in\mathbb{Z}_N$. The integer lift is needed because for generic $\alpha$ the phase is not compatible with the mod $N$ identification on $m_{\rho,\rho'}$. We interpret eq.~\eqref{YM_triplelinking_2} as the assignment of a charge to the configuration of two linked 't~Hooft surfaces under the extended $U(1)^{(0)}_I$. This charge is valued in $\frac{1}{N}\mathbb{Z}$ and is given by
\begin{equation}\label{fractional_instanton}
    \frac{\widetilde{m}_\rho \widetilde{m}_{\rho'}}{N} \ .
\end{equation}
Here we see explicitly that the action of the extended $U(1)^{(0)}_I$ depends on the integer lifts $\widetilde{m}_{\rho}$ and $\widetilde{m}_{\rho'}$, unlike the action of the 2-form symmetry and the $\mathbb{Z}^{(0)}_N$ subgroup that only depend on their $\text{mod}\,N$ reduction, i.e.~the $N$-ality of the representation. The uplift $\widetilde{m}_\rho$ is a function of the representation $\rho$ defining the 't~Hooft surface. One cannot determine this function using only the symmetry constraint above. A natural conjecture is that $\widetilde{m}_\rho$ coincides with the number of boxes of the Young diagram for the representation $\rho$ of $SU(N)$.
 
Under the shift $\widetilde{m}_{\rho,\rho'}\to \widetilde{m}_{\rho,\rho'} + N$, the phase in \eqref{YM_triplelinking_2} changes by an integer multiple of $i \alpha L_3(\Sigma_4,\gamma_2,\gamma_2')$. This additional contribution has a physical interpretation. Whenever the defects have a non-trivial triple-linking, a couple of points $P$ and $P'$ on $\gamma_2$ and $\gamma_2'$ necessarily has a non-trivial double-linking with $\Sigma_4$ (i.e.~the two points are necessarily on ``opposite sides'' of $\Sigma_4$). Adding two instanton operators $\hat{I}^q(P)$, $\hat{I}^{-q}(P')$ of opposite charge $\pm q\in\mathbb{Z}$ in such a couple of points shifts the phase in \eqref{YM_triplelinking_2} precisely by $i\alpha q L_3(\Sigma_4,\gamma_2,\gamma_2')$, while at the same time keeping a zero net instanton charge, as measured by an $S^4$ enclosing both surfaces. Making explicit the insertion of the instanton operators, we can rewrite \eqref{YM_triplelinking_2} as 
\begin{align}
\begin{split}
   & U^{\alpha}_I[\Sigma_4]  H^\rho[\gamma_2] \hat{I}^q(P) H^{\rho'}[\gamma_2'] 
     \hat{I}^{-q}(P') 
    \\
    \ & = \exp\left(-i \alpha \left(\frac{\widetilde{m}_\rho \widetilde{m}_{\rho'}}{N} -q\right)\, L_3(\Sigma_4,\gamma_2,\gamma_2') \right)
     H^\rho[\gamma_2] \hat{I}^q(P) H^{\rho'}[\gamma_2'] 
     \hat{I}^{-q}(P') \ ,
\label{YM_triplelinking_3}
\end{split}
\end{align}
where we took $\gamma_2$ to lie inside $\Sigma_4$, as in figure \ref{fig:triplelinking}, and used that in this configuration  $L_3(\Sigma_4,\gamma_2,\gamma_2')=L_2(\Sigma_4,P)$. The label that fully specifies the charges of the configuration of two linked 't~Hooft surfaces and instanton operators is then given by the triplet $(\widetilde{m}_\rho, \widetilde{m}_{\rho'},q)$. The identification of charges $\widetilde{m}_\rho\sim \widetilde{m}_\rho+N$ is not completely lost, but rather it requires to also shift the instanton charge $q$. Indeed, suppose we shift both magnetic labels as $\widetilde{m}_\rho\to \widetilde{m}_\rho+pN$ and $\widetilde{m}_{\rho'}\to \widetilde{m}_{\rho'}+p'N$ with $p,p'\in\mathbb{Z}$. Then
\begin{align}
\begin{split}
   \frac{\alpha}{N}  \widetilde{m}_\rho \widetilde{m}_{\rho'}  
    & \to
   \frac{\alpha}{N}  \widetilde{m}_\rho \widetilde{m}_{\rho'}
   +\alpha (p'\widetilde{m}_{\rho}+ p \,\widetilde{m}_{\rho'} +pp'N)~,  
\end{split}
\end{align}
which means that the correct identification is in fact
\begin{equation}\label{eq:identcharge}
    (\widetilde{m}_\rho+pN,\widetilde{m}_{\rho'}
    +p'N,q) \sim
    (\widetilde{m}_\rho,\widetilde{m}_{\rho'},q-p' \widetilde{m}_\rho -p\, \widetilde{m}_{\rho'}-pp'N) \ .
\end{equation}
To avoid possible confusion, we stress that we are not claiming that the configurations of 't~Hooft surfaces and instanton operators associated to the left- and right-hand side of \eqref{eq:identcharge} should be identified as operators, we are simply claiming that they behave in the same way in correlation functions with the topological operators, and as such cannot be distinguished by selection rules or any other symmetry constraint (a possible exception comes from considering intersections with the topological operators, as we comment below).

Let us comment further on the interpretation of this identification of charges. In general, in $SU(N)$ or $PSU(N)$ gauge theories, higher-form charges are only sensitive to the $N$-ality of the representations that label the Wilson/'t~Hooft operators. From the point of view of the states created by these extended operators, this can be ascribed to the existence of dynamical processes that allow states labeled by different representations with the same $N$-ality to transition between each other. Here we are seeing that for the 't~Hooft surfaces of $5d$ $PSU(N)$ Yang-Mills, the dynamical processes that transition between different representations with the same $N$-ality also require to nucleate an instanton charge, if the states are created by two surfaces linking with each other.

It would be interesting to study also configurations of the instantonic charge operator that intersect the 't~Hooft surfaces. The intersecting configuration defines a defect operator supported on $\Sigma_1 =\Sigma_4 \cap \gamma_2$, which is a natural candidate to define a $U(1)^{(0)}_I$ symmetry operator $\widehat{U}^\alpha[\Sigma_1]$ on the defect $H^\rho[\gamma_2]$. If this intersection is topological, it provides constraints analogous to those of a $0$-form symmetry in two dimensions, e.g.~selection rules associated to charge conservation for pointlike operators supported on $H^\rho[\gamma_2]$. In the terminology of \cite{Antinucci:2024izg}, in this case we would say that the 't~Hooft surface is a {\it symmetric defect} for the $U(1)^{(0)}_I$ symmetry. If the 't~Hooft surface is a symmetric defect, the instanton charge carried by pointlike operators on $H^\rho[\gamma_2]$ would be conserved, and this would break the identification \eqref{eq:identcharge}. Another more likely possibility is that the 't~Hooft surface is not a symmetric defect, but rather the $U(1)^{(0)}_I$ symmetry is broken on the defect to the $\mathbb{Z}^{(0)}_N$ subgroup that provides the extension, consistently with the identification \eqref{eq:identcharge}. We leave this as an open question for the future. 

\subsection{\texorpdfstring{$\mathbb{Z}^{(0)}_N$}{ZN} generator as a condensation defect}  
\label{subsection:Znascondensation}

In this section, we show how the defect $\mathcal{C}[\Sigma_4]$ defined in \eqref{D-ext-2} can be obtained as a condensation defect via higher gauging of $\mathbb{Z}_N^{(2)}$ on $\Sigma_4$ \cite{Roumpedakis:2022aik}. This is similar to the case of Maxwell theory, with the difference that in the Abelian case the whole $U(1)_I^{(0)}$ is a composite of the magnetic symmetry, whereas in the non-Abelian $PSU(N)$ case only its $\mathbb{Z}_N^{(0)}$ subgroup can be interpreted in this way. Indeed, to the extent that we can view $\mathscr{b}_2$ as a discrete analogue of a current for the $\mathbb{Z}_N^{(2)}$ symmetry, we can also view $\mathcal{C}$ as a discrete analogue of a composite symmetry, being generated by the ``square'' $\mathcal{P}(\mathscr{b}_2)$ of $\mathscr{b}_2$.

Let us consider a condensation defect obtained by higher gauging $\mathbb{Z}_N^{(2)}$ on $\Sigma_4$, in the presence of a discrete torsion \cite{Shao:2023gho}. This is achieved by summing over all inequivalent insertions of defects $U_M[\Sigma_2]$ along $\Sigma_4$, weighted by a discrete torsion term,
\begin{align}
\begin{split}\label{eq:condth} 
    N^{-\chi(\Sigma_4)/2} \frac{|H^0(\Sigma_4,\mathbb{Z}_N)|}{|H^1(\Sigma_4,\mathbb{Z}_N)|} 
    \sum_{\mathscr{c}_2}\,
    \exp \oint_{\Sigma_4}\left(\frac{2\pi i}{N} \mathscr{c}_2 \cup \mathscr{b}_2 + \frac{2\pi i}{2N} \mathcal{P}(\mathscr{c}_2) \right)~,
\end{split}
\end{align}
where $\mathscr{c}_2\in H^2(\Sigma_4,\mathbb{Z}_N)$ and $\chi(\Sigma_4)$ is the Euler characteristic of $\Sigma_4$. In addition to the standard normalization factors of $|H^k(\Sigma_4,\mathbb{Z}_N)|$ for the gauging of a discrete symmetry, we have added for later convenience the factor of $N^{-\chi(\Sigma_4)/2}$. Since $\chi(\Sigma_4)$ is the integral of a local density, this factor can be interpreted as the choice of a local counterterm. The sum over $\mathscr{c}_2$ imposes $\mathscr{c}_2 = -\mathscr{b}_2$, so that the defect in \eqref{eq:condth} gives (see appendix B of \cite{Gaiotto:2014kfa})
\begin{align}
\begin{split}\label{eq:CD}
    N^{-\chi(\Sigma_4)/2} \frac{|H^0(\Sigma_4,\mathbb{Z}_N)|}{|H^1(\Sigma_4,\mathbb{Z}_N)|}
    |H^2(\Sigma_4,\mathbb{Z}_N)|^{1/2} e^{2\pi i \sigma(\Sigma_4)/8}
    \exp\left(-\frac{2\pi i}{2N} \oint_{\Sigma_4} \mathcal{P}(\mathscr{b}_2)\right) = \mathcal{C}[\Sigma_4]
    \ ,
\end{split}
\end{align}
where we have used that the signature of a spin manifold $\Sigma_4$ satisfies $\sigma(\Sigma_4)=0$ mod $16$, and $\chi(\Sigma_4)=2b_0-2b_1+b_2$, where $b_p$ are the Betti numbers of $\Sigma_4$.  

We have thus shown that the condensation defect \eqref{eq:condth} is precisely the generator of the $\mathbb{Z}_N^{(0)}$ composite symmetry that extends $U(1)_I^{(0)}$. Note that the addition of the discrete torsion term $\mathcal{P}(\mathscr{c}_2)$ is crucial to ensure that this condensation defect is invertible: generically, higher-gauging symmetries is achieved by a sum over defect insertions, without any weight, and the result is a non-invertible defect. One needs the additional phase factors due to the term $\mathcal{P}(\mathscr{c}_2)$ in \eqref{eq:condth} to obtain invertibility.\footnote{Another instance of invertible condensation defects comes from 1-gauging a $\mathbb{Z}_2^{(1)}$ 1-form symmetry in a $3d$ QFT \cite{Roumpedakis:2022aik}: if the generating line is fermionic, the symmetry is anomalous in the bulk and the resulting condensation defect is invertible.}

Actually, one can obtain any defect of the form $\mathcal{C}^k[\Sigma_4]$, with $k\in\mathbb{Z}_N$, via a similar higher gauging construction. Indeed, consider the defect
\begin{align}
\begin{split}\label{eq:condth:double}
    \mathcal{D}[k;\Sigma_4] \equiv 
    \frac{1}{|H^2(\Sigma_4,\mathbb{Z}_N)|}
    \sum_{\mathscr{c}_2,\mathscr{b}_2'}
    \exp\oint_{\Sigma_4} \left(
    \frac{2\pi i}{N} 
    \mathscr{c}_2 \cup \mathscr{b}_2
    +\frac{2\pi i}{N} 
    \mathscr{b}_2' \cup \mathscr{c}_2
    -\frac{2\pi ik}{2N}  \mathcal{P}(\mathscr{b}_2') 
    \right)~,
\end{split}
\end{align}
where $\mathscr{c}_2,\mathscr{b}_2'\in H^2(\Sigma_4,\mathbb{Z}_N)$.
The first term corresponds to higher gauging $\mathbb{Z}_N^{(2)}$ on $\Sigma_4$ by coupling to $\mathscr{c}_2$, which produces a dual $\widehat{\mathbb{Z}}^{(1)}_N$ symmetry. The second term is the higher gauging of the latter, by coupling it to $\mathscr{b}_2'$, in presence of a discrete torsion parametrized by $k \in \mathbb{Z}_N$, given by the third term. Summing over $\mathscr{c}_2$ imposes $\mathscr{b}_2'=-\mathscr{b}_2$, yielding
\begin{equation}
    \mathcal{D}[k;\Sigma_4]= 
    \exp \left(-\frac{2\pi ik}{2N} \oint_{\Sigma_4} \mathcal{P}(\mathscr{b}_2)\right) = \mathcal{C}^k[\Sigma_4] \ ,
\end{equation}
as sought. This equality shows that, perhaps surprisingly, this whole set of topological defects obtained by higher gauging have invertible fusion rules.  

We conclude that the mixed anomaly between the electric 1-form symmetry and the instantonic symmetry in the $SU(N)$ theory maps, in the $PSU(N)$ theory, to the extension of the instantonic symmetry by an invertible condensation defect of the magnetic 2-form symmetry. As we already mentioned, a more complete characterization of the symmetries of the $PSU(N)$ theory could be achieved by determining the higher-category, specifically a 4-category, which includes the interrelated 0- and 2-form charges, and may include additional non-invertible defects and junctions \cite{Sheckler:2025rlk}.\footnote{For instance, by analogy with the Maxwell case \cite{Arbalestrier:2024oqg}, we expect that the condensation defect $\mathcal{C}[X_4]$ on an open surface $X_4$ gives rise to a non-invertible topological defect on $\partial X_4$. However, this non-invertible defect does not participate in the fusion of the symmetry defects of the invertible 0-form symmetry $U^\alpha[\Sigma_4]$, that are defined on a closed $\Sigma_4$.} We leave this problem for future investigation.

\subsection{Generalization to \texorpdfstring{$SU(N)/\mathbb{Z}_k$}{SU(N)/Zk}}
\label{subsection:SU(N)/Zk}

In this section we discuss the global variants of the $\mathfrak{su}(N)$ theory with gauge group $SU(N)/\mathbb{Z}_k$, where $k$ is a divisor of $N$. This problem was already discussed in \cite{Gukov:2020btk}, whose approach we follow, providing more details and highlighting a few subtleties. Unlike the limiting cases $k=1$ and $k=N$, these theories have both an electric and a magnetic global symmetry. The electric 1-form symmetry is $\mathbb{Z}_{N/k}^{(1)}$, generated by topological Gukov-Witten operators labeled by the center of the gauge group, $Z(SU(N)/\mathbb{Z}_{k})=\mathbb{Z}_{N/k}$. The charged operators are Wilson lines associated to representations of $SU(N)$ with $N$-ality multiple of $k$. The magnetic 2-form symmetry is $\mathbb{Z}_{k}^{(2)}$, generated by topological operators analogous to the ones in \eqref{SDO-M}, with the difference that now $\mathscr{b}_2$ is a $\mathbb{Z}_k^{(1)}$ gauge field. The charged operators are 't~Hooft surfaces associated to representations of $SU(N)$ with $N$-ality multiple of $N/k$.

These theories also have a $U(1)_I^{(0)}$ instantonic 0-form symmetry, with associated current $\star J_I^{(1)}=\frac{1}{8\pi^2}\text{Tr}(f\wedge f)$. We show below that the periodicity of this $U(1)$ is fixed by the normalization of the charge to be $2\pi k/L$, where 
\begin{equation}
L \equiv \text{gcd}(k,N/k)~.    
\end{equation}
For $k=1$ and $k=N$ we have $L=1$ and we recover the correct periodicities of the $SU(N)$ and $PSU(N)$ global variants, respectively. The charged operators are both pointlike instanton operators, labeled by integer charges, and links of 't~Hooft surfaces, that generically carry a fractional instanton charge. We show below that for $L\neq 1$ the electric, magnetic, and instantonic symmetries together form an anomalous 3-group, in analogy with the Maxwell case.

To derive these results, let us start from the $SU(N)$ theory coupled to background gauge fields. Recall that in the $SU(N)$ theory the instantonic current is quantized in integer multiples of $1/N$ in the presence of a $\mathbb{Z}^{(1)}_N$ background gauge field $\mathcal{B}_2^N\in H^2(\mathcal{M}_5,\mathbb{Z}_N)$,
see \eqref{chargefractionalization}. This was our starting point to derive the extension of the instantonic symmetry in the $PSU(N)$ theory. In order to perform a similar derivation for the $SU(N)/\mathbb{Z}_k$ theory, we need to decompose  $\mathcal{B}_2^N$ in a $\mathbb{Z}^{(1)}_k$ and a $\mathbb{Z}^{(1)}_{N/k}$ part. In order to do so, let us consider the short exact sequence
\begin{equation}\label{Zses}
    1\rightarrow \mathbb{Z}_k \rightarrow \mathbb{Z}_N \rightarrow \mathbb{Z}_{N/k} \rightarrow 1~.
\end{equation}
For $L\neq 1$ the sequence does not split, expressing that $\mathbb{Z}_N$ is a non-trivial extension of $\mathbb{Z}_{N/k}$ by $\mathbb{Z}_k$, whereas for $L=1$ it splits and $\mathbb{Z}_N$ can be expressed simply as the direct product $\mathbb{Z}_k\times \mathbb{Z}_{N/k}$. As we show in appendix \ref{appendix-A}, the sequence \eqref{Zses} implies that the decomposition of the $\mathbb{Z}_N$ gauge field can be written in the form 
\begin{equation}\label{eq:decexp}
\mathcal{B}^N_2 = \frac{N}{k}\widehat{\mathcal{B}}^{k}_2 + \widehat{\mathcal{B}}_2^{N/k}~,
\end{equation}
where $\mathcal{B}_2^r$ is an element of $C^2(\mathcal{M}_5,\mathbb{Z}_r)$, where $C^n(\mathcal{M}_5,\mathbb{Z}_r)$ is the set of $n$-cochains with coefficients in $\mathbb{Z}_r$. The hat denotes the uplift to $\mathbb{Z}_N$ obtained by taking the value mod $N$ between $0$ and $k-1$ and between $0$ and $N/k-1$ for $\mathcal{B}^{k}_2$ and $\mathcal{B}_2^{N/k}$, respectively.
Using that $\delta\mathcal{B}^N_2 = 0$, where $\delta : C^n(\mathcal{M}_5,\mathbb{Z}_r) \rightarrow C^{n+1}(\mathcal{M}_5,\mathbb{Z}_r) $ is the discrete differential, we obtain (see appendix \ref{appendix-A} for details)
\begin{equation}\label{eq:Bockback}
\delta\mathcal{B}^{k}_2 = -\mathrm{Bock}(\mathcal{B}^{N/k}_2)~,
\end{equation}
where $\text{Bock} : H^n(\mathcal{M}_5,\mathbb{Z}_{N/k}) \rightarrow H^{n+1}(\mathcal{M}_5,\mathbb{Z}_k)$ is the Bockstein homomorphism, that characterizes the obstruction to lifting a class in $H^n(\mathcal{M}_5,\mathbb{Z}_{N/k})$ to  a class in $H^n(\mathcal{M}_5,\mathbb{Z}_{N})$.
Equivalently, the $\mathbb{Z}_N$ uplifts satisfy
\begin{equation}\label{eq:Bockbacklift}
    \delta \widehat{\mathcal{B}}_2^{N/k}=\frac{N}{k}\omega~,~~\delta \widehat{\mathcal{B}}_2^k=-\omega~,~~\text{with}~[\omega]_k = \text{Bock}(\mathcal{B}_2^{N/k})~.
\end{equation}

If we activate only the $\mathbb{Z}^{(1)}_k$ background $\mathcal{B}_2^k$, setting ${\mathcal{B}}_2^{N/k}$ to zero, the anomaly inflow \eqref{BGT-anomaly} becomes
\begin{align}\label{bubk}
 \begin{split}
\mathcal{A}_6 = \exp\left(\frac{2\pi i N}{k^2} \int_{\mathcal{Y}_6} 
     \left[\frac{dA_1}{2\pi}\right]_N \cup\frac{\mathcal{P}(\widehat{\mathcal{B}}_2^k)}{2}\right)~.
\end{split} 
\end{align}
Note that the expression $\mathcal{P}(\widehat{\mathcal{B}}_2^k)$ is well-defined because $\delta \widehat{\mathcal{B}}_2^k = 0$ for $\widehat{\mathcal{B}}_2^{N/k}=0$.
From eq.~\eqref{bubk} we derive, analogously to \eqref{D-ext}, that
\begin{equation}\label{extendedU_SUNZk}
 U^{\alpha+ 2\pi}_I[ \Sigma_4,\mathcal{B}_2^k] = 
    U^{\alpha}_I[\Sigma_4,\mathcal{B}_2^k]
    \exp\left(-\frac{2\pi iN}{k^2}
    \oint_{\Sigma_4} \frac{\mathcal{P}(\widehat{\mathcal{B}}_2^k)}{2}\right)  \ .
\end{equation}
Noticing that
\begin{equation}
U^{2\pi \ell}_I[ \Sigma_4,\mathcal{B}_2^k]=\exp\left(-\frac{2\pi i \ell N}{k^2}\oint_{\Sigma_4}\frac{\mathcal{P}(\widehat{\mathcal{B}}_2^k)}{2}\right)=1~,
\end{equation}
if and only if $\ell$ is a multiple of $k/L$, the periodicity of $\alpha$ is now $\alpha \sim \alpha + 2\pi k/L$.
Therefore, eq.~\eqref{extendedU_SUNZk} implies that in the presence of a $\mathbb{Z}_k^{(1)}\subset\mathbb{Z}_N^{(1)} $ electric background the instantonic charge of the $SU(N)$ theory is quantized as
\begin{equation}\label{eq:perk}
Q_I[\Sigma_4,\mathcal{B}_2^k] \in \frac{1}{k/L} \mathbb{Z}~. \end{equation}
This is a generalization of \eqref{chargefractionalization}, and reduces to it when $k=N$. Notice that if $N$ is a multiple of $k^2$ (namely, if $L=k$) there is no extension of the original $2\pi$ periodicity of the $U(1)_I^{(0)}$ symmetry of the $SU(N)$ theory, even in the presence of a $\mathbb{Z}_k^{(1)}$ background field. This happens because the mixed 't~Hooft anomaly \eqref{BGT-anomaly} trivializes when restricting the $\mathbb{Z}^{(1)}_N$ background field to the one for the $\mathbb{Z}^{(1)}_k$ subgroup, when $k^2$ divides $N$.

We now turn to the analysis of the $SU(N)/\mathbb{Z}_k$ theory by promoting $\mathcal{B}_2^k$ to a dynamical gauge field $\mathscr{b}_2^k$. An immediate consequence of the considerations above is that, if we ignore the $\mathbb{Z}^{(1)}_{N/k}$ symmetry by setting to zero the background gauge field ${\mathcal{B}}_2^{N/k}$, the $U(1)_I^{(0)}$ instantonic symmetry of the $SU(N)/\mathbb{Z}_k$ theory is a $\mathbb{Z}_{k/L}$ extension of that of the $SU(N)$ theory. The elements of $\mathbb{Z}_{k/L}$ are powers of the condensation defect of the 2-form symmetry
\begin{equation}
\mathcal{C}[\Sigma_4] = \exp\left(- \frac{2\pi i N}{k^2}\oint_{\Sigma_4}\frac{\mathcal{P}(\widehat{\mathscr{b}}_2^k)}{2}\right)~.
\end{equation}
This is completely analogous to the discussion in the $PSU(N)$ theory. Note that, as a consequence of the extension of $U(1)^{(0)}_I$ by $\mathbb{Z}_{k/L}$, the fluxes of $A_1$ are integer multiples of $k/L$.

The symmetry structure becomes more interesting if we also take into account the electric $\mathbb{Z}^{(1)}_{N/k}$ 1-form symmetry. The partition function of the $SU(N)/\mathbb{Z}_k$ theory coupled to all the background gauge fields, $A_1, \mathcal{B}_2^{N/k}$, and $\mathcal{C}_3$, the latter being the one associated to the $\mathbb{Z}^{(2)}_k$ magnetic 2-form symmetry, is
\begin{align}
\begin{split}\label{ZofSU(N)Zk_old}
     Z_{SU(N)/\mathbb{Z}_k}[ A_1,\mathcal{B}_2^{N/k}, \mathcal{C}_3] 
    & =
    \frac{|H^0(\mathcal{M}_5,\mathbb{Z}_k)|}{|H^1(\mathcal{M}_5,\mathbb{Z}_k)|} \sum_{\mathscr{b}_2^k} Z_{SU(N)}\left[A_1,\frac{N}{k}\widehat{\mathscr{b}}^{k}_2 + \widehat{\mathcal{B}}_2^{N/k}\right]\ \times \phantom{\left[\frac{dA_1}{2\pi}\right]_N}\\
    &\exp\left(\frac{2\pi i}{k}\oint_{\mathcal{M}_5} \mathscr{b}_2^k\cup\mathcal{C}_3-\frac{2\pi i}{k}\int_{\mathcal{M}_5}(\widehat{\mathcal{B}}_2^{N/k}\cup_1 \widehat{\mathscr{b}}_2^k)\cup \left[\frac{dA_1}{2\pi}\right]_N \right ) \ ,
\end{split}
\end{align}
where, in the argument of $Z_{SU(N)}[A_1,\mathcal{B}^N_2]$, we have substituted the decomposition \eqref{eq:decexp}. From \eqref{eq:Bockback} we get that the sum over $\mathscr{b}_2^k$ now runs over configurations satisfying
\begin{equation}\label{eq:delb2}
\delta \mathscr{b}_2^k = -\mathrm{Bock}(\mathcal{B}_2^{N/k})~.
\end{equation}
The partition function \eqref{ZofSU(N)Zk_old} is such that (1) the sum over $\mathscr{b}_2^k$ is well-defined, namely the summand is gauge invariant under $\mathscr{b}_2^k$ gauge transformations; and (2) the summand only picks up a $\mathscr{b}_2^k$-independent phase under background gauge transformations, ensuring that we can consistently couple to the background gauge fields for the global symmetries. In order to satisfy these conditions we need to include the coupling of the dynamical gauge field $\mathscr{b}_2^k$ to background gauge fields via the non-conventional term with the higher cup product $\cup_1$ (see e.g.~\cite{Benini:2018reh} for a review). Moreover, we need to modify the gauge transformation of the background gauge field $\mathcal{C}_3$ to include a $\mathbb{Z}^{(1)}_{N/k}$ gauge transformation, and modify consistently the flatness constraint to ensure gauge invariance, obtaining
\begin{equation}\label{H4_SU(N)/Zk}
\delta \mathcal{C}_3 - \left[ \widehat{\mathcal{B}}^{N/k}_2 \cup \left[\frac{d A_1}{2\pi}\right]_N  \right]_k = 0~.
\end{equation}
This equation signals the existence of the 3-group involving the 0-, 1-, and 2-form symmetries of the theory. Note that the additional term in \eqref{H4_SU(N)/Zk} is trivial when $L=1$, because in that case the flux of $A_1$ is a multiple of $k$. As a result the 3-group structure trivializes for $L=1$. One can further compute the phase picked by the partition function \eqref{ZofSU(N)Zk_old} under background gauge transformations, and see that there is a specific anomaly for the 3-group symmetry, that can be canceled by the following anomaly-inflow action\footnote{This anomaly inflow differs from the one found in \cite{Gukov:2020btk} in the coefficient of the mixed 't~Hooft anomaly between the electric and the instantonic symmetry. This is perhaps due to a typo in \cite{Gukov:2020btk}.}
\begin{equation}\label{SU/N)/Zk_anomaly_old}
    \mathcal{A}^{SU(N)/\mathbb{Z}_k}_6 =   \text{exp}\left(
   \frac{ 2\pi i}{N}\int_{\mathcal{Y}_6} \left[\frac{dA_1}{2\pi}\right]_{N}
    \cup
    \frac{\mathcal{P}(\widehat{\mathcal{B}}^{N/k}_2)}{2} +\frac{2\pi i}{k} \int_{\mathcal{Y}_6} \text{Bock}(\mathcal{B}_2^{N/k})\cup \mathcal{C}_3 \right)~,
\end{equation}
where $\mathcal{P}(\widehat{\mathcal{B}}^{N/k}_2)$ is now the affine Pontryagin square \cite{Benini:2018reh}, since $\widehat{\mathcal{B}}^{N/k}_2$ is not closed. This anomaly inflow can be checked to satisfy the consistency condition of \cite{Benini:2018reh}, namely the cochain appearing in the inflow action is $\delta$-closed. We provide a detailed calculation in appendix \ref{appendix-A} that verifies all of the properties of \eqref{ZofSU(N)Zk_old} mentioned above. Note that the mixed 't~Hooft anomaly between the electric and instantonic symmetries, captured by the first term in \eqref{SU/N)/Zk_anomaly_old}, correctly reduces to \eqref{BGT-anomaly} for $k=1$ and vanishes for $k=N$. Instead, setting $A_1 = 0$ we are left only with a mixed electric-magnetic 't~Hooft anomaly. When $L=1$ this anomaly can be reabsorbed by a counterterm and thus trivializes.\footnote{This is the non-Abelian version of the fact that when restricting the $U(1)_{E}\times U(1)_M$ symmetry of Maxwell theory to a $\mathbb{Z}_p \times \mathbb{Z}_q$ subgroup, the usual mixed 't~Hooft anomaly trivializes when gcd$(p,q) = 1$ \cite{Niro:2022ctq,Cordova:2023ent}.}

\subsection{Supersymmetric theory and the global forms of the \texorpdfstring{$E_1$}{E1} SCFT}
\label{E1_hat_FT}

In this section we discuss the supersymmetric version of the theory, i.e.~$\mathcal{N}=1$ pure SYM with gauge algebra $\mathfrak{su}(N)$. This theory can be UV-completed by an SCFT \cite{Seiberg:1996bd, Morrison:1996xf,Intriligator:1997pq,Jefferson:2017ahm,Jefferson:2018irk}. While for $N>2$ the algebra of the instantonic symmetry is expected to remain $\mathfrak{u}(1)_I$  \cite{Tachikawa:2015mha,Cremonesi:2015lsa}, for $N=2$ it enhances to $\mathfrak{su}(2)_I$ at the UV fixed point, which is described by the $E_1$ theory \cite{Seiberg:1996bd}.\footnote{In addition to the instantonic 0-form symmetry there is also an $R$-symmetry with algebra $\mathfrak{su}(2)_R$, and it is believed that in the $E_1$ theory the whole 0-form symmetry enhances to $SO(4)$ (see e.g.~\cite{Akhond:2024nyr}). We will ignore the $R$-symmetry in our considerations and we will then focus only on the  $\mathfrak{su}(2)_I$ instantonic symmetry.}  We will use our previous results to constrain the global structure of the symmetries of the SCFTs which UV complete $\mathfrak{su}(N)$ SYM. 

Let us first consider $N>2$. Since in this case the UV SCFT has a $\mathfrak{u}(1)_I$ instantonic symmetry like the IR gauge theory, if we further assume that none of the 1- and/or 2-form symmetries are emergent, the most natural possibility is that the symmetry structure in the UV matches the one of the IR gauge theory. 
For instance, for the $SU(N)$ global form this means that the anomaly \eqref{BGT-anomaly} involving the instantonic $U(1)^{(0)}$ and the electric $\mathbb{Z}_N^{(1)}$ is also present in the SCFT. For $PSU(N)$, a $\mathbb{Z}_N^{(0)}$ subgroup of $U(1)^{(0)}$ generated by a condensation defect of the magnetic $\mathbb{Z}_N^{(2)}$ symmetry persists at the UV fixed point. Similar remarks apply to $SU(N)/\mathbb{Z}_k$ with $k \not = 1,N$. 

For $N=2$ the analysis needs to be refined due to the enhancement of the instantonic symmetry. As shown in \cite{BenettiGenolini:2020doj}, assuming that the $\mathbb{Z}_2^{(1)}$ symmetry is not emergent, the global form of the $\mathfrak{su}(2)^{(0)}_I$ instantonic symmetry of the $E_1$ theory must be $SO(3)$ to match the anomaly \eqref{BGT-anomaly} of the IR theory (similar conclusions were reached in  \cite{Kim:2012gu, Bashkirov:2012re, Rodriguez-Gomez:2013dpa, Cremonesi:2015lsa, Apruzzi:2021vcu} by different arguments). This is because only if the global form is $SO(3)_I^{(0)}$ one can write the anomaly polynomial
\begin{equation}\label{eq:anoinfUV} 
   \mathcal{A}_6^{\text{UV}} = \exp \left( \frac{2\pi i}{2} \int_{\mathcal{Y}_6} 
    w_2(SO(3)_I)
    \cup
    \frac{\mathcal{P}(\mathcal{B}_2)}{2} \right) 
    \ ,
\end{equation}
where $w_2(SO(3)_I)$ is the second Stiefel-Whitney class of the $SO(3)_I^{(0)}$ bundle. Under the symmetry breaking pattern $SO(3)^{(0)}_I\to U(1)_I^{(0)}$ triggered by the relevant deformation that drives the SCFT to the IR gauge theory phase, we have
\begin{equation}
    w_2(SO(3)_I) = \left[\frac{d A_1}{2\pi}\right]_2~.
\end{equation}
This implies that $\mathcal{A}_6^{\text{UV}}$ reduces to \eqref{BGT-anomaly} and anomalies match. We can use the inflow \eqref{eq:anoinfUV} to import the discussion around \eqref{extension_IR} to the UV fixed point. Consider the defects $U^{\alpha \vec{n}\cdot \vec{T}}_I[ \Sigma_4]$ parametrized by a unit 3-vector $\vec{n}$ ($\vec{n}\cdot\vec{n}=1$) and the real number $\alpha$, where $\vec{T}$ are the generators of $\mathfrak{su}(2)$. With the standard normalization the generator has integer eigenvalues, giving the identification $\alpha\sim \alpha+2\pi$. In the presence of a non-trivial 1-form symmetry background we have
\begin{align}\label{extension_UV-1}
\begin{split}
U^{(\alpha+2\pi)\vec{n}\cdot \vec{T}}_I[ \Sigma_4,\mathcal{B}_2] 
&  = U^{\alpha \vec{n}\cdot \vec{T}}_I[\Sigma_4,\mathcal{B}_2]  \exp\left(-\frac{2\pi i}{2}
    \oint_{\Sigma_4} \frac{\mathcal{P}(\mathcal{B}_2) }{2}\right)
~.
\end{split}    
\end{align}
With reference to figure \ref{fig:S1M5}, we are using that the ratio between $U^{(\alpha+2\pi)\vec{n}\cdot \vec{T}}_I$ and $U^{\alpha\vec{n}\cdot \vec{T}}_I$ gives an $SO(3)$ bundle with two patches on a 2-sphere, related by a gauge transformation with parameter $\alpha$ that winds once around the equator, i.e.~$\alpha\to\alpha + 2\pi$. As a result, the bundle has $w_2 = 1~\text{mod}\,2$ on the 2-sphere. Note that in the SCFT the use of the anomaly polynomial is the only way to derive this equation, unlike in the IR gauge theory where we could provide a simpler derivation based on the quantization condition of the instantonic current. Despite \eqref{extension_UV-1}, in the $E_1$ theory the periodicity of the parameter appearing in the topological operator is still $\alpha\sim \alpha +2\pi$, as appropriate for an $SO(3)$ global symmetry, because the violation in   \eqref{extension_UV-1} is only a local functional of the background source $\mathcal{B}_2$.

We now gauge $\mathbb{Z}_2^{(1)}$ to obtain the SCFT that UV completes SYM with gauge group $SO(3)$. We use the symbol $\widehat{E}_1$ for this theory. The $\widehat{E}_1$ SCFT has a $\mathbb{Z}_2^{(2)}$ 2-form symmetry, corresponding to the Pontryagin dual to $\mathbb{Z}_2^{(1)}$. In this theory \eqref{extension_UV-1} reads
\begin{align}\label{extension_UV-2}
\begin{split}
    U^{(\alpha+2\pi)\vec{n}\cdot \vec{T}}_I[ \Sigma_4] 
    &  = U^{\alpha \vec{n}\cdot \vec{T}}_I[\Sigma_4]  \exp\left(-\frac{2\pi i}{2}
    \oint_{\Sigma_4} \frac{\mathcal{P}(\mathscr{b}_2) }{2}\right)
    \equiv U^{\alpha \vec{n}\cdot \vec{T}}_I[\Sigma_4] \mathcal{C}[\Sigma_4]
     \ ,
\end{split}    
\end{align}
where $\mathscr{b}_2$ is the gauge field of $\mathbb{Z}_2^{(1)}$ which is now dynamical. Therefore, the identification $\alpha\sim\alpha+2\pi$ is now violated by the operator $\mathcal{C}[\Sigma_4]$ rather than a $c$-number. The same derivation as for the low-energy SYM theory shows that  $\mathcal{C}[\Sigma_4]$ is a condensation defect (with discrete torsion) of the $\mathbb{Z}_2^{(2)}$ symmetry. As a result of \eqref{extension_UV-2} the correct periodicity becomes $\alpha\sim\alpha + 4\pi$, which implies that now 
the global form of the $\mathfrak{su}(2)_I$ symmetry algebra is $SU(2)_I^{(0)}$. Thus the symmetry $SO(3)_I^{(0)}$ of the $E_1$ theory is extended to $SU(2)_I^{(0)}$ in the $\widehat{E}_1$ theory, namely these groups take part in the non-split short exact sequence
\begin{equation}\label{}
    1\rightarrow \mathbb{Z}_2 
    \xrightarrow{} SU(2)_I
    \xrightarrow{} SO(3)_I 
    \rightarrow 1 \ ,
\end{equation}
where the $\mathbb{Z}_2$ central subgroup of $SU(2)$ is generated by $\mathcal{C}[\Sigma_4]$. 

In analogy to what we discussed in the IR theory, the symmetry extension, being generated by a condensation defect, does not act on any pointlike operators. An important consequence is that one cannot argue for this symmetry enhancement using the superconformal index. To detect it with supersymmetric partition functions one would need to consider manifolds that, unlike $S^1\times S^4$, support fractional instantons.  

We can  use \eqref{extension_UV-2} to derive the action of $\mathcal{C}[\Sigma_4]$ on (linked) surface operators. In $\widehat{E}_1$ there are surface operators $\widehat{H}[\gamma_2]$ charged under $\mathbb{Z}^{(2)}_2$. Under the relevant deformation that makes $\widehat{E}_1$ flow to $SO(3)$ SYM, these operators flow to 't~Hooft surfaces in representations $\rho$ of $SU(2)$ with integer/half-integer spin depending on whether they are even/odd under $\mathbb{Z}^{(2)}_2$.  If we take $\mathcal{C}[\Sigma_4]$ and two magnetic operators $\widehat{H}[\gamma_2],\widehat{H}'[\gamma'_2]$ as in figure \ref{fig:triplelinking} we have
\begin{equation}\label{E1Hat-activated}
    \mathcal{C}[\Sigma_4] \widehat{H}[\gamma_2]\widehat{H}'[\gamma'_2] = e^{i\pi m m'}
    \widehat{H}[\gamma_2]\widehat{H}'[\gamma'_2]~,
\end{equation}
where $m,m'$ are their respective $\mathbb{Z}^{(2)}_2$ charges. If we take instead $\Sigma_4=S^4$ we have
\begin{equation}\label{E1Hat-S4}
    \mathcal{C}[S^4] \widehat{H}[\gamma_2]\widehat{H}'[\gamma'_2] = \widehat{H}[\gamma_2]\widehat{H}'[\gamma'_2] \ .
\end{equation}
Equations \eqref{E1Hat-activated} and \eqref{extension_UV-2} express that a 0-form symmetry defect can detect a projective representation of $SO(3)_I$, i.e.~an $SU(2)_I$ representation, if it is triple-linking with the surface operators as in figure \ref{fig:triplelinking}, while it can only detect ordinary (non-projective) representations if the triple-linking is not activated. To satisfy both of these constraints, we assign irreducible representations of $SU(2)_I$ to $\gamma_2$ and $\gamma_2'$ with spin $j,j'\in \mathbb{N}/2$ such that
\begin{equation}\label{E1-hat-charges}
    [2j]_2 = [2j']_2 = m m' \ . 
\end{equation}
The positive integers $2j$ and $2j'$ play a role similar to $\mathbb{Z}$ uplifts $\widetilde{m}_\rho$ and $\widetilde{m}_{\rho'}$ in the $PSU(N)$ case of section \ref{gaug-PSU}.\footnote{More precisely, one can think of a configuration of 't Hooft surfaces in $PSU(N)$ with insertions of instanton operators of different charges $q$ and $q'$ on their support. Then, $2j$ and $2j'$ are the analogue of $\widetilde{m}_\rho\widetilde{m}_{\rho'} + N q$ and $\widetilde{m}_\rho\widetilde{m}_{\rho'} + N q'$, respectively. Here we are considering generic $q,q'$ because there is no analogue of having a trivial net instanton charge in the non-Abelian case, i.e.~the tensor product of two representations of half-integer spin will always contain also non-trivial representations.} When both surfaces are odd under $\mathbb{Z}^{(2)}_2$, i.e.~$m=m'=1~\text{mod }2$, both $2j$ and $2j'$ are odd and we reproduce the minus sign under a $2\pi$ rotation in \eqref{E1Hat-activated}. When instead either one of the surfaces is even under $\mathbb{Z}^{(2)}_2$, both $2j$ and $2j'$ are even, corresponding to representations of $SO(3)_I$, and consistently there is no sign under a $2\pi$ rotation in \eqref{E1Hat-activated}. Irrespectively of the parity of the surfaces, when $U_I^{\alpha \vec{n}\cdot\vec{T}}[\Sigma_4]$ wraps both surfaces, e.g.~with an $S^4$ topology, it acts in the representation $j\otimes j'$ of integer spin, consistently with the absence of a sign under a $2\pi$ rotation in \eqref{E1Hat-S4}. 

In section \ref{gaug-PSU} we discussed how in the $PSU(N)$ theory the identification $\widetilde{m}_{\rho,\rho'} \sim \widetilde{m}_{\rho,\rho'}+ N$ is only recovered at the price of inserting pointlike instanton operators on the 't~Hooft surfaces. Similarly here, different choices of the spins $j_1,j'_1$ and $j_2,j_2'$, both satisfying \eqref{E1-hat-charges}, are related to each other by picking points $P\in \gamma_2$, $P'\in \gamma'_2$, and inserting there instanton operators in representations $J$ and $J'$ of integer spin, such that

\begin{equation}
j_1 \otimes J \supseteq j_2~,~~j'_1 \otimes J' \supseteq j_2'~.
\end{equation}

\subsubsection{Comments about the 2-group structure in the $E_1$ theory} \label{sect:2-group-E1}
In \cite{Apruzzi:2021vcu} it was proposed that in the $E_1$ theory the $\mathbb{Z}_2^{(1)}$ 1-form symmetry and the $SO(3)_I^{(0)}$ 0-form symmetry together form a 2-group, which can be encoded in the relation
\begin{equation}\label{eq:E12g}
\delta \mathcal{B}_2 = w_3(SO(3)_I)~,
\end{equation}
between the background gauge fields of the two symmetries. Here $w_3$ is the third Stiefel-Whitney class of the $SO(3)$ bundle.

However, it is not immediately clear how to reconcile this 2-group structure with the 't~Hooft anomaly \eqref{eq:anoinfUV}. One possibility is that the two structures fit together giving rise to an anomalous 2-group. A non-trivial constraint on the anomalies of discrete 2-groups is that the anomaly-inflow action should be closed \cite{Benini:2018reh}. While the 0-form symmetry is continuous in this example, only discrete data of the $SO(3)_I^{(0)}$ bundles enter the anomaly \eqref{eq:anoinfUV}, and therefore the constraint should apply in this case as well. Clearly, the anomaly-inflow action in \eqref{eq:anoinfUV} is not closed if the background field $\mathcal{B}_2$ satisfies the 2-group relation \eqref{eq:E12g}. However, there might be extra terms that can be built out of $\mathcal{B}_2$ and the discrete data of the $SO(3)_I^{(0)}$ bundle, and can be added to the anomaly-inflow action, such that: (1) it is still possible to match the IR anomaly \eqref{BGT-anomaly} when we flow to the IR gauge theory, breaking $SO(3)_I^{(0)}$ to $U(1)_I^{(0)}$, and (2) they make the total inflow action closed. We could not find any suitable extension of the anomaly-inflow action with these features, but it would be worth to investigate this issue further. Note that, if there is an anomalous 2-group, the symmetry structure of the $\widehat{E}_1$ theory would also be affected and the considerations in the previous section would need to be amended to take this into account.

An independent indication of this tension is that the presence of the 't~Hooft anomaly may invalidate some steps in the derivation of the 2-group structure in \cite{Apruzzi:2021vcu}, as we now explain. As a starting point, the authors of \cite{Apruzzi:2021vcu} consider the Coulomb branch of the $E_1$ theory. They argue that at a generic point on the Coulomb branch the effective theory is given by a $U(1)$ gauge theory with massive particles of charge 2 in the fundamental representation of $\mathfrak{su}(2)_I$. In this theory, gauge invariant pointlike operators only carry $SO(3)_I$ representations. Moreover the Wilson lines of charge $4$ can be screened by making them end on pointlike operators in $SO(3)_I$ representations, but those of charge $2$ can only end on operators which also carry charge under the $\mathbb{Z}_2$ center of $SU(2)_I$, i.e.~they are in $SU(2)_I$ representations which are not $SO(3)_I$ ones (in particular, they cannot be singlets). One therefore finds a short exact sequence 
\begin{equation}\label{2-group-sequence}
    1\rightarrow \mathbb{Z}_2 
    \xrightarrow{} \mathbb{Z}_4 
    \xrightarrow{}\mathbb{Z}_2  
    \rightarrow 1 \ ,
\end{equation}
expressing that the $\mathbb{Z}_2$ charges of the unscreened Wilson lines, associated to the unbroken $\mathbb{Z}^{(1)}_2$ 1-form symmetry, are extended by the center of $SU(2)_I$ to form the $\mathbb{Z}_4$ charges of Wilson lines that are unscreened by operators in representations of $SO(3)_I$. This state of affairs signals the existence of a 2-group in which \eqref{eq:E12g} holds, as explained in \cite{Bhardwaj:2021wif,Lee:2021crt,DelZotto:2022joo}. Given that the 1-form and the 0-form symmetries exist both on the Coulomb branch and at its origin, where the $E_1$ SCFT resides, it is then natural to assume that this structure persists in the SCFT.

The way the anomaly affects this argument is by providing us with additional configurations that carry $SU(2)_I$ representations. In the previous section we have discussed how in the $\widehat{E}_1$ global form, obtained from $E_1$ by gauging $\mathbb{Z}_2^{(1)}$, there are configurations of linking surface operators which are required, by consistency with the anomaly in $E_1$, to carry generic $SU(2)_I$ representations. These surface operators are not gauge invariant by themselves in the original $E_1$ theory with $\mathbb{Z}_2^{(1)}$ symmetry, but they still exists also in $E_1$ as non-genuine operators that are attached to 3-dimensional Gukov-Witten operators. Assuming that moving along the Coulomb branch does not change the properties of these non-genuine surface operators, we can terminate the charge-2 Wilson line on a point of one of the two linked magnetic surfaces, where we also insert the scalar in the fundamental representation of $\mathfrak{su}(2)_I$. In this way we can build a gauge invariant junction that is also uncharged under the center of $SU(2)$, thus screening the charge-2 Wilson line.
Therefore, as a consequence of the anomaly, it is no longer necessary to use operators that carry charge under the center of $SU(2)_I$ to screen the line of charge 2, and the aforementioned argument for the 2-group structure in the SCFT does not apply.

\section{Symmetry Topological Field Theory}
\label{Section:SymTFT}

The Symmetry Topological Field Theory is a $(d + 1)$-dimensional topological theory which encodes the RG-invariant data related to the (generalized) global symmetries of all possible global variants of a given $d$-dimensional QFT \cite{Freed:2012bs,Gaiotto:2020iye,Apruzzi:2021nmk,Freed:2022qnc}. The SymTFT is placed on a slab between two $d$-dimensional boundaries: a physical boundary, where it is coupled to the QFT, and a topological boundary, which imposes topological boundary conditions. These boundary conditions determine which bulk operators can end at the topological boundary -- defining charged operators after the slab compactification -- and which ones can be pushed to the boundary -- defining the symmetry defects of the QFT. From the fusion and the braiding between these operators, one can extract the symmetry structure of the QFT and all possible charges carried by its dynamical, non-topological operators for any given global variant. 

In this section, we show how our previously described results fit within the SymTFT framework. We begin by analyzing ordinary Yang-Mills theory with gauge algebra $\mathfrak{su}(N)$, and then turn to the case of $\mathcal{N}=1$ $\mathfrak{su}(2)$ supersymmetric Yang-Mills and its UV fixed point.

\subsection{SymTFT for \texorpdfstring{$\mathfrak{su}(N)$}{su(2)} Yang-Mills}

For five-dimensional Yang-Mills theory with gauge algebra $\mathfrak{su}(N)$, we propose the six-dimensional SymTFT described by the (Euclidean) action
\begin{equation}\label{symtftaction}
S_{\rm SymTFT} = \int_{\mathcal{Y}_6} \left(-\frac{iN}{2\pi} B_2 \wedge dC_3 +  \frac{i}{2\pi} b_4 \wedge dA_1 + \frac{iN}{8\pi^2} dA_1 \wedge B_2 \wedge B_2 \right) \ .     
\end{equation}
The dynamical connections $A_1$, $B_2$, and $C_3$ are $U(1)$ gauge fields, whose fluxes are quantized in $2\pi\mathbb{Z}$, whereas $b_4$ is an $\mathbb{R}$ gauge field, whose fluxes identically vanish. The first term is the usual $U(1)$ BF term that describes $\mathbb{Z}_N$ (generalized) symmetries, and subgroups thereof, on the boundary QFT. The second term is a mixed $U(1)/\mathbb{R}$ BF term that describes $U(1)$ symmetries \cite{Brennan:2024fgj,Antinucci:2024zjp}. Finally, the third term is the gauged version of the anomaly-inflow action that describes the mixed 't~Hooft anomaly \eqref{BGT-anomaly}. Roughly, $\frac{NB_2}{2\pi}$ and $\frac{NC_3}{2\pi}$ should be thought of as the continuous versions -- dynamical in the bulk -- of the discrete background gauge fields $\mathcal{B}_2$ and $\mathcal{C}_3$ of the electric $\mathbb{Z}^{(1)}_N$ and the magnetic $\mathbb{Z}^{(2)}_N$ symmetries of the $SU(N)$ and $PSU(N)$ global variant of the boundary QFT, respectively, whereas $A_1$ is the bulk version of the $U(1)_I^{(0)}$ background gauge field.%
\footnote{Interestingly, one can consider the theory on $\mathcal{M}_4 \times S^2$, where $\mathcal{M}_4$ is a spin manifold and $S^2$ is a sphere with unit volume form $\Omega_2$. In presence of a non-trivial flux $\oint_{S^2} dA_1 = 2\pi \ell$, one can perform a dimensional reduction on $S^2$, by taking $C_3=C_1 \wedge \Omega_2$ and both $B_2$ and $C_1$ to have only components along $\mathcal{M}_4$, to obtain the continuous description of a $\mathbb{Z}_N$ Dijkgraaf–Witten theory with twist $\ell$, studied in \cite{Kapustin:2014gua}. This corresponds to the 4$d$ SymTFT describing the global variants of a 3$d$ Yang-Mills-Chern-Simons theory with gauge algebra $\mathfrak{su}(N)$ and Chern-Simons level $k$ such that $\ell=k \text{ mod }N$, studied in \cite{Argurio:2024oym}.\label{footnote::30}}
The gauge transformations are\footnote{Notice that the gauge transformations of $C_3$ are well-defined. Indeed, the ambiguity of $A_1$ is a closed form $d\Lambda_0$ with periods that are integer multiples of $2\pi$. This ambiguity produces a shift $C_3 \rightarrow C_3 + \frac{1}{2\pi} d\Lambda_1 \wedge d\Lambda_0$, which is correctly quantized and can be reabsorbed by a $d\Lambda_2$ gauge transformation. Moreover, under gauge transformations $dC_3$ shifts as $dC_3 \rightarrow dC_3 + \frac{1}{2\pi}d\Lambda_1 \wedge dA_1$, preserving  the correct flux quantization condition.}
\begin{equation}\label{symtftgaugetransf}
B_2 \rightarrow B_2+d\Lambda_1 \,, \quad
C_3 \rightarrow C_3+d\Lambda_2  + \frac{1}{2\pi}d\Lambda_1 \wedge A_1 \,, \quad
A_1 \rightarrow A_1+d\Lambda_0 \,, \quad
b_4 \rightarrow b_4+d\lambda_3 \ ,
\end{equation}
where $\oint d\Lambda_i \in 2\pi\mathbb{Z}$ and $\oint d\lambda_3 =0$. Under these gauge transformations the action \eqref{symtftaction} transforms as a local total derivative,
\begin{equation}\label{symtftactionvariation}
S_{\rm SymTFT} \, \rightarrow \, S_{\rm SymTFT} + \int_{\mathcal{Y}_6} \left(-\frac{iN}{2\pi} d\Lambda_1 \wedge dC_3 +  \frac{i}{2\pi} d\lambda_3 \wedge dA_1 - \frac{iN}{8\pi^2} dA_1 \wedge d\Lambda_1 \wedge d\Lambda_1 \right) \ .
\end{equation}
Recalling the quantization conditions of the gauge fields and their transformations, it follows that the variation above is an integer multiple of $2\pi$ on a closed manifold $\mathcal{Y}_6$.\footnote{This is always true for even $N$. As we want to connect this theory to the 5$d$ analysis, we can assume for simplicity that any oriented four-cycle $\Sigma_4$ is spin, such that the last term of the gauge variation \eqref{symtftactionvariation} is a multiple of $2\pi$ also for odd $N$.\label{footnote:evenoddN}} Hence, the theory is gauge invariant. All gauge fields have gauge invariant field strengths with the exception of $C_3$, for which the naive field strength needs to be modified to the gauge invariant combination $dC_3-\frac{1}{2\pi}B_2 \wedge dA_1$. The non-trivial gauge transformation of $dC_3$ signal a 3-group bundle which can potentially produce a higher-group symmetry after the slab compactification (this will indeed be the case for some global variants discussed below). Note that all field strengths vanish on-shell.  

Topological operators are given by the Wilson surfaces of $B_2$, $A_1$, and $b_4$, defined on closed manifolds $\Sigma_i$ as
\begin{equation}\label{W's-bulk}
W_2^n[\Sigma_2] = \exp\left(in \oint_{\Sigma_2} B_2\right) \,, \
W_1^p[\Sigma_1] = \exp\left(ip \oint_{\Sigma_1} A_1\right) \,, \
W_4^\alpha[\Sigma_4] = \exp\left(\frac{i\alpha}{2\pi} \oint_{\Sigma_4} b_4\right) \,,
\end{equation}
where gauge invariance requires $m,p\in\mathbb{Z}$, whereas $\alpha\in\mathbb{R}$. Instead, the Wilson surface of $C_3$ is not gauge invariant and needs to be dressed as
\begin{equation} \label{W3-bulk}
\widetilde{W}_3^m[\Sigma_3,D_4] = \exp\left(im \left(\oint_{\Sigma_3}  C_3 - \frac{1}{2\pi} \int_{D_4} B_2 \wedge dA_1 \right)\right) \ ,
\end{equation}
where $D_4$ is an open surface with $\partial D_4=\Sigma_3$ -- which is thus a homologically trivial closed manifold -- and gauge invariance requires $m\in\mathbb{Z}$. Notice that the dependence of $\widetilde{W}_3^m$ on $D_4$ is only activated by insertion of $W_4^\alpha[\Sigma_4]$, which imposes $dA_1=\alpha \,\delta^{(2)}(\Sigma_4)$. Without insertion of $W_4^\alpha[\Sigma_4]$ we have that $dA_1=0$ on-shell, and thus $\widetilde{W}_3^m$ only depends on $\Sigma_3$. 

The non-trivial bulk correlation functions of the theory can be computed to be
\begin{equation}\label{SymTFT_linkings}
\begin{split}
\langle W_2^n[\Sigma_2]\widetilde{W}_3^m[\Sigma_3,D_4] \rangle &= \exp\left(-\frac{2 \pi i m n}{N}  L_2(\Sigma_2,\Sigma_3)\right)
\ , \\
\langle W_1^p[\Sigma_1] W_4^\alpha[\Sigma_4] \rangle &= \exp\left( i \alpha p \, L_2(\Sigma_1,\Sigma_4) \right)\ , \\
\langle W_4^\alpha[\Sigma_4] \widetilde{W}_3^m[\Sigma_3,D_4] \widetilde{W}_3^{m'}[\Sigma_3',D_4']  \rangle &= \exp\left(\frac{i\alpha mm'}{N} L_3(\Sigma_4,\Sigma_3,\Sigma'_3)\right)\ .
\end{split}
\end{equation}
From the correlation functions, we conclude that the labels of the operators in \eqref{W's-bulk} are subject to the identifications
\begin{equation}
n \sim n+N \ , \qquad \alpha \sim \alpha + 2\pi N \ ,    
\end{equation}
whereas no identification holds for $p$. In fact, note that the second correlator in \eqref{SymTFT_linkings} would only give an identification $\alpha\sim \alpha+2\pi$, but the third correlator shows that the periodicity of $\alpha$ is actually extended. This can also be understood by performing the sum over fluxes of the $U(1)$ gauge field $A_1$ in \eqref{symtftaction}, which gives
\begin{equation}\label{sumoverfluxesA1}
\exp\left(i\oint_{\Sigma_4} \left( b_4 + \frac{N}{4\pi} B_2 \wedge B_2 \right)\right) = 1 \ ,   
\end{equation}
together with the fact that
\begin{equation}\label{SymTFT_BuB}
\exp\left(\frac{iN}{4\pi}\oint_{\Sigma_4} B_2 \wedge B_2 \right) =
\exp\left(\frac{2\pi i}{2N}\oint_{\Sigma_4} \frac{N B_2}{2\pi} \wedge \frac{N B_2}{2\pi} \right) = \exp\left(\frac{2\pi is}{N}\right) \ , \qquad s \in \mathbb{Z}_{N} \ .
\end{equation}
The operator in the exponent is the continuous version of $\frac{2\pi i}{2N}\mathcal{P}(\mathcal{B}_2)$. Thus, we get
\begin{equation}\label{instanton_u(1)generators}
W_4^{2\pi N}[\Sigma_4] = 1 \ ,
\end{equation}
namely that $\alpha \in U(1) = \mathbb{R}/\mathbb{Z} = [0,2\pi N)$. The naive periodicity of $W_4^\alpha$, $\alpha \sim \alpha + 2\pi$, is thus extended by the cubic term in the SymTFT to be $\alpha \sim \alpha + 2\pi N$.

Concerning the label of the operator $\widetilde{W}_3^m[\Sigma_3,D_4]$, we see that the identification $m \sim m+N$ from the first correlator in \eqref{SymTFT_linkings} is broken by the third correlator. This is the manifestation in the SymTFT of a fact already observed in section \ref{Section:Yang-Mills}, namely that one needs to consider an integer lift of the magnetic charges of 't~Hooft surfaces in order to define their correlator with the defect of the instantonic symmetry. Similarly to what we discussed there, the arbitrariness in the integer lift corresponds to the arbitrariness in adding operators that are charged under the instantonic symmetry. In the context of the SymTFT, these are the line operators $W_1^p$.

\begin{figure}[t]
    \centering
    \includegraphics[width=0.43\linewidth]{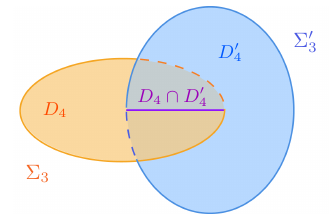}
\caption{Three-dimensional depiction of the two 3-surfaces with intersecting fillings representing the insertion of the magnetic operators $\widetilde{W}_3^m[\Sigma_3,D_4]$ and $\widetilde{W}_3^{m'}[\Sigma_3',D_4']$ in six dimensions. When we shift their labels, we generate two Wilson lines for $A_1$, with opposite charges, at the two boundaries of the intersection of their fillings $D_4\cap D'_4$.} 
\label{fig:instanton-shift}
\end{figure}

Indeed, the equation of motion for $C_3$ implies that inserting the operators $\widetilde{W}_3^m[\Sigma_3,D_4]$ and $\widetilde{W}_3^{m'}[\Sigma_3',D_4']$ is equivalent to turning on
\begin{equation}
    B_2 = \frac{2\pi }{N} 
    \left(
    m\,\delta^{(2)}(D_4) + m'\delta^{(2)}(D'_4)
    \right)
    \ .
\end{equation}
If we shift $m\to m+pN$ and $m'\to m'+p'N$, then $B_2$ varies by $2\pi\left(p\,\delta^{(2)}(D_4) + p'\delta^{(2)}(D'_4) \right)$.
Consequently, the on-shell action in the presence of the two magnetic operators, after $C_3$ is integrated out, varies by
\begin{equation}\label{eq:WilsonLineGeneration}
-i (mp'+pm'+pp'N) \int_{D_4\cap D_4'} dA_1 \ .    
\end{equation}
This corresponds to the insertion, on each of the two disconnected components of $\partial (D_4\cap D_4')$ that are $\Sigma_3 \cap D_4'$ and $\Sigma_3' \cap D_4$, of a Wilson line of $A_1$ with integer charge $\mp(mp'+pm'+pp'N)$, respectively, as depicted in figure \ref{fig:instanton-shift}. Whenever we insert two magnetic operators that activate the triple-linking and whose fillings intersect, we can also insert a Wilson line for $A_1$ with charge $\pm q\in \mathbb{Z}$, respectively, on each component of the boundary of their intersection. Thus, the correct label identification is
\begin{equation}\label{bulk-id}
    (m+pN,m'
    +p'N,q) \quad \sim \quad
    (m,m',q-mp' -pm'-pp'N) \ ,
\end{equation}
as in \eqref{eq:identcharge}. As we will discuss below, the two Wilson lines for $A_1$ correspond to the insertion of two pointlike instanton operators with opposite charges on the worldvolume of the magnetic operators (that represent the 't~Hooft surfaces of the theory), in agreement with the discussion in section \ref{Section:Yang-Mills}. 

\subsubsection{Global variants and boundary conditions}
In the SymTFT framework, the discussion of the global variants of the $\mathfrak{su}(N)$ Yang-Mills theory can be done in (at least) two equivalent ways.

One possibility is to classify the possible topological boundary conditions of the theory in \eqref{symtftaction}, once ${\cal Y}_6$ is taken to be a slab manifold with a topological and a physical boundary. At the physical boundary one puts the physical (non-topological) theory, whose operators couple to the dynamical bulk fields. At the topological boundary, one can introduce localized fields (usually referred to as \textit{topological edge modes}) that cancel the variation of the action under a gauge transformation in the presence of a boundary. The set of possible topological boundary conditions is in one-to-one correspondence with the global variants of the physical theory.

Another possibility is to classify the possible Lagrangian algebras of the bulk theory. A Lagrangian algebra $\mathcal{L}$ is a set of bulk topological operators of the theory that are closed (i.e.~the fusion of any two elements gives another element of the set), mutually transparent with respect to both double- and triple-linking (i.e.~all correlators in \eqref{SymTFT_linkings} between elements of the set are trivial), and maximal (i.e.~no other operator can be added to the set). The bulk operators in a given choice of $\mathcal{L}$ correspond to the ones that can end on the boundary, so that each choice of $\mathcal{L}$ is a choice of charges in the boundary theory. Instead, the bulk operators of the theory modulo $\mathcal{L}$ correspond to the ones that can lie on the boundary, thus giving the symmetry defects in the boundary theory.

Let us explore these two possibilities in turn.

\paragraph{\color{black} Topological boundary conditions.} If ${\cal Y}_6$ is a manifold with a boundary $\partial {\cal Y}_6$, the gauge variation of the action in \eqref{symtftactionvariation} is given by the boundary term 
\begin{equation}\label{symtftgaugevariationboundary}
    S_{\rm SymTFT} \, \rightarrow \, S_{\rm SymTFT} + \int_{\partial\mathcal{Y}_6} \left(-
    \frac{iN}{2\pi} d\Lambda_1 \wedge C_3 +  
    \frac{i}{2\pi} d\lambda_3 \wedge A_1 - 
    \frac{iN}{8\pi^2} A_1 \wedge d\Lambda_1 \wedge d\Lambda_1 
    \right) \ .
\end{equation}
To cancel this variation at the topological boundary $\partial{\cal Y}^T_6$, we can write the boundary action
\begin{align}
\begin{split}
    S_{\partial} = \int_{\partial\mathcal{Y}^T_6} \bigg(  - \frac{ik}{2\pi} B_2 \wedge 
    \left( dC_2 \right. & \left.-\, r C_3 \right) + \frac{i}{2\pi} (dA_0 - A_1) \wedge b_4 \\
    &+ \frac{iN}{8\pi^2} (dA_0 - A_1) \wedge B_2 \wedge B_2
    \bigg) \ , 
\end{split}
\end{align}
where $k,r\in \mathbb{Z}$, whereas $C_2$ and $A_0$ are $U(1)$ edge modes, with gauge transformations
\begin{equation}
    C_2 \rightarrow C_2 + d\Lambda_C +r \Lambda_2 + \frac{r}{2\pi} A_0 \wedge d\Lambda_1\,, \qquad
    A_0 \rightarrow A_0 + 2\pi n_A + \Lambda_0 \ ,
\end{equation}
where $\oint d\Lambda_C \in 2\pi \mathbb{Z}$ and $n_A \in \mathbb{Z}$, to be considered together with the ones in \eqref{symtftgaugetransf}. Under these gauge transformations, the boundary action transforms (modulo $2\pi\mathbb{Z}$) as
\begin{align}
\begin{split}
    S_{\partial} \, \rightarrow \, S_{\partial} +  \int_{\partial\mathcal{Y}^T_6}  \bigg(
    \frac{ikr}{2\pi} d\Lambda_1 \wedge C_3 
    & - \frac{i}{2\pi} d\lambda_3 \wedge A_1 +
    \frac{i(2kr-N)}{8\pi^2} 
        A_1 \wedge d\Lambda_1 \wedge 
     d\Lambda_1
    \\
    &  
    - \frac{i(kr-N)}{4\pi^2} (dA_0-A_1) \wedge B_2 \wedge d\Lambda_1
    \bigg) \ ,
\end{split}
\end{align}
which cancels the variation in \eqref{symtftgaugevariationboundary} if and only if
\begin{equation}\label{conditionsoncoeff}
N = kr \ .   
\end{equation}
This implies that the choice of the above boundary action corresponds to the choice of a divisor $k$ of $N$.

We now determine the topological boundary conditions for each $k$. To this end, we compute the variation of the bulk action (on-shell on the bulk equations of motion), which gives the boundary term
\begin{equation}\label{var_bulk}
    \delta S_{\rm SymTFT} \big|_\partial = \int_{\partial{\cal Y}^T_6} \left(-\frac{iN}{2\pi} B_2 \wedge \delta C_3 + \frac{i}{2\pi} \left( b_4 + \frac{N}{4\pi} B_2 \wedge B_2 \right) \wedge \delta A_1\right) \ .  
\end{equation}
Instead, the variation coming from the boundary action (we only write the terms involving variations of the bulk fields) is
\begin{align}\label{var_bdy}
\begin{split}
    \delta S_{\partial} = \int_{\partial\mathcal{Y}^T_6} \bigg( & 
    \frac{iN}{2\pi} B_2 \wedge \delta C_3 - \frac{ik}{2\pi} \left(
    dC_2 - r C_3 - \frac{r}{2\pi} (dA_0-A_1)\wedge B_2
    \right) \wedge \delta B_2 \\
    & + \frac{i}{2\pi} (dA_0-A_1) \wedge \delta b_4
    - \frac{i}{2\pi} \left( b_4 + \frac{N}{4\pi} B_2 \wedge B_2 \right) \wedge \delta A_1
   \bigg) \ . 
\end{split}
\end{align}
Imposing that \eqref{var_bulk} and \eqref{var_bdy} cancel and using \eqref{conditionsoncoeff}, we find
\begin{equation}\label{eom-edge-modes}
    \frac{N}{k}C_3|_\partial = dC_2  \,, \qquad  
    A_1|_\partial = dA_0 \ .
\end{equation}
Summing over the fluxes of the $U(1)$ edge modes gives the relation \eqref{sumoverfluxesA1} on the boundary, but also
\begin{equation}\label{fluxes-edge-modes}
\exp\left(ik\oint_{\Sigma_2} B_2\right) = 1 \,,  \qquad \Sigma_2 \subset \partial \mathcal{Y}_6^T \ .  
\end{equation}
Instead, the terms involving variations of the edge modes give equations that are trivially satisfied. The equations \eqref{eom-edge-modes} and \eqref{fluxes-edge-modes} (together with \eqref{sumoverfluxesA1} on the boundary) constitute the set of topological boundary conditions.

\begin{figure}[ht]
    \centering
    \includegraphics[width=0.43\linewidth]{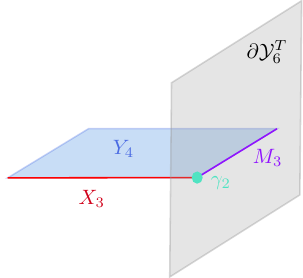}
\caption{The configuration  \eqref{W3-ending} of the operators $\widetilde{W}_3^{\widetilde{m}N/k}[X_3,Y_4]$ ending at the topological boundary $\partial\mathcal{Y}_6^T$.}
\label{fig:endingthooft}
\end{figure}

We now show that the topological boundary conditions found above give rise to the $SU(N)/\mathbb{Z}_k$ global variant of the theory. The operators $W_2^n$ in \eqref{W's-bulk}, with $n=\tilde n k$ and $\tilde{n}\in\mathbb{Z}_{N/k}$, can end at the boundary due to \eqref{fluxes-edge-modes}, and furnish the Wilson lines that are charged under the $\mathbb{Z}_{N/k}^{(1)}$ electric 1-form symmetry of the $SU(N)/\mathbb{Z}_k$ theory. The latter is generated by $\widetilde{W}_3^{m}$, defined in \eqref{W3-bulk}, with $m \in \mathbb{Z}_{N/k}$: these operators are the ones that, when laid on the boundary, are not trivialized due to the first of \eqref{eom-edge-modes}. The latter equation also implies that the operators $\widetilde{W}_3^{m}[X_3,Y_4]$, with $m=\widetilde{m}N/k$  and $\widetilde{m}\in\mathbb{Z}_k$, can end at the boundary on the lines of $C_2$. In this case, we also have to consider the dressing of such operators, so that the full configuration that ends at the boundary -- depicted in figure \ref{fig:endingthooft} -- is given by
\begin{equation}\label{W3-ending}
    \exp \left(i\widetilde{m}\left(
    \frac{N}{k} \int_{X_3} C_3 -
    \frac{N/k}{2\pi} \int_{Y_4} B_2\wedge dA_1 +
    \oint_{\gamma_2} C_2 -
    \frac{N/k}{2\pi} \int_{M_3} B_2 \wedge (dA_0-A_1)
    \right)\right) \ ,
\end{equation}
where $\gamma_2,M_3\subset\partial \mathcal{Y}_6^T$, $\partial Y_4=X_3\cup M_3$, and $\partial M_3 = - \partial X_3 = \gamma_2$.
They furnish the 't~Hooft surfaces that are charged under the $\mathbb{Z}_k^{(2)}$ magnetic 2-form symmetry of the $SU(N)/\mathbb{Z}_k$ symmetry, generated by $W_2^n$ with $n \in \mathbb{Z}_k$ -- which are the ones that, when laid on the boundary, are not trivialized due to \eqref{fluxes-edge-modes}. Notice that the dependence on $M_3$ and $Y_4$ trivializes on-shell, but it is important when considering configurations that activate the triple-linking, or more generally in the presence of a non-trivial instantonic background.

Due to the second of \eqref{eom-edge-modes}, in all global variants we can end the lines $W_1^p[X_1]$, for all $p\in\mathbb{Z}$, on a point $P$ at the boundary,
\begin{equation}\label{W1-ending}
    \exp \left( ip  \left(\int_{X_1} A_1 + A_0(P)   
    \right)\right) \,,
\end{equation}
where $\partial X_1=-P$. These correspond to the instantonic operators with integer charge, measured by the operators $W_4^\alpha$ that can always be laid on the boundary and are the generators of $U(1)_I^{(0)}$. Notice that if $\alpha \in 2\pi\mathbb{Z}$, then the corresponding $W_4^\alpha$ does \textit{not} act on the instanton operators, but it acts instead on two linked 't~Hooft surfaces. The $U(1)_I^{(0)}$ periodicity in each global variant can be simply determined by using \eqref{fluxes-edge-modes} -- rewritten as $kB_2|_\partial=dB_1$, where $B_1$ is an ordinary $U(1)$ gauge field -- in the relation \eqref{sumoverfluxesA1}. We get that
\begin{equation}
\exp\left(\frac{iN}{4\pi}\oint_{\Sigma_4} B_2 \wedge B_2 \right) =
\exp\left(\frac{2\pi i N}{8\pi^2 k^2}\oint_{\Sigma_4} dB_1 \wedge dB_1 \right) = \exp\left(\frac{2\pi is}{k/L}\right) \,, \qquad s \in \mathbb{Z}_{k/L} \ ,
\end{equation}
which implies that $\alpha \sim \alpha + 2\pi k/L$, consistently with the discussion in section \ref{subsection:SU(N)/Zk}.

Let us now specify to the boundary conditions associated to the $SU(N)$ and $PSU(N)$ theories. If $k=1$, we are describing a Dirichlet boundary condition for $B_2$, corresponding to the $SU(N)$ global variant. The relation \eqref{fluxes-edge-modes} implies that all Wilson surfaces of $B_2$ are trivial when laid on the boundary, so that there is no 2-form symmetry. The first of \eqref{eom-edge-modes} implies that there is a $\mathbb{Z}_N^{(1)}$ 1-form symmetry, acting on the Wilson surfaces of $B_2$ that end at the boundary.

If $k=N$, we are describing a Dirichlet boundary condition for $C_3$, corresponding to the $PSU(N)$ global variant. The relations \eqref{eom-edge-modes} imply that all (suitably dressed) Wilson surfaces of $C_3$ are trivial when laid on the boundary, so that there is no 1-form symmetry. Instead, the relation \eqref{fluxes-edge-modes} implies that there is a $\mathbb{Z}_N^{(2)}$ 2-form symmetry, acting on the Wilson surfaces of $C_3$ that end at the boundary. In this case, the configuration ending at the topological boundary shown in \eqref{W3-ending} becomes
\begin{equation}\label{eq:C3PSU(N)_boundary}
     \begin{split}
     \widetilde{W}_3^m[X_3,Y_4,\gamma_2,M_3]&
    \\= \exp \left(im\left(
     \int_{X_3} C_3\right.\right.&\left.\left. -
    \frac{1}{2\pi} \int_{Y_4} B_2 \wedge dA_1 +
    \oint_{\gamma_2} C_2 -
    \frac{1}{2\pi} \int_{M_3} B_2 \wedge (dA_0-A_1)
    \right)\right) \ ,
    \end{split}
\end{equation}
where $m \in \mathbb{Z}$. The fact that Wilson surfaces of $B_2$ do not trivialize at the boundary implies that $U(1)_I^{(0)}$ has the whole $2\pi N$ periodicity. 
The extension of the instantonic symmetry is only activated when the symmetry defect acts on a suitable configuration of the operators in \eqref{eq:C3PSU(N)_boundary} via the triple-linking. Such configuration is realized when $\widetilde{W}_3^m,\widetilde{W}_3^{m'}$ end on surfaces $\gamma_2,\gamma_2'$, respectively, that double-link at the boundary. Then, it is possible to pick the bulk support of the defects $X_3,X_3'$ such that in the bulk they form a triple-linking configuration with the instantonic defect. By construction, the instantonic defect also triple-links at the boundary with the surfaces $\gamma_2,\gamma_2'$. Thanks to the mechanism described in \eqref{eq:WilsonLineGeneration}, under the shift $m \to m+pN$ and $m'\to m'+p'N$ Wilson lines of $A_1$ are inserted on the two disconnected components of  $\partial(Y_4\cap Y_4')$. Given the specified configuration, these lines end on two points at the boundary of  $M_3\cap M_3'$, i.e.~in $M_3\cap \gamma_2'$ and in $M_3'\cap \gamma_2$. The endpoints of the Wilson lines furnish instanton operators inserted on the support of the 't~Hooft surfaces. This discussion bridges the identification of operators \eqref{bulk-id} in the bulk with the identification of charge labels \eqref{eq:identcharge} in the boundary theory.

\paragraph{Lagrangian algebras.}
The $SU(N)$ global form is associated to the Lagrangian algebra
\begin{equation}
\label{Lag_alg_1}
\mathcal{L} = \lbrace W_1^p \,, W_2^n \,, W_4^\alpha \, | \, p \in \mathbb{Z} \,; n \in \mathbb{Z}_N \,; \alpha = 2\pi q \,, q \in \mathbb{Z}_N \rbrace \ .   
\end{equation}
This Lagrangian algebra is equivalent to choosing Dirichlet boundary conditions for $B_2,A_1$ and Neumann boundary conditions for $C_3,b_4$. The operators $W_2^n$ can end on a curve $\gamma_1 \in \partial {\cal Y}_6$ on the boundary. They represent Wilson lines of the 5$d$ theory supported on $\gamma_1$, which carry charge $n\in\mathbb{Z}_N$ under a $\mathbb{Z}_N^{(1)}$ electric 1-form symmetry.
The operators $\widetilde{W}_3^m$ can lie on the boundary and represent the codimension-two symmetry defects of the electric 1-form symmetry of the boundary 5$d$ QFT. Note that these operators lose their dependence on $D_4$ if no instantonic background is turned on. They measure the $\mathbb{Z}_N$ electric charge of the Wilson lines via the first correlator in \eqref{SymTFT_linkings}.

Since $B_2$ vanishes on $\partial{\cal Y}_6$, the operator in \eqref{SymTFT_BuB} always trivializes on the boundary. Thus, when laid on the boundary, the $U(1)^{(0)}_I$ symmetry defects $W_4^\alpha$ satisfy $W_4^{2\pi}=1$. This means that the periodicity of $\alpha$ is $\alpha \sim \alpha + 2\pi$.

The cubic term in the SymTFT action is constructed only with fields that have Dirichlet boundary conditions, and thus indicates a mixed 't~Hooft anomaly between $U(1)^{(0)}_I$ and $\mathbb{Z}_N^{(1)}$. This is also clear from the fact that the generators of this symmetry have non-trivial mutual braiding, as is evident from the third correlator of \eqref{SymTFT_linkings}. The anomaly inflow after the slab compactification is correctly reproduced by the identification
\begin{equation}
    \frac{iN}{8\pi^2}\int_{\mathcal{Y}_6} dA_1\wedge B_2 \wedge B_2 \quad \mapsto \quad \frac{iN}{8\pi^2}\int_{\mathcal{Y}_6} 2\pi\left[\frac{dA_1}{2\pi}\right]_N\cup \left(\frac{2\pi}{N}\right)^2\mathcal{P}( \mathcal{B}_2) \,.
\end{equation}
Moreover, the 3-group structure encoded in the bulk gauge transformations of $C_3$ is trivialized by the boundary condition for $B_2$.

We then see that Dirichlet boundary conditions for $A_1$ and $B_2$ give a 5$d$ theory with a $\mathbb{Z}_N^{(1)}$ electric 1-form symmetry, a $U(1)^{(0)}_I$ 0-form symmetry with periodicity $2\pi$, and a mixed 't~Hooft anomaly between them. Thus, these boundary conditions correspond to the $SU(N)$ global form of the gauge group.

The $PSU(N)$ global form is associated to the Lagrangian algebra
\begin{equation}
\label{Lag_alg_2}
\mathcal{L} = \lbrace W_1^p \,, \widetilde{W}_3^m \, | \, p \in \mathbb{Z} \,; m \in \mathbb{Z}  \rbrace \ ,    
\end{equation}
which corresponds to Dirichlet boundary conditions for $C_3,A_1$ and Neumann boundary conditions for $B_2,b_4$.
The operators $\widetilde{W}_3^m$ can end on a two-surface $\gamma_2 \in \partial {\cal Y}_6$ on the boundary. They represent 't~Hooft surfaces of the 5$d$ theory supported on $\gamma_2$, which carry charge $m \text{ mod } N \in\mathbb{Z}_N$ under a $\mathbb{Z}_N^{(2)}$ magnetic 2-form symmetry. The operators $W_2^n$ can lie on the boundary and represent the codimension-three symmetry defects of the magnetic 2-form symmetry of the boundary 5$d$ QFT. They measure the $\mathbb{Z}_N$ magnetic charge of the 't~Hooft surfaces via the first correlator in \eqref{SymTFT_linkings}.

Since $B_2$ does not vanish on $\partial{\cal Y}_6$, the operator in \eqref{SymTFT_BuB} does not necessarily trivialize on the boundary. Thus, when laid on the boundary, the $U(1)^{(0)}_I$ symmetry defects $W_4^{2\pi}$ can be non-trivial. This means that the periodicity of $\alpha$ is now extended to be $\alpha \sim \alpha + 2\pi N$. However, we see that the symmetry defect $W_4^{2\pi}$ does \textit{not} act on pointlike operators, given that the second correlator in \eqref{SymTFT_linkings} vanishes when $\alpha \in 2\pi\mathbb{Z}$. As we discussed in the previous section, this is related to the fact that there is no $PSU(N)$ fractional instanton on $S^4$ -- the four-cycle that wraps a point in 5$d$. However, this symmetry measures a non-trivial $\mathbb{Z}_N$ instanton charge of operators that are supported on geometric configurations that can be wrapped by a four-cycle $\Sigma_4$, such that the charge of the corresponding $PSU(N)$ instanton can be fractional. In our case, these are given by two linked 't~Hooft surfaces. As shown by the third correlator in \eqref{SymTFT_linkings}, this configuration has a non-trivial charge, measured by $W_4^{2\pi}$.

The cubic term in the SymTFT action does not correspond now to a mixed 't~Hooft anomaly, as $B_2$ has Neumann boundary conditions. In other words, the third correlator in \eqref{SymTFT_linkings} has now the interpretation of a symmetry defect measuring the charge of an operator which is not a symmetry defect itself, as it was instead in the previous case. Here the 3-group structure is absent as well, now due to the boundary condition of $C_3$.

We then see that Dirichlet boundary conditions for $A_1$ and $C_3$ give a 5$d$ theory with a $\mathbb{Z}_N^{(2)}$ magnetic 2-form symmetry and a $U(1)^{(0)}_I$ 0-form symmetry with periodicity $2\pi N$. With respect to the previous case, the $U(1)^{(0)}_I$ symmetry is extended by $\mathbb{Z}_N^{(0)}$, with this extension acting on a particular class of extended operators, only. Thus, these boundary conditions correspond to the $PSU(N)$ global form of the gauge group.

Finally, in the generic case with global form $SU(N)/\mathbb{Z}_k$, the Lagrangian algebra is given by
\begin{align}\label{symtftLagrAlb_k}
\begin{split}
\mathcal{L} = \lbrace
W_1^p \,, W_2^n \,, \widetilde{W}_3^m \,, W_4^\alpha \, | \,
p \in \mathbb{Z} \,;
& \, n=k\tilde{n} \,, \tilde{n}\in \mathbb{Z}_{N/k} \,;\\
& \, m=\frac{N}{k}\widetilde{m} \,, \widetilde{m} \in \mathbb{Z} \,;
\alpha= 2\pi \frac{k}{L} q \,, q \in \mathbb{Z}_{\frac{N}{k}L}  \rbrace \ ,  
\end{split}
\end{align}
where $L=\text{gcd}(k,N/k)$.
It is easy to check that the operators in the Lagrangian algebra are mutually transparent. Indeed, the two-component braiding phases between any two elements of the Lagrangian algebra are trivial, 
\begin{equation}
\exp\left( \frac{2\pi imn}{N} \right) = \exp\left( 2\pi i \widetilde{m} \tilde{n} \right) = 1 \,, \quad \exp(i\alpha p) = \exp(2\pi i\frac{k}{L}qp) = 1 \ , 
\end{equation}
as well as the three-component braiding phase,
\begin{equation}
\exp\left(\frac{i\alpha m m'}{N} \right)  = \exp\left(2\pi i \frac{N/k}{L} q \widetilde{m}\widetilde{m}' \right)=1 \ .   
\end{equation}
Therefore, for general $k$ the periodicity of $U(1)_I^{(0)}$ is $\alpha \sim \alpha + 2\pi k/L $. In each global variant, the electric symmetry is $\mathbb{Z}_{N/k}^{(1)}$ and the magnetic symmetry is $\mathbb{Z}_k^{(2)}$. 

Let us now analyze the different symmetry structures in each global variant to show how the results of section \ref{subsection:SU(N)/Zk} are matched by the SymTFT \eqref{symtftaction}.

\begin{itemize}

\item The mixed 't~Hooft anomaly between the electric and the instantonic symmetries, given by the cubic term in the action, is present whenever there are non-trivial operators of both $A_1$ and $B_2$ that can terminate on the boundary, that is when $k\neq N$, namely for all global forms but $PSU(N)$. To show this, consider the triple correlator between the defects of the instantonic symmetry and the electric symmetries after the slab compactification.
These are given by the bulk topological operators $W_4^\alpha$ and $\widetilde{W}_3^m$, modulo the ones that are included in the Lagrangian algebra \eqref{symtftLagrAlb_k}, and thus they are labeled by $\alpha+2\pi ka/L$ and $m-bN/k$, respectively, where $a,b\in\mathbb{Z}$ are arbitrary, $\alpha\in [0,2\pi k/L)$ and $m\in \mathbb{Z}_{N/k}$. Their correlator, signaling a mixed 't~Hooft anomaly when it is not trivialized, is
\begin{equation}\label{mixedanomalycondition}
\exp\left(\frac{i}{N}\left(\alpha+ \frac{2\pi ak}{L}\right)\left(m-b\frac{N}{k}\right)\left(m'-b'\frac{N}{k}\right)\right) \ .    
\end{equation}
There is a choice of $a$, $b$, and $b'$ such that this braiding phase is trivial for all $\alpha$, $m$, and $m'$ if and only if $k=N$. Indeed, if $k=N$ then $m=m'=0$ and one can simply choose $b=b'=0$ to trivialize the phase. Conversely, to show that $k=N$ is a necessary condition, it is enough to specify to the case where $m=m'=1$ and $\alpha$ is irrational. In this case, in order to trivialize \eqref{mixedanomalycondition}, $a$ can be chosen to vanish without loss of generality and
\begin{equation}\label{mixedanomalycondition2}
bb'\left(\frac{N}{k}\right)^2 - (b+b') \frac{N}{k} + 1 = 0 \ . 
\end{equation}
It is easy to show that if this relation is satisfied then $k=N$,\footnote{Indeed, the solutions of \eqref{mixedanomalycondition2} are $N/k=1/b$ and $N/k=1/b'$, which imply that $N/k=1$, given that $b$, $b'$, and $N/k$ are all integers.} so that the anomaly is trivialized in this case.

Finally, notice that the anomaly-inflow action is correctly reproduced, after the slab compactification, by the identification\footnote{We convert the $U(1)$ gauge fields $B_2$ and $C_3$ into $\mathbb{Z}_N$ ones, that we further decompose in a $\mathbb{Z}_k$ and a $\mathbb{Z}_{N/k}$ part, so that 
\begin{equation}
    B_2 \mapsto \frac{2\pi}{N} \mathcal{B}_2^{N}=\frac{2\pi}{N} \left(\frac{N}{k}\widehat{\mathscr{b}}_2^{k}+\widehat{\mathcal{B}}_2^{N/k}\right) \ , \qquad
    C_3 \mapsto \frac{2\pi}{N} \mathcal{C}_3^{N}=\frac{2\pi}{N} \left(k\,\widehat{\mathscr{c}}_3^{\frac{N}{k}}+\widehat{\mathcal{C}}_3^{k}\right) \ ,
\end{equation}
where $\widehat{\mathscr{b}}_2^{k}$ and $\widehat{\mathscr{c}}_3^{N/k}$ are gauged while $\widehat{\mathcal{B}}_2^{N/k}$ and $\widehat{\mathcal{C}}_3^{k}$ are the background gauge fields (from the boundary viewpoint).\label{footnote:fromU1toZN}}
\begin{equation}
    \frac{iN}{8\pi^2}\int_{\mathcal{Y}_6} dA_1\wedge B_2 \wedge B_2 \quad \mapsto \quad \frac{iN}{8\pi^2}\int_{\mathcal{Y}_6} 2\pi \left[\frac{dA_1}{2\pi}\right]_N\cup \left(\frac{2\pi}{N}\right)^2\mathcal{P}( \widehat{\mathcal{B}}_2^{N/k}) \ ,  
\end{equation}
which gives the first term in \eqref{SU/N)/Zk_anomaly_old}.

\item The mixed 't~Hooft anomaly between the electric and the magnetic symmetries, given by the $U(1)$ BF term in the action, is present whenever $L\neq 1$. To show this, consider the mutual braiding between the electric and the magnetic symmetry operators after the slab compactification. These are given by the bulk topological operators $W_2^n$ and $\widetilde{W}_3^m$, modulo the ones that are included in the Lagrangian algebra \eqref{symtftLagrAlb_k}, and thus they are labeled by $n-ak$ and $m-bN/k$, respectively, where $a,b\in\mathbb{Z}$ are arbitrary, $n\in \mathbb{Z}_{k}$ and $m\in \mathbb{Z}_{N/k}$. Their braiding phase, signaling a mixed 't~Hooft anomaly when it is not trivialized, is
\begin{equation}
\exp\left(-\frac{2\pi i}{N}(n-ak)\left(m-b\frac{N}{k}\right)\right) \ ,     
\end{equation}
which is trivial for all $m$ and $n$ if and only if there is a choice of $a$ and $b$ such that
\begin{equation}\label{anomalycondition}
akm + bn\frac{N}{k} = nm \quad \text{mod } N \ . 
\end{equation}
It is easy to show that this condition is satisfied if and only if $L=1$,\footnote{The fact that $L=1$ is necessary follows by choosing $m=n=1$. In this case, the condition \eqref{anomalycondition} simply reads $ak+bN/k=1$ mod $N$. Using B\'ezout lemma, there exist $a,b\in\mathbb{Z}$ such that the left-hand side is $cL$, where $c$ can be an arbitrary integer (which depends on the choice of $a,b$). Thus, we have $cL=1$ mod $N$ and clearly $cL=0$ mod $L$. Since $L$ divides $N$, this is only possible if $cL=1$ which implies $L=1$. The fact that $L=1$ is sufficient follows by rewriting the left-hand side of \eqref{anomalycondition} as $L'(au+bv)$ where $L'=\text{gcd}(n,m)$, $u=k\cdot m/L'$ and $v=N/k \cdot n/L'$. Using again B\'ezout lemma, there exist $a,b\in\mathbb{Z}$ such that $au+bv=c\,\text{gcd}(u,v)=c\,\text{gcd}(k,n/L')\,\text{gcd}(N/k,m/L')$, where $c$ can be an arbitrary integer (which depends on the choice of $a,b$) and in the second step we used that $\text{gcd}(k,N/k)=\text{gcd}(n/L',m/L')=1$. In this way, the left-hand side is $c\,L'\text{gcd}(k,n/L')\,\text{gcd}(N/k,m/L')$. By suitably choosing $c$, this coincides with $nm$, as the latter is a multiple of $L'\text{gcd}(k,n/L')\,\text{gcd}(N/k,m/L')$. Hence, \eqref{anomalycondition} is satisfied.}  so that the anomaly is trivialized in this case.

Finally, notice that the anomaly inflow is reproduced by the identification (after integrating by parts, using the dictionary in footnote \ref{footnote:fromU1toZN} and the relation \eqref{eq:Bockbacklift}) 
\begin{equation}
\begin{split}
     \frac{iN}{2\pi}\int_{\mathcal{Y}_6} dB_2\wedge C_3  \ \mapsto \  \frac{iN}{2\pi}\int_{\mathcal{Y}_6} \frac{2\pi}{N}\delta\widehat{\mathcal{B}}_2^{N/k}\cup\frac{2\pi}{N}\widehat{\mathcal{C}}^k_3=  \frac{2\pi i}{k}\int_{\mathcal{Y}_6} \text{Bock}(\mathcal{B}_2^{N/k})\cup\mathcal{C}_3  \ ,
\end{split}
\end{equation}
where $\mathcal{C}_3\in\mathbb{Z}_k$, which gives the second term in \eqref{SU/N)/Zk_anomaly_old}.

\item The 3-group structure survives after the slab compactification when there are symmetry defect operators that need to be dressed by other operators, in presence of non-trivial background fields. In our case, the boundary symmetry operators that need dressing are $\widetilde{W}_3$, labeled by $m-bN/k$, where $b\in\mathbb{Z}$ is arbitrary and $m\in\mathbb{Z}_{N/k}$. Their operator-valued dressing is obtained by plugging $kB_2|_\partial=dB_1$ (where $B_1$ is a properly quantized $U(1)$ gauge field) in presence of a non-trivial background for the instantonic symmetry,
\begin{equation}
\exp\left(-2\pi i\left(m-b\frac{N}{k} \right) \frac{1}{k}  \int_{D_4} \frac{dB_1}{2\pi} \wedge \frac{dA_1}{2\pi}\right) \,,  \qquad D_4 \subset \partial \mathcal{Y}_6 \ ,
\end{equation}
which is trivialized when
\begin{equation}\label{3groupcondition}
m-b\frac{N}{k} = 0 \text{ mod } k \ .  
\end{equation}
It is easy to show that this condition is satisfied for all $m$ if and only if $L=1$.\footnote{The condition \eqref{3groupcondition} is equivalent to finding arbitrary integers $a$ and $b$ such that any $m$ can be rewritten as $m=ak+bN/k$. By B\'ezout lemma, this is $m=cL$, where $c$ is an arbitrary integer. Thus, we can write any $m$ in this form if and only if $L=1$.} Thus, there is a non-trivial 3-group structure whenever $L \neq 1$.

Finally, notice that the gauge invariant magnetic field strength is reconstructed via the identification (using again the dictionary in footnote \ref{footnote:fromU1toZN})
\begin{equation}
    dC_3 - B_2\wedge \frac{dA_1}{2\pi} \quad\mapsto\quad  \frac{2\pi}{N}\delta\widehat{\mathcal{C}}^k_3 -\frac{2\pi}{N} \widehat{\mathcal{B}}_2^{N/k}\cup \left[\frac{dA_1}{2\pi}\right]_N~.
\end{equation}
Since $\delta\mathcal{C}_3 \in \mathbb{Z}_k$, collecting from the right-hand side $2\pi/N$, reducing to $\mathbb{Z}_k$ and using the equations of motion, we reproduce the vanishing gauge invariant field strength of \eqref{H4_SU(N)/Zk}.
\end{itemize}

\subsection{SymTFT for the \texorpdfstring{$E_1$}{E1} theory and its variants}

In section \ref{E1_hat_FT} we discussed the symmetry structure and the extended operators of the $\widehat{E}_1$ SCFT, that is obtained by gauging the $\mathbb{Z}_2^{(1)}$ symmetry of the $E_1$ SCFT. We now want to exploit the framework of the SymTFT to analyze how these different global variants can be obtained from different boundary conditions at the topological boundary.

Following recent work on how to generalize the SymTFT framework to non-Abelian 0-form symmetries \cite{Brennan:2024fgj,Antinucci:2024zjp,Bonetti:2024cjk,Jia:2025jmn,Apruzzi:2025hvs,Bonetti:2025dvm}, building on earlier mathematical work \cite{Horowitz:1989ng,Cattaneo:2000mc,Cattaneo:2002tk}, we propose the action 
\begin{equation}\label{SymTFT:nonAb}
S_{\rm SymTFT} = \int_{\mathcal{Y}_6}\left( i\pi\, \mathcal{B}_2 \cup \delta \mathcal{C}_3  + \frac{i}{2\pi} \left(b_4,F_2\right) + i\pi\, \frac{\mathcal{P}(\mathcal{B}_2)}{2} \cup w_2(SO(3)) \right) \,,    
\end{equation}
where $\mathcal{B}_2$ and $\mathcal{C}_3$ are cochains in $C^2(\mathcal{Y}_6,\mathbb{Z}_2)$ and $C^3(\mathcal{Y}_6,\mathbb{Z}_2)$, respectively, $F_2=dA_1+iA_1\wedge A_1$ is the field strength of the non-Abelian gauge field $A_1=\vec{A}_1 \cdot \vec{T}$, $b_4=\vec{b}_4 \cdot\vec{T}$ is an adjoint-valued 4-form real gauge field, and $(\cdot,\cdot)$ is the Killing form on $\mathfrak{su}(2)$.
Notice that the cubic term is the inflow action \eqref{eq:anoinfUV}  associated to the mixed 't~Hooft anomaly in the $E_1$ theory (up to a five-dimensional counterterm).

As we discussed in section \ref{E1_hat_FT}, the 0-form symmetry in the physical theory can be either $SO(3)_I$ or $SU(2)_I$ depending on the global structure. However, the SymTFT is formulated with gauge group $SO(3)$ for the bulk gauge field $A_1$. This is consistent because in all the global structures that we consider there are only pointlike operators in $SO(3)_I$ representations, and this choice of gauge group ensures that we only include the corresponding bulk Wilson lines. 

One can show that the action \eqref{SymTFT:nonAb}, on a closed 6-manifold, is invariant under the gauge transformations 
\begin{align}\label{gt:nonab}
\begin{split}
\mathcal{B}_2 &\rightarrow \mathcal{B}_2 + \delta\lambda_1 ~,\\   \mathcal{C}_3 &\rightarrow \mathcal{C}_3 + \delta\lambda_2 - \lambda_1 \cup w_2(SO(3))~, \\
A_1 & \rightarrow g^{-1}A_1 g + ig^{-1}dg ~, \\
b_4 &\rightarrow g^{-1} b_4 g + D\lambda_3~,
\end{split}
\end{align}
where $g$ is a group element of $SO(3)$ and $D(\cdot)=(d+iA_1 \wedge)(\cdot)$ is the gauge-covariant derivative with respect to $A_1$. The equations of motion are
\begin{equation}
\begin{split}
\delta \mathcal{B}_2 = 0 \,, \qquad \delta \mathcal{C}_3 + \mathcal{B}_2 \cup w_2(SO(3)) = 0 \,, \qquad F_2=0 \,, \qquad Db_4=0 \ .
\end{split}    
\end{equation}

Let us now consider the bulk topological operators. Using $\mathcal{B}_2$ and $\mathcal{C}_2$, we can build operators analogous to those of the previous section,
\begin{align}\label{operators_E1}
\begin{split}
W_2^n[\Sigma_2] & = \exp\left(i\pi n \oint_{\Sigma_2} \mathcal{B}_2\right) ~, \\   \widetilde{W}_3^m[\Sigma_3,D_4] & = \exp\left(i\pi m \left(\oint_{\Sigma_3}  \mathcal{C}_3 + \int_{D_4} \mathcal{B}_2 \cup w_2(SO(3)) \right)\right) \ ,  
\end{split}
\end{align}
with $m,n\in\mathbb{Z}_2$ and $\partial D_4=\Sigma_3$. Using $A_1$ we can build the Wilson line operators
\begin{equation}\label{nonabelianlines}
W_1^{j}[\gamma_1] = \text{Tr}_{\mathcal{R}_j} P\exp\left(i \oint_{\gamma_1} A_1 \right) \,,
\end{equation}
in a representation $\mathcal{R}_j$ of $SO(3)$ with spin $j\in \mathbb{N}$ and dimension $2j+1$. 

Before discussing the operators we can build with $b_4$, let us recall some basic facts about $SU(2)$ and $SO(3)$ groups. We can build any element of $SU(2)$ as $U_{\theta,\vec{n}}=\exp(i\theta \frac{\vec{n} \cdot \vec{\sigma}}{2})$ where $\theta \in [0,4\pi)$ and $\vec{n}$ is a unit vector. Conjugacy classes of $SU(2)$ are in one-to-one correspondence with the value of the trace,
\begin{equation}\label{coniug-SU(2)}
\Tr U_{\theta,\vec{n}} 
= \Tr U_{\theta,\hat{z}}
= \Tr \begin{pmatrix}
    e^{\frac{i\theta}{2}} & 0 \\
    0 & e^{-\frac{i\theta}{2}}
\end{pmatrix}
= 2 \cos \frac{\theta}{2} \ ,   
\end{equation}
and so are parametrized by $\theta \in [0,2\pi]$.
For $SO(3)$ we can write any element as $R_{\theta,\vec{n}}=\exp(i\theta \vec{n} \cdot \vec{T})$ where now $\theta \in [0,2\pi)$. Again conjugacy classes are in one-to-one correspondence with the value of the trace,
\begin{equation}\label{coniug-SO(3)}
\Tr R_{\theta,\vec{n}} = \Tr R_{\theta,\hat{z}} = \Tr
\begin{pmatrix}
    \cos \theta & -\sin \theta & 0 \\
    \sin \theta & \cos \theta & 0 \\
    0 & 0 & 1 
\end{pmatrix}
=
1 + 2 \cos \theta \ ,   
\end{equation}
 and so are parametrized by $\theta \in [0,\pi]$.

The topological operators for $b_4$ are labeled in the bulk by a parameter $\theta$ and are given by
\begin{equation}\label{nonabeliangeneratorsbulk}
U_{\theta}[\Sigma_4] = \int dg \exp\left( i \oint_{\Sigma_4} \left(  g \, \theta \, T_3 \,g^{-1} ,b_4 \right)\right) \ ,   
\end{equation}
where the integral is over all $g\in SO(3)$ using the Haar measure. One can prove \cite{Jia:2025jmn} that
\begin{equation}\label{eq:2link_nonAb}
\langle U_{\theta}[\Sigma_4] W_1^j[\gamma_1] \rangle = \frac{\chi_j\left( \theta L_2(\Sigma_4,\gamma_1)\right)}{\text{dim}\mathcal{R}_j} \langle W_{1}^j[\gamma_1] \rangle \ ,
\end{equation}
where $\chi_j$ is the character of the representation $\mathcal{R}_j$ and $L_2(\Sigma_4,\gamma_1)$ is the double-linking defined in \eqref{doublelink}. Recall that the character\footnote{The traces \eqref{coniug-SU(2)} and \eqref{coniug-SO(3)} coincide respectively with $\chi_{\frac{1}{2}}(\theta)$ and $\chi_{1}(\theta)$.} is given by
\begin{equation}\label{character-E1}
\chi_j(\theta) = \Tr_{\mathcal{R}_j} \exp (i\theta T_3)
=
\sum_{m=-j}^{j} e^{i\theta m}
=
\frac{\sin\left((2j+1)\theta/2\right)}{\sin(\theta/2)}~. 
\end{equation}
Therefore, if we only allow for lines of spin $j\in \mathbb{N}$, the symmetry of the correlation function \eqref{eq:2link_nonAb} under $\theta \rightarrow -\theta$ and $\theta \rightarrow \theta + 2\pi$ allows to restrict the range of the parameter $\theta$ to the interval $[0,\pi]$. This is the range of $\theta$ corresponding to the conjugacy classes of $SO(3)$. Notice that if we plug a half-integer $j$ in \eqref{character-E1} we obtain the character for the corresponding representation of $SU(2)$, and consistently in this case the range of $\theta$ is $[0,2\pi]$.

Let us now consider two magnetic operators $\widetilde{W}_3^m[\Sigma_3,D_4]$ and $\widetilde{W}_3^{m'}[\Sigma'_3,D'_4]$ whose fillings $D_4$ and $D_4'$ intersect. If we sum over $\mathcal{C}_3$, this corresponds to the insertion of the operator
\begin{equation}
    \exp\left(-i\pi mm' \int_{D_4\cap D'_4} w_2(SO(3))\right) \ .
\end{equation}
The consistency of this insertion on an open manifold requires to specify an additional label $j$, namely its boundary behaves in correlation functions like a Wilson line of spin $j$, with $[2 j]_2=m m'$. Since $\partial(D_4\cap D_4')$ has two disconnected components, we actually need to specify two labels $j$ and $j'$, with $[2 j]_2=[2 j']_2=m m'$, one for each connected components. If either $m$ or $m'$ are $0~\text{mod} ~2$ this is consistent with trivial lines at the boundary. However, if both $m$ and $m'$ are $1~\text{mod} ~2$, the lines at the boundary have half-integer spins $j$ and $j'$, i.e.~they are in $SU(2)_I$ representations. Different choices of $j,j'$ satisfying $[2 j]_2=[2 j']_2=m m'$ are related by the insertion of Wilson lines of $A_1$ with integer spin at the disconnected boundary of $D_4\cap D_4'$. These lines are the non-Abelian version of the line \eqref{eq:WilsonLineGeneration} in the previous section. Thus we can write the action of $U_{\theta}[\Sigma_4]$ on this configuration as
\begin{align}\label{TLK-E1-SymTFT}
\begin{split}
    \langle U_{\theta}[\Sigma_4]& \widetilde{W}_3^m[\Sigma_3,D_4]
    \widetilde{W}_3^{m'}[\Sigma'_3,D'_4] \rangle \\
    & = 
    \frac{\chi_j\left( \theta L_2(\Sigma_4,\Sigma_3\cap D'_4)\right)}{\text{dim}\mathcal{R}_j}
    \frac{\chi_{j'}\left( \theta L_2(\Sigma_4,\Sigma_3'\cap D_4)\right)}{\text{dim}\mathcal{R}_{j'}}
    \langle
    \widetilde{W}_3^m[\Sigma_3,D_4]
    \widetilde{W}_3^{m'}[\Sigma'_3,D'_4] 
    \rangle \ ,
\end{split}
\end{align}
where $[2 j]_2=[2 j']_2=m m'$. Note that $\chi_j(0)=\text{dim} \mathcal{R}_j$, so if $\Sigma_4$ links with only one of the two boundaries and $mm'=1$, the correlation function \eqref{TLK-E1-SymTFT} contains a single character of half-integer spin.\footnote{\label{footnote:tlasdl}The argument of the character of half-integer spin can always be recast in terms of the triple-linking. Indeed using \eqref{eq:tl=dl} we can rewrite $ L_2(\Sigma_4,\Sigma_3\cap D'_4) = L_3(\Sigma_4,\Sigma_3,\Sigma_3')-L_2(\Sigma_4,\Sigma_3'\cap D_4)$. From the properties of characters it then follows that $\chi_j\left( \theta L_2(\Sigma_4,\Sigma_3\cap D'_4)\right) \chi_{j'}\left( \theta L_2(\Sigma_4,\Sigma_3'\cap D_4)\right)=\chi_j\left( \theta L_3(\Sigma_4,\Sigma_3,\Sigma_3')\right)\chi_{j'\otimes j}\left( \theta L_2(\Sigma_4,\Sigma_3'\cap D_4)\right) $. The residual dependence on the double-linking is only through characters of representations with integer spin.} As a result, in this case the range of $\theta$ cannot be restricted to $[0,\pi]$ as for the correlation function \eqref{eq:2link_nonAb}, but is rather extended to $[0,2\pi]$, which is the one of the conjugacy classes of $SU(2)$. Thus, the relation above encodes in the bulk the possibility of the enhancement of the global symmetry of the boundary theory from $SO(3)_I$ to $SU(2)_I$, which happens in a specific global variant, as we now show. 

The symmetry structure in the physical theory is specified by the choice of topological boundary conditions. Since the 0-form instantonic symmetry is a global symmetry in the 5$d$ theory, the boundary conditions for $A_1$ are always Dirichlet. Pointlike operators transforming in representations $\mathcal{R}$ of $SO(3)_I^{(0)}$ are given, in the slab picture, by open Wilson line operators that are suspended between a point on the physical boundary and a point on the topological boundary, where they can end. What distinguishes between $E_1$ and $\widehat{E}_1$ are instead the boundary conditions of $\mathcal{B}_2$ and $\mathcal{C}_3$. Dirichlet boundary conditions for $\mathcal{B}_2$ provide line operators charged under a $\mathbb{Z}_2^{(1)}$ symmetry, constructed via suspended surface operators of the latter gauge field. Moreover, the cubic term in \eqref{SymTFT:nonAb} reduces in this case to the anomaly inflow of the theory after slab compactification. Hence, we conclude that this choice of topological boundary conditions produces the $E_1$ global variant. Dirichlet boundary conditions for $\mathcal{C}_3$, instead, provide surface operators charged under a $\mathbb{Z}_2^{(2)}$ symmetry. This second global variant corresponds to the $\widehat{E}_1$ theory. 

Notice that the topological bulk operators $U_\theta[\Sigma_4]$ are labeled by conjugacy classes and not by group elements. The $g$-dependent integrand in \eqref{nonabeliangeneratorsbulk} seems to define a bulk operator labeled by a group element, but actually it is not gauge invariant \cite{Jia:2025jmn, Apruzzi:2025hvs}. However, after the latter operator is pushed to the boundary and the slab is compactified (with Dirichlet boundary conditions for $A_1$), it becomes a topological gauge invariant operator that, acting on the endpoint of the open Wilson lines of $A_1$, generates the non-Abelian global symmetry of the boundary theory \cite{Jia:2025jmn,Apruzzi:2025hvs}. 

Let us recall that the configuration of operators $\widetilde{W}_3$ in eq.~\eqref{TLK-E1-SymTFT} is responsible for the extension of the parameter $\theta$. This configuration is not possible when the 3-dimensional operators are pushed to the boundary ($E_1$ theory). As a result, in ${E}_1$ the global form of the instantonic symmetry is $SO(3)_I$. When $\widetilde{W}_3$ are allowed to end at the boundary ($\widehat{E}_1$ theory), instead, the same bulk configuration produces linking surface operators at the boundary.  In this case, from the bulk operator labeled by $SU(2)$ conjugacy classes one  gets boundary topological operators labeled by the group $SU(2)$. The linking surfaces are ``charged'' under the extension. Therefore, for $\widehat{E}_1$ the instantonic symmetry is $SU(2)_I$. 

Finally, let us comment on the compatibility between the anomaly and the possible 2-group structure, that we already discussed in section \ref{sect:2-group-E1}, from the point of view of the SymTFT. Including the 2-group structure \eqref{eq:E12g} requires to supplement the SymTFT action \eqref{SymTFT:nonAb} with the extra term
\begin{equation}\label{eq:twogrouSym}
    \int_{\mathcal{Y}_6} i\pi \ \mathcal{C}_3 \cup w_3(SO(3)) \ .
\end{equation}
If one would ignore the cubic term responsible for the 't~Hooft anomaly, the term \eqref{eq:twogrouSym} would imply a 2-group for the $E_1$ global variant, and a mixed 't~Hooft anomaly for the $\widehat{E}_1$ global variant. However, in the presence of the cubic term, due to the gauge transformations \eqref{gt:nonab} this additional extra term spoils the gauge invariance of the action on a closed 6-manifold, the variation of the action being $- \int_{\mathcal{Y}_6} i\pi \, \lambda_1 \cup w_2 \cup w_3$.
We were not able to find a different version of the SymTFT action consistent with both the 2-group structure and the 't~Hooft anomaly. We leave a detailed treatment of this issue for future investigations.  

\section*{Acknowledgements}
We thank Bobby Acharya, Mohammad Akhond, Andrea Antinucci, Francesco Benini, Christian Copetti, Michele Del Zotto, Nicola Dondi, Giovanni Galati, Eduardo Garc\'ia-Valdecasas, Francesco Mignosa, Pavel Putrov, Rajath Radhakrishnan, Giovanni Rizi, Diego Rodriguez-Gomez, Matteo Sacchi, Andrea Sangiovanni, Luigi Tizzano, and Roberto Valandro for comments and discussions. We are especially grateful to Riccardo Argurio for valuable feedback on the draft. All the authors acknowledge support by INFN Iniziativa Specifica ST\&FI. M.B. is also supported by MIUR PRIN Grant 2020KR4KN2 “String Theory as a bridge between Gauge Theories and Quantum Gravity”. P.N. is also supported by the ERC-COG grant NP-QFT No.~864583 “Non-perturbative dynamics of quantum fields: from new deconfined phases of matter to quantum black holes” and by the MUR-FARE2020 grant No.~R20E8NR3HX “The Emergence of Quantum Gravity from Strong Coupling Dynamics”.

\appendix
\section{Anomalous 3-group for \texorpdfstring{$SU(N)/\mathbb{Z}_k$}{SU(N)/Zk}}\label{appendix-A}
In this appendix we provide details about the results presented in section \ref{subsection:SU(N)/Zk} for the $SU(N)/\mathbb{Z}_k$ gauge theory. First, we prove that the decomposition \eqref{eq:decexp} and the relation \eqref{eq:Bockback} hold. Then, we show that the expression \eqref{ZofSU(N)Zk_old} for the $SU(N)/\mathbb{Z}_k$ partition function is consistent with gauge transformations of the dynamical field $\mathscr{b}_2^k$, and that the symmetry structure of the $SU(N)/\mathbb{Z}_k$ gauge theory is that of an (anomalous) 3-group.
To simplify the discussion, we will assume no torsion cycles so that the definition of the square operation of $\mathbb{Z}_N$ gauge fields takes analogous expressions for $N$ even or odd. 

\subsection{\texorpdfstring{$\mathbb{Z}_N$}{ZN} gauge fields and group extensions}
Our task is to decompose $\mathcal{B}_2^N \in H^2(\mathcal{M}_5,\mathbb{Z}_N)$ in a $\mathbb{Z}_k^{(1)}$ and a $\mathbb{Z}_{N/k}^{(1)}$ part.
This cannot be done just at the level of the cohomology, but requires to uplift to the cochains. The short exact sequence \eqref{Zses}, which we repeat here,
\begin{equation}\label{Zses2}
    1\rightarrow \mathbb{Z}_k \xrightarrow{i}\mathbb{Z}_N\xrightarrow{\pi} \mathbb{Z}_{N/k} \rightarrow 1~,
\end{equation}
translates into a short exact sequence of cochain complexes,
\begin{displaymath}
\begin{tikzcd}
 &...\arrow[d, "\delta"]&...\arrow[d, "\delta"]&...\arrow[d, "\delta"]&\\
1 \arrow[r]  & C^n(\mathcal{M}_5,\mathbb{Z}_k)\arrow[r, "i"]\arrow[d, "\delta"] & C^{n}(\mathcal{M}_5,\mathbb{Z}_N)\arrow[r, "\pi"] \arrow[d, "\delta"] & C^{n}(\mathcal{M}_5,\mathbb{Z}_{N/k})\arrow[d, "\delta"]\arrow[r]&1\\
1 \arrow[r] & C^{n+1}(\mathcal{M}_5,\mathbb{Z}_k)\arrow[r,"i"]\arrow[d, "\delta"] & C^{n+1}(\mathcal{M}_5,\mathbb{Z}_N) \arrow[r,"\pi"]\arrow[d, "\delta"] &C^{n+1}(\mathcal{M}_5,\mathbb{Z}_{N/k})\arrow[r]\arrow[d, "\delta"]&1\\
 &...&...&...&~.
\end{tikzcd}
\end{displaymath}
Given an element $\mathcal{B}^N_2$ of $C^2(\mathcal{M}_5,\mathbb{Z}_N)$, denote with $\mathcal{B}^{N/k}_2 = \pi(\mathcal{B}^N_2)$ its image under the projection $\pi$. To obtain the decomposition, we need to make a choice of a pre-image of $\pi$ for any cochain in $C^2(\mathcal{M}_5,\mathbb{Z}_{N/k})$. Denoting with $\widetilde{\mathcal{B}}^{N/k}_2$ the fixed choice of pre-image for $\mathcal{B}^{N/k}_2$, we have $\pi(\widetilde{\mathcal{B}}^{N/k}_2) = \pi(\mathcal{B}^N_2) =\mathcal{B}^{N/k}_2$ and therefore $\widetilde{\mathcal{B}}^{N/k}_2-\mathcal{B}^N_2 \in \mathrm{Ker}\,\pi = \mathrm{Im}\,i$. As a result
\begin{equation}\label{eq:decgen}
\mathcal{B}^N_2 = i(\mathcal{B}^{k}_2) + \widetilde{\mathcal{B}}^{N/k}_2~,
\end{equation}
for a certain $\mathcal{B}^{k}_2\in C^2(\mathcal{M}_5,\mathbb{Z}_{k})$, that is fixed once the pre-image $\widetilde{\mathcal{B}}^{N/k}_2$ is. A convenient explicit choice for the pre-image, that we will use below, consists in uplifting $\mathcal{B}_2^{N/k}$ to $\mathbb{Z}_N$ by assigning values between $0$ and $N/k-1$ mod $N$. This leads to the decomposition
\begin{equation}\label{eq:decexp2}
\mathcal{B}^N_2 = \frac{N}{k}\widehat{\mathcal{B}}^{k}_2 + \widehat{\mathcal{B}}_2^{N/k}~,
\end{equation}
where the hats denote the uplift to $\mathbb{Z}_N$ for $\mathcal{B}^{k}_2$ and $\mathcal{B}_2^{N/k}$. As explained in section \ref{subsection:SU(N)/Zk}, they are uplifted by taking the value mod $N$ between $0$ and $k-1$ and between $0$ and $N/k-1$, respectively.\footnote{Making an explicit choice for the pre-image is equivalent to making an explicit choice for how to decompose an element $g^N$ of $\mathbb{Z}_N$ into pairs $(g^{k},g^{N/k})$ of elements of $\mathbb{Z}_k$ and $\mathbb{Z}_{N/k}$. Since the exact sequence \eqref{Zses2} does not split, the extension is non-trivial and it is labeled by a non-trivial class in group cohomology $H^2(\mathbb{Z}_{N/k},\mathbb{Z}_k)$. Choosing a particular decomposition is equivalent to picking a representative of this cohomology class.} This proves eq.~\eqref{eq:decexp}.

Going back to the cohomology, the short exact sequence above induces the long exact sequence in cohomology
\begin{equation}\label{Zles}
    ...\rightarrow H^n(\mathcal{M}_5,\mathbb{Z}_k)\xrightarrow{i^*} H^n(\mathcal{M}_5,\mathbb{Z}_N)\xrightarrow{\pi ^*} H^n(\mathcal{M}_5,\mathbb{Z}_{N/k}) \xrightarrow{\text{Bock}} H^{n+1}(\mathcal{M}_5,\mathbb{Z}_k) \rightarrow ... \ ,
\end{equation}
where $i^*$ and $\pi ^*$ are the maps in cohomology induced by $i$ and $\pi$, and $\text{Bock}$ is the Bockstein homomorphism. In eq.~\eqref{eq:decgen} we now take $\mathcal{B}^N_2$ to be the representative of a cohomology class in $H^2(\mathcal{M}_5,\mathbb{Z}_N)$, and $\mathcal{B}^{N/k}_2 = \pi(\mathcal{B}^N_2)=\pi(\widetilde{\mathcal{B}}^{N/k}_2)$ to be the representative of $\pi^*$ of that class in $H^2(\mathcal{M}_5,\mathbb{Z}_{N/k})$. Applying the discrete differential $\delta$ to \eqref{eq:decgen} and using that $\delta\mathcal{B}^N_2 = 0$ we obtain
\begin{equation}\label{eq:decgen-2}
i(\delta\mathcal{B}^{k}_2) = -\delta \widetilde{\mathcal{B}}^{N/k}_2~.
\end{equation}
By definition of the Bockstein homomorphism this means that\footnote{Here, with a slight abuse of notation, we are using the same symbol to denote the representative and the associated cohomology class. Note that when $\text{Bock}$ is a non-trivial class necessarily $\mathcal{B}^{k}_2$ is not globally well-defined as a cochain. We discuss the redundancy on $\mathcal{B}^{k}_2$ in equations \eqref{eq:idnonlift}-\eqref{eq:idlifts} below.}
\begin{equation}\label{eq:Bockback2}
\delta\mathcal{B}^{k}_2 = -\mathrm{Bock}(\mathcal{B}^{N/k}_2)~.
\end{equation}
This proves eq.~\eqref{eq:Bockback}. 

\subsection{Gauge transformations}

In order to check the behavior of \eqref{ZofSU(N)Zk_old} under gauge transformations we need to discuss the redundancy affecting the decomposition \eqref{eq:decgen}-\eqref{eq:decexp2}. The most general transformation that acts on $\widetilde{\mathcal{B}}^{N/k}_2$ and $\mathcal{B}_2^k$ leaving \eqref{eq:decgen} invariant is 
\begin{align}\label{eq:idnonlift}
\begin{split}
\mathcal{B}_2^k & \to \mathcal{B}_2^k + \zeta^k_2~,\\
\widetilde{\mathcal{B}}^{N/k}_2 & \to \widetilde{\mathcal{B}}^{N/k}_2 + \zeta^{N/k}_2~,
\end{split}
\end{align}
with
\begin{equation}\label{eq:gaugepar}
i(\zeta^k_2) + \zeta^{N/k}_2 = \delta\lambda_1^N~,
\end{equation}
where on the right-hand side we allowed for a gauge transformation on the parent background gauge field $\mathcal{B}_2^N \to \mathcal{B}_2^N + \delta \lambda_1^N$. For the explicit choice \eqref{eq:decexp2} of $\mathbb{Z}_N$ uplifts the identifications \eqref{eq:idnonlift} become 
\begin{align}\label{eq:idlifts}
\begin{split}
\widehat{\mathcal{B}}^k_2 & \to \widehat{\mathcal{B}}^k_2+ \widehat{\zeta}^k_2 \\
\widehat{\mathcal{B}}^{N/k}_2 & \to \widehat{\mathcal{B}}^{N/k}_2 + \widehat{\zeta}^{N/k}_2 ~,
\end{split}
\end{align}
with
\begin{equation}\label{eq:gtdec}
\widehat{\zeta}^{N/k}_2 = - \frac{N}{k}\widehat{\zeta}_2^k + \delta \lambda^N_1~. 
\end{equation}
Here $\widehat{\zeta}^k_2$ and $\widehat{\zeta}^{N/k}_2$ are cochains in $C^2(\mathcal{M}_5, \mathbb{Z}_N)$, expressing the ambiguity in the decomposition. Note that only a subset of these transformations can be interpreted as background gauge transformations acting on the background gauge field $\mathcal{B}_2^k$, namely
\begin{align}
\begin{split}\label{eq:Z2kgt}
\text{$\mathbb{Z}_k^{(1)}$ gauge transformations:}& \quad\zeta^k_2 = \delta\lambda_1^k~,~~\zeta^{N/k}_2 = 0~,~~\lambda_1^N = i(\lambda_1^k) \\
{\text{i.e.}}&\quad \widehat{\zeta}^k_2 = \delta\widehat{\lambda}_1^k~,~~\widehat{\zeta}^{N/k}_2 = 0~,~~  \lambda^N_1 = \frac{N}{k} \widehat{\lambda}_1^k~.
\end{split}
\end{align}
On the other hand, a generic $\zeta^{N/k}_2$ when projected gives
\begin{equation}
\pi(\zeta^{N/k}_2) = \delta(\pi(\lambda_1^N))~.
\end{equation}
This can be interpreted as a $\mathbb{Z}^{(1)}_{N/k}$ background gauge transformation for $\mathcal{B}^{N/k} =\pi(\widetilde{\mathcal{B}}^{N/k}) = \pi(\mathcal{B}^N)$. The gauge parameter $\pi(\lambda_1^N)$ is not affected by a shift of $\lambda_1^N$ by $i(\lambda_1^k)$, expressing that the $\mathbb{Z}^{(1)}_{N/k}$ transformations are only defined up to $\mathbb{Z}^{(1)}_k$ ones.  For the $\mathbb{Z}_N$ uplift $\widehat{\mathcal{B}}^{N/k}_2$ only the special case $\widehat{\zeta}^{N/k}_2 = \delta \widehat{\lambda}^{N/k}_1$ corresponds to a gauge transformation, with $\widehat{\lambda}^{N/k}_1 = \lambda_1^N - \frac{N}{k}\widehat{\lambda}^k_1$, while the more general transformation also includes a change of uplift. 

\subsection{Gauge invariance}

Given the anomaly inflow of the $SU(N)$ gauge theory \eqref{BGT-anomaly}, we fix a choice of local counterterms such that the $SU(N)$ partition function picks the phase
\begin{align}
\begin{split}
    & \frac{Z_{SU(N)}[\mathcal{B}_2+\delta \lambda_1,A_1+d\lambda_0]}{Z_{SU(N)}[\mathcal{B}_2,A_1]} \\ 
    & \hspace{25mm} = \exp\left(-\frac{2\pi i}{2N}\int_{\mathcal{Y}_6} \left(\mathcal{B}_2\cup \delta \lambda_1 + \delta \lambda_1 \cup \mathcal{B}_2 + \delta \lambda_1 \cup \delta \lambda_1\right) \cup \left[\frac{dA_1}{2\pi}\right]_N\right)\ , 
\end{split}
\end{align}
where the combination $\mathcal{B}_2\cup \delta \lambda_1 + \delta \lambda_1 \cup \mathcal{B}_2$ gives an even integer, being the difference of even integers. We derive the five-dimensional expression\footnote{Here and in the rest of the calculation we use the Leibniz rule for higher cup products \cite{Steenrod:1947}, which reads
\begin{equation}\label{Leibniz_cup}
    \delta(f\cup_i g)=\delta f \cup_i g + (-1)^p f\cup_i \delta g + (-1)^{p+q-i}f \cup_{i-1} g + (-1)^{pq +p+q} g \cup_{i-1} f \ ,
\end{equation}
where $f$ is a $p$-cochain and $g$ is a $q$-cochain.}
\begin{equation}
\begin{split}\label{eq:phaseptfn}
    & \frac{Z_{SU(N)}[\mathcal{B}_2+\delta \lambda_1,A_1+d\lambda_0]}{Z_{SU(N)}[\mathcal{B}_2,A_1]}\\
    &\hspace{25mm}= \exp\left(-\frac{2\pi i}{2N}\int_{\mathcal{Y}_6}\left(2 \delta \lambda_1 \cup \mathcal{B}_2 - \delta( \mathcal{B}_2 \cup_1 \delta \lambda_1) + \delta \lambda_1 \cup \delta \lambda_1\right) \cup \left[\frac{dA_1}{2\pi}\right]_N \right)\\
    &\hspace{25mm}=\exp\left(-\frac{2\pi i}{2N}\int_{\mathcal{M}_5}\left(2  \lambda_1 \cup \mathcal{B}_2 - ( \mathcal{B}_2 \cup_1 \delta \lambda_1) + \lambda_1 \cup \delta \lambda_1\right) \cup \left[\frac{dA_1}{2\pi}\right]_N \right) \ ,
\end{split}
\end{equation}
where, for the same reason as above, the term $ \mathcal{B}_2 \cup_1 \delta \lambda_1$ also gives an even integer. Notice that the parentheses in the above expression are necessary since higher cup products are not associative.

We first show that the proposed expression for the partition function of $SU(N)/\mathbb{Z}_k$ coupled to background gauge fields, eq.~\eqref{ZofSU(N)Zk_old}, is gauge invariant under $\mathscr{b_2^k} \mapsto \mathscr{b_2^k} + \delta \lambda_1^k$. Following \eqref{eq:Z2kgt} we plug $\lambda_1=\frac{N}{k}\widehat{\lambda}_1^k$ in \eqref{eq:phaseptfn}. Using also that the fluxes of $A_1$ are integer multiples of $k/L$, some of the terms in the variation are found to give integers and therefore trivialize. The remaining terms are
\begin{equation}
    \exp\left(-\frac{2\pi i}{k}\int_{\mathcal{M}_5}  \widehat{\lambda}_1^k\cup \widehat{\mathcal{B}}_2^{N/k} \cup \left[\frac{dA_1}{2\pi}\right]_N+\frac{2\pi i}{2k}\int_{\mathcal{M}_5}(\widehat{\mathcal{B}}_2^{N/k}\cup_1 \delta\widehat{\lambda}_1^k) \cup \left[\frac{dA_1}{2\pi}\right]_N\right) \ .
\end{equation}
This is exactly canceled by the variation of the action that couples $\mathscr{b}_2^k$ to the background gauge fields,
\begin{equation}
    \exp\left(\frac{2\pi i}{k}\int_{\mathcal{M}_5} \delta \lambda_1^k \cup \mathcal{C}_3 - \frac{2\pi i}{2k}\int_{\mathcal{M}_5}(\widehat{\mathcal{B}}_2^{N/k}\cup_1 \delta\widehat{\lambda}_1^k)\cup\left[\frac{dA_1}{2\pi}\right]_N \right) \ ,
\end{equation}
provided the modified constraint \eqref{H4_SU(N)/Zk} on $\mathcal{C}_3$. This modified constraint is invariant under $\mathbb{Z}_{N/k}^{(1)}$ transformations if they also act on $\mathcal{C}_3$ as follows,
\begin{equation}\label{eq:3group}
\mathcal{C}_3 \to \mathcal{C}_3 + \delta \lambda_2 + \left[\lambda^N_1\cup\left[\frac{d A_1}{2\pi}\right]_N \right]_k~.
\end{equation}
We will see below that this modified gauge transformation is also required to derive the anomaly. As we said in the main text, this gauge transformation signals a 3-group structure.

\subsection{Anomaly-inflow action}
We now want to look at the effect of gauge transformations of the background gauge fields $\mathcal{B}_2^{N/k}$ and $\mathcal{C}_3$. Let us start from $\mathbb{Z}_{N/k}^{(1)}$ transformations. They act not only on $\mathcal{B}_2^{N/k}$ but also on $\mathcal{C}_3$ according  to \eqref{eq:3group}. The phase \eqref{eq:phaseptfn} now produces both $c$-numbers and operator-valued terms
\begin{equation}\label{Zvariation_N/k}
\begin{split}
    &\exp\left(-\frac{2\pi i }{N}\int_{\mathcal{M}_5}\widehat{\lambda}_1^{N/k}\cup \left( \frac{N}{k} \widehat{\mathscr{b}}_2^k+\widehat{\mathcal{B}}_2^{N/k}\right) \cup \left[\frac{dA_1}{2\pi}\right]_N\right) \times\\
    &\exp\left(\frac{2\pi i}{2N}\int_{\mathcal{M}_5}\left( \left(\left( \frac{N}{k} \widehat{\mathscr{b}}_2^k+\widehat{\mathcal{B}}_2^{N/k}\right) \cup_1 \delta\widehat{ \lambda}_1^{N/k} \right)-\widehat{ \lambda}_1^{N/k}\cup \delta \widehat{ \lambda}_1^{N/k}\right)\cup \left[\frac{dA_1}{2\pi}\right]_N\right) \ ,
\end{split}
\end{equation}
while the coupling to $\mathscr{b}_2^k$ gives
\begin{equation}
    \exp\left(\frac{2\pi i}{k}\int_{\mathcal{M}_5}\widehat{\mathscr{b}}_2^k\cup \widehat{\lambda}_1^{N/k} \cup \left[\frac{dA_1}{2\pi}\right]_N-\frac{2\pi i}{2k}\int_{\mathcal{M}_5}(\delta \widehat{\lambda}_1^{N/k}\cup_1  \widehat{\mathscr{b}}_2^k)\cup \left[\frac{dA_1}{2\pi}\right]_N\right) \ .
\end{equation}
Grouping together all operator-valued variations, and using  \eqref{Leibniz_cup}, we can see that they combine into pure $c$-numbers
\begin{equation}\label{Zvariation_N/k:operators}
    \begin{split}
        &\exp\left(\frac{2\pi i}{k}\int_{\mathcal{M}_5}(\widehat{\mathscr{b}}_2^k\cup \widehat{\lambda}_1^{N/k}-\widehat{\lambda}_1^{N/k}\cup\widehat{\mathscr{b}}_2^k)\cup\left[\frac{dA_1}{2\pi}\right]_N\right)\times \\&\hspace{3cm}\exp\left(\frac{2\pi i}{2k}\int_{\mathcal{M}_5}(\widehat{\mathscr{b}}_2^k\cup_1 \delta \widehat{\lambda}_1^{N/k} -\delta \widehat{\lambda}_1^{N/k}\cup_1  \widehat{\mathscr{b}}_2^k)\cup \left[\frac{dA_1}{2\pi}\right]_N\right)\\
        =&\exp\left(\frac{2\pi i}{k}\int_{\mathcal{M}_5}(\widehat{\lambda}_1^{N/k}\cup_1 \mathrm{Bock}(\mathcal{B}_2^{N/k}))\cup\left[\frac{dA_1}{2\pi}\right]_N \right.\\
        & \hspace{3cm}\left. +\frac{2\pi i}{2k}\int_{\mathcal{M}_5}(\mathrm{Bock}(\mathcal{B}_2^{N/k})\cup_2 \delta \widehat{\lambda}_1^{N/k})\cup\left[\frac{dA_1}{2\pi}\right]_N\right).
    \end{split}
\end{equation}
Using \eqref{eq:Bockbacklift} we can combine the $c$-number phase from the first line in \eqref{Zvariation_N/k} with the second term in \eqref{Zvariation_N/k:operators} giving
\begin{equation}
\begin{split}
       &\exp\left(\frac{2\pi i}{2N}\int_{\mathcal{M}_5}(\delta\widehat{\mathcal{B}}_2^{N/k}\cup_2 \delta \widehat{\lambda}_1^{N/k}+\widehat{\mathcal{B}}_2^{N/k}\cup_1 \delta \widehat{\lambda}_1^{N/k})\cup\left[\frac{dA_1}{2\pi}\right]_N\right) \\
       =&\exp\left(-\frac{2\pi i}{2N}\int_{\mathcal{M}_5}(\delta \widehat{\lambda}_1^{N/k} \cup_1 \widehat{\mathcal{B}}_2^{N/k})\cup\left[\frac{dA_1}{2\pi}\right]_N\right) \ .
\end{split}
\end{equation}
We can rewrite the phase linear in $\lambda_1^{N/k}$ in a convenient form as 
\begin{equation}
    \begin{split}
        &\exp\left(\frac{2\pi i }{N}\int_{\mathcal{M}_5}\left((\widehat{\lambda}_1^{N/k}\cup_1 \delta\widehat{\mathcal{B}}_2^{N/k})-\frac{1}{2}(\delta \widehat{\lambda}_1^{N/k} \cup_1 \widehat{\mathcal{B}}_2^{N/k})-(\widehat{\lambda}_1^{N/k}\cup\widehat{\mathcal{B}}_2^{N/k})\right)\cup\left[\frac{dA_1}{2\pi}\right]_N\right)\\
        =&\exp\left(-\frac{2\pi i}{N}\int_{\mathcal{M}_5}\widehat{\mathcal{B}}_2^{N/k}\cup\widehat{\lambda}_1^{N/k}\cup\left[\frac{dA_1}{2\pi}\right]_N
        +\frac{2\pi i}{2N}\int_{\mathcal{M}_5}(\delta \widehat{\lambda}_1^{N/k} \cup_1 \widehat{\mathcal{B}}_2^{N/k})\cup\left[\frac{dA_1}{2\pi}\right]_N\right) \ .
    \end{split}
\end{equation}
So, the total variation, including the quadratic term in $\lambda_1^{N/k}$, is
\begin{equation}
    \begin{split}
        &\exp\left(-\frac{2\pi i}{N}\int_{\mathcal{M}_5}\widehat{\mathcal{B}}_2^{N/k}\cup\widehat{\lambda}_1^{N/k}\cup\left[\frac{dA_1}{2\pi}\right]_N
        +\frac{2\pi i}{2N}\int_{\mathcal{M}_5}(\delta \widehat{\lambda}_1^{N/k} \cup_1 \widehat{\mathcal{B}}_2^{N/k})\cup\left[\frac{dA_1}{2\pi}\right]_N\right) \times\\
        &\exp\left(\frac{2\pi i}{2N}\int_{\mathcal{M}_5}\widehat{\lambda}_1^{N/k} \cup \delta \widehat{\lambda}_1^{N/k} \cup \left[\frac{dA_1}{2\pi}\right]_N \right) \ .
    \end{split}
\end{equation}
The phase produced by $\mathbb{Z}_k^{(2)}$ transformations, that only act on $\mathcal{C}_3$, is instead given by
\begin{equation}
    \begin{split}
        \exp\left(\frac{2\pi i}{k}\int_{\mathcal{M}_5} \mathscr{b}_2^k \cup \delta \lambda_2\right)=\exp\left(\frac{2\pi i}{k}\int_{\mathcal{M}_5} \mathrm{Bock}(\mathcal{B}_2^{N/k}) \cup \lambda_2\right) \ .
    \end{split}
\end{equation}
Using the anomaly inflow \eqref{SU/N)/Zk_anomaly_old} one can now easily check that $Z_{SU(N)/\mathbb{Z}_k}\mathcal{A}_6^{SU(N)/\mathbb{Z}_k}$ is gauge invariant under background gauge transformations, as expected. Assuming that there are no torsion cycles in the homology with integer coefficient, both for $N/k$ even or odd we can express the affine Pontryagin square as
\begin{equation}
\mathcal{P}(\widehat{\mathcal{B}}_2^{N/k})= \widehat{\mathcal{B}}_2^{N/k} \cup \widehat{\mathcal{B}}_2^{N/k}-\widehat{\mathcal{B}}_2^{N/k} \cup_1 \delta \widehat{\mathcal{B}}_2^{N/k}~.
\end{equation}

Finally, we check that the anomaly-inflow action \eqref{SU/N)/Zk_anomaly_old} is a closed cochain when extended to a $7d$ manifold \cite{Benini:2018reh}. Indeed we have
\begin{equation}
    \begin{split}
        \delta & \left(\frac{2\pi i}{N}\right.  \left.\mathcal{P}(\widehat{\mathcal{B}}_2^{N/k})\cup \left[\frac{dA_1}{2\pi}\right]_N +\frac{2\pi i }{k}\mathrm{Bock}(\mathcal{B}_2^{N/k}) \cup \mathcal{C}_3\right)\\
         = &\,\frac{2\pi i }{2N}\left(\delta\widehat{\mathcal{B}}_2^{N/k} \cup \widehat{\mathcal{B}}_2^{N/k}+\widehat{\mathcal{B}}_2^{N/k} \cup\delta \widehat{\mathcal{B}}_2^{N/k}  -\delta(\widehat{\mathcal{B}}_2^{N/k} \cup_1 \delta \widehat{\mathcal{B}}_2^{N/k})\right)\cup \left[\frac{dA_1}{2\pi}\right]_N \\ & -\frac{2\pi i}{k}\mathrm{Bock}(\mathcal{B}_2^{N/k})\cup \delta \mathcal{C}_3\\
         = &\,\frac{2\pi i N/k }{k}\left(\mathrm{Bock}(\mathcal{B}_2^{N/k})\cup_1 \mathrm{Bock}(\mathcal{B}_2^{N/k})\right)\cup \left[\frac{dA_1}{2\pi}\right]_N \ ,
    \end{split}
\end{equation}
which is trivial because the fluxes of $A_1$ are multiples of $k/L$ and $\frac{N/k}{L}$ is an integer. 


\bibliography{refs}
\bibliographystyle{JHEP}

\end{document}